\documentclass[12pt]{article}
\usepackage{amsmath,amssymb,amsthm,amsxtra,overpic,bbm,bm,epsfig,ulem}
\usepackage{color}
\usepackage{cite}
\usepackage{multirow}
\usepackage{makecell}

\usepackage{hyperref}
\usepackage{url}

\usepackage[title]{appendix}

\def\thefootnote{\fnsymbol{footnote}}

\def\thefootnote{\fnsymbol{footnote}}

\begin{document}

\vspace{0.2cm}

\begin{center}
	{\large\bf Integral solutions to the one-loop renormalization-group equations for lepton flavor mixing parameters and the Jarlskog invariant}
\end{center}

\vspace{0.2cm}

\begin{center}
	{\small \bf Di Zhang$^{a,b}$}\footnote{Email: zhangdi@ihep.ac.cn} \\
	{$^a$Institute of High Energy Physics, Chinese Academy of Sciences, Beijing 100049, China \\
	$^b$School of Physical Sciences, University of Chinese Academy of Sciences, Beijing 100049, China}
\end{center}

\vspace{1.5cm}

\begin{abstract}
	Working in the basis where the charged-lepton Yukawa matrix is diagonal and making the $\tau$-dominance approximations, we analytically derive integral solutions to the one-loop renormalization-group equations (RGEs) for neutrino masses, flavor mixing angles, CP-violating phases and the Jarlskog invariant under the standard parametrization of the PMNS matrix in the standard model or its minimal supersymmetric extension for both Majorana and Dirac neutrinos. With these integral solutions, we carry out numerical calculations to investigate the RGE running of lepton flavor mixing parameters and the Jarlskog invariant, and also compare these integral solutions with the exact results obtained by numerically solving the one-loop RGEs. It is shown that these integral solutions coincide with the exact results and can well describe the evolution of lepton flavor mixing parameters and the Jarlskog invariant in most cases. Some important features of our integral solutions and the evolution behaviors of relevant flavor parameters are also discussed in detail both analytically and numerically.
\end{abstract}

\newpage

\def\thefootnote{\arabic{footnote}}
\setcounter{footnote}{0}
\section{Introduction}

In the last two decades, compelling evidences obtained from a number of successful neutrino oscillation experiments have proved that neutrinos are massive and lepton flavor mixing exists~\cite{Tanabashi:2018oca}, and this demonstrates that the standard model (SM) of particle physics is incomplete. In order to understand the origin of small neutrino masses and the observed lepton flavor mixing pattern, many models with extra flavor symmetries (see reviews~\cite{Altarelli:2010gt,Ishimori:2010au,King:2013eh,Petcov:2017ggy,Xing:2019vks} and references therein) or some new degrees of freedom (e.g., the right-handed neutrino fields in the type-I seesaw mechanism~\cite{Minkowski:1977sc,Yanagida:1979as,GellMann:1980vs,Glashow:1979nm,Mohapatra:1979ia}) have been put forward at superhigh energy scales, as well as many phenomenological textures of lepton mass matrices (e.g., texture zeros of the neutrino mass matrix~\cite{Xing:2019vks,Frampton:2002yf,Xing:2002ta,Xing:2002ap}). With the help of the renormalization-group equations (RGEs), one can confront phenomenological consequences of those models or textures with current experimental data at the low energy scale. Thus, it is very important and useful to investigate the evolution of relevant flavor parameters or the stability of some specific textures against the energy scale by means of the RGEs, especially in the cases where nearly degenerate neutrino masses or large $\tan\beta$ in the minimal supersymmetric standard model (MSSM) is taken into account, so as to establish some correlations between physical phenomena at high and low energy scales and reveal some underlying structures of lepton mass matrices or flavor mixing pattern which are instructive for model building.

So far, whether neutrinos are Dirac or Majorana particles (which is referred to as the Dirac case or the Majorana case) is still an open question. If the neutrinoless double-beta ($0\nu\beta\beta$) decay is observed, we shall conclude that neutrinos are Majorana particles thanks to the Schechter-Valle theorem~\cite{Schechter:1981bd}. However, we can not claim that neutrinos are Dirac particles even though the $0\nu\beta\beta$ decay is not observed in experiments\cite{Xing:2019vks,Xing:2003jf,Xing:2003ez}. There still exist some rooms and interesting models for Dirac neutrinos, and it is worth considering these two possibilities before the nature of massive neutrinos is convincingly determined by future experiments. In the Majorana case, the small neutrino masses can be generated by the unique dimension-5 Weinberg operator~\cite{Weinberg:1979sa} in an effective field theory. This operator can be obtained by integrating out heavy degrees of freedom in some extended models~\cite{Xing:2011zza}, such as the type-I seesaw mechanism. Then the one-loop RGE for the effective coupling matrix $\kappa$ of Majorana neutrinos is given by~\cite{Chankowski:1993tx,Babu:1993qv,Antusch:2001ck,Antusch:2001vn}
\begin{eqnarray}
16 \pi^2 \frac{{\rm d} \kappa}{{\rm d}t} = \alpha^{}_\kappa \kappa + C \left[ \left(Y^{}_l Y^\dagger_l \right) \kappa + \kappa \left( Y^{}_l Y^\dagger_l \right)^T \right] \;,
%	(1)
\end{eqnarray}
where $t \equiv \ln \left( \mu/\Lambda^{}_{\rm EW} \right)$ with $\mu$ being an arbitrary renormalization scale between the electroweak scale $\Lambda^{}_{\rm EW} \sim 100$ GeV and the cutoff scale $\Lambda$, $Y^{}_l$ is the charged-lepton Yukawa coupling matrix, and 
\begin{eqnarray}
C = \left\{\begin{array}{cl} -3/2 \;, & {\rm in~SM}  \\ 1 \;, & {\rm in~MSSM} \end{array} \right. \;, \quad 
\alpha^{}_\kappa \simeq \left\{ \begin{array}{cl}
-3g^2_2 + 6y^2_t + \lambda \;, & {\rm in~SM}  \\ -6g^2_1/5 -6g^2_2 + 6y^2_t \;, & {\rm in~MSSM} \end{array}\right. 
%	(2)
\end{eqnarray}	 
with $g^{}_{1,2}$, $y^{}_t$ and $\lambda$ being the gauge couplings, the top-quark Yukawa coupling and the Higgs self-coupling constant respectively. After spontaneous gauge symmetry breaking, the effective Majorana neutrino mass matrix is given by $M^{}_\nu = \kappa v^2$ (SM) or $M^{}_\nu = \kappa v^2 \tan^2\beta/(1 + \tan^2\beta)$ (MSSM) with $v \simeq 174$ GeV being the vacuum expectation value of the SM Higgs field and $\tan\beta$ denoting the ratio of the vacuum expectation values of two Higgs doublets in the MSSM. In the Dirac case, the one-loop RGE for the Yukawa coupling matrix $Y^{}_\nu$ of Dirac neutrinos is~\cite{Cheng:1973nv,Machacek:1983fi,Grzadkowski:1987tf,Lindner:2005as}
\begin{eqnarray}
16\pi^2  \frac{{\rm d} Y^{}_\nu}{ {\rm d}t } = \left[ \alpha^{}_\nu + C^\prime \left( Y^{}_\nu Y^\dagger_\nu \right) + C \left( Y^{}_l Y^\dagger_l \right) \right] Y^{}_\nu \;,
%	(3)
\end{eqnarray}
where $C$ has been given in Eq.~(2), and 
\begin{eqnarray}
C^\prime = \left\{\begin{array}{cl} 3/2 \;, & {\rm in~SM}  \\ 3 \;, & {\rm in~MSSM} \end{array} \right. \;, \quad 
\alpha^{}_\nu \simeq \left\{ \begin{array}{cl}
-9g^2_1/20 - 9g^2_2/4 + 3y^2_t \;, & {\rm in~SM}  \\ -3g^2_1/5 - 3g^2_2 + 3y^2_t \;, & {\rm in~MSSM} \end{array}\right. \;.
%	(4)
\end{eqnarray}
Note that in this case, the Yukawa coupling matrix $Y^{}_\nu$ must be much smaller so as to be accordant with the smallness of neutrino masses. Thus $Y^{}_\nu \ll Y^{}_l$ holds pretty well and it is quite safe to ignore $Y^{}_\nu Y^\dagger_\nu$ in Eq.~(3). The Dirac neutrino mass matrix is given by $M^{}_\nu = Y^{}_\nu v$ (SM) or $M^{}_\nu = Y^{}_\nu v \tan\beta/\sqrt{1+\tan^2\beta}$ (MSSM). In the basis where $Y^{}_l$ is diagonal, namely $Y^{}_l = {\rm Diag}\{ y^{}_e, y^{}_\mu, y^{}_\tau \}$, the neutrino mass matrix can be diagonalized by the Pontecorvo-Maki-Nakagawa-Sakata (PMNS) lepton flavor mixing matrix $U$~\cite{Pontecorvo:1957cp,Maki:1962mu,Pontecorvo:1967fh}, i.e., $U^\dagger M^{}_\nu U^\ast = D^{}_\nu \equiv {\rm Diag} \{m^{}_1, m^{}_2, m^{}_3 \}$ in the Majorana case or $U^\dagger H^{}_\nu U = D^2_\nu$ in the Dirac case with $m^{}_i$ (for $i=1,2,3$) being the neutrino masses and $H^{}_\nu$ being defined as $H^{}_\nu \equiv M^{}_\nu M^{\dagger}_\nu$. A popular parametrization of the PMNS matrix $U$ is given by~\cite{Tanabashi:2018oca}
\begin{eqnarray}
U = P^{}_l \left(\begin{matrix} c^{}_{12}c^{}_{13} & s^{}_{12}c^{}_{13} & s^{}_{13}e^{-{\rm i}\delta} \cr -s^{}_{12}c^{}_{23} - c^{}_{12}s^{}_{13} s^{}_{23}e^{{\rm i}\delta} & c^{}_{12}c^{}_{23} - s^{}_{12}s^{}_{13}s^{}_{23} e^{{\rm i}\delta} & c^{}_{13}s^{}_{23} \cr s^{}_{12}s^{}_{23} - c^{}_{12} s^{}_{13}c^{}_{23}e^{{\rm i}\delta} & -c^{}_{12}s^{}_{23} - s^{}_{12}s^{}_{13} c^{}_{23}e^{{i}\delta} & c^{}_{13}c^{}_{23} \end{matrix}\right) P^{}_\nu \;,
%	(5)
\end{eqnarray}
in which $c^{}_{ij} \equiv \cos \theta^{}_{ij}$ and $s^{}_{ij} \equiv \sin
\theta^{}_{ij}$ (for $ij=12,13,23$) with $\theta^{}_{ij}$ lying in the first
quadrant, $\delta$ is the so-called Dirac CP phase, $P^{}_l \equiv {\rm Diag}
\{ e^{{\rm i}\phi^{}_e}, e^{{\rm i}\phi^{}_\mu}, e^{{\rm i}\phi^{}_\tau} \}$
with $\phi^{}_e$, $\phi^{}_\mu$ and $\phi^{}_\tau$ being the unphysical phases
associated with the charged-lepton fields, and $P^{}_\nu \equiv {\rm Diag}\{ e^{{\rm i} \rho}, e^{{\rm i} \sigma}, 1\}$ with $\rho$ and $\sigma$ being the Majorana phases which become unphysical in the Dirac case.

Based on Eqs.~(1)---(4), the one-loop RGE running effects on neutrino masses and flavor mixing parameters have been extensively studied, which can be seen in the review~\cite{Ohlsson:2013xva} and references therein. With some specific parametrizations of the PMNS matrix $U$, the individual RGEs for neutrino masses, flavor mixing angles and CP-violating phases have been derived in Refs.~\cite{Lindner:2005as,Casas:1999tg,Antusch:2003kp,Mei:2003gn,Xing:2005fw} and threshold effects in seesaw models have also been discussed, such as those in the type-I seesaw mechanism~\cite{Antusch:2002rr,Mei:2004rn,Antusch:2005gp,Mei:2005qp}. Furthermore, the running effects on leptonic CP-violating phases have been studied in detail~\cite{Luo:2005sq,Xing:2006sp,Luo:2006tb,Luo:2012ce,Ohlsson:2012pg}, showing that the evolution of three CP-violating phases is entangled in the Majorana case so that the Dirac CP-violating phase can be radiatively generated even if it is initially assumed to be zero (in the Dirac case, the Dirac CP-violating phase keeps vanishing during the RGE evolution if it is initially zero). And in particular, some recent works~\cite{Xing:2020erm,Zhao:2020bzx,Xing:2020ghj} find that the RGE effects can play a significant role in establishing a direct link between the low energy CP violation and the CP-violating asymmetries at a superhigh energy scale, and making the leptogenesis mechanism~\cite{Fukugita:1986hr} work successfully.

In the present work, our main purpose is to analytically derive integral solutions to the one-loop RGEs for neutrino masses, flavor mixing angles, CP-violating phases and the Jarlskog invariant of $U$ both in the Majorana case and in the Dirac case, with the $\tau$-dominance approximations. Some previous attempts in this connection have been made to some extent~\cite{Bergstrom:2010id,Zhou:2014sya,Xing:2015fdg,Xing:2017mkx,Huan:2018lzd,Zhu:2018dvj,Huang:2020kgt,Xing:2020ezi}. But our work differs from them at least in the following aspects:
\begin{itemize}
	\item Ref.~\cite{Bergstrom:2010id} mainly focuses on the seesaw threshold effects in the low-scale seesaw model, and has only derived analytical results for neutrino masses and flavor mixing angles with a special parametrization of $U$ by assuming that CP is preserved. While in our work, we consider the most general case with the popular parametrization of $U$ given in Eq.~(5) below the cutoff scale, and the integral results for neutrino masses, flavor mixing angles, CP-violating phases and the Jarlskog invariant are exhaustively derived. In particular, the case for Dirac neutrinos is also taken into account in this work.
	\item Unlike those works done in Refs.~\cite{Zhou:2014sya,Xing:2015fdg,Xing:2017mkx,Huan:2018lzd,Huang:2020kgt,Xing:2020ezi} where some special situations are considered, such as the $\mu$-$\tau$ reflection symmetry leading to $\theta^{}_{23} = \pi/4$ and $\delta = \pm \pi/2$~\cite{Harrison:2002et,Xing:2015fdg} or the lightest neutrino being massless, our work is essentially independent of the specific textures of lepton Yukawa coupling matrices or flavor mixing patterns, and gives the most general results for relevant flavor parameters without further assumptions. Thus, the corresponding results in Refs.~\cite{Zhou:2014sya,Xing:2015fdg,Xing:2017mkx,Huan:2018lzd,Huang:2020kgt,Xing:2020ezi} can be easily reproduced from our results under some further constraints on flavor mixing parameters or neutrino masses.
	\item The general integral results for the Jarlskog invariant in the Majorana and Dirac cases have been acquired in Ref.~\cite{Zhu:2018dvj} without any specific parametrization of $U$. Instead, our work takes the widely used parametrization of $U$ given in Eq.~(5) so that one can clearly see properties of the evolution of the Jarlskog invariant from one scale to another, especially its dependence on flavor mixing angles and CP-violating phases. In the Majorana case, our result clearly shows that there is an additional term mainly dominated by two Majorana CP-violating phases, from which the Jarlskog invariant can be radiatively generated even if it is initially vanishing at a specific energy scale. This observation is very intuitive to understand CP violation in a long-baseline neutrino oscillation experiment, whose strength is uniquely governed by the Jarlskog invariant.
	\item Moreover, in our work, we discuss the evolution behaviors of lepton flavor parameters and the Jarlskog invariant both analytically and numerically in great detail by using the latest experimental data and global-fit inputs, including the T2K measurement of CP violation~\cite{Abe:2019vii}. This is also a merit of our work.
\end{itemize}
Compared with the differential forms of RGEs for lepton flavor parameters, the integral solutions can explicitly reveal their evolution behaviors, since they are only dominated by two quantities $I^{}_\beta$ (for $\beta = \kappa$ or $\nu$) and $\Delta^{}_\tau$ which are integrals of $\alpha^{}_\beta$ (for $\beta = \kappa$ or $\nu$) and $y^2_\tau$ to the energy scale and almost independent of the initial inputs. Therefore, given the values of these two quantities against the energy scale, we can easily obtain the RGE running of relevant flavor parameters with some inputs. That is why our integral solutions are expected to be very useful to study radiative corrections to some interesting flavor mixing patterns or textures of lepton mass matrices, shed light on some possible underlying flavor symmetries at a superhigh energy scale, and establish a direct link between physical phenomena at the low and high energy scales. 

The remainder of this paper is organized as follows. In section 2, the integral solutions to RGEs for lepton flavor mixing parameters and the Jarlskog invariant in the SM or MSSM for both Majorana and Dirac neutrinos are analytically derived. The numerical calculations are carried out to illustrate the evolution behaviors of relevant flavor parameters in section 3. We summarize our main results in section 4.

\section{Integral solutions to one-loop RGEs}
Working in the basis where $Y^{}_l$ is diagonal, we find that $Y^{}_l$ remains diagonal
%%%%%%%%%%%%%%%%%%%%%%%%%%%%%%%%%%% footnote 1 %%%%%%%%%%%%%%%%%%%%%%%%%%%%%%%%
\footnote{$Y^{}_l$ keeps diagonal strictly in the Majorana case but approximately in the Dirac case since in the Dirac case, the RGE of $Y^{}_l$ contains $Y^{}_\nu Y^\dagger_\nu$ which can make $Y^{}_l$ deviate from the diagonal form during the RGE running. Fortunately, due to $Y^{}_\nu \ll Y^{}_l$ in this case, the off-diagonal elements induced by RGE effects are much smaller than the diagonal ones of $Y^{}_l$, namely, $Y^{}_l$ approximately remains diagonal. The complete one-loop RGEs of Yukawa coupling matrices, gauge couplings and the Higgs self-coupling constant in the Majorana and Dirac cases can be found in the latest review~\cite{Xing:2019vks}.}
%%%%%%%%%%%%%%%%%%%%%%%%%%%%%%%%%%%%%%%%%%%%%%%%%%%%%%%%%%%%%%%%%%%%%%%%%%%%%%%
during the RGE running so that we can integrate Eq.~(1) or (3) from an arbitrary lower energy scale $\mu$ to the superhigh energy scale $\Lambda$ and obtain
\begin{eqnarray}
M^{}_\nu \left(\mu\right) = I^{}_\kappa  T^{}_l  M^{}_\nu \left(\Lambda\right) T^{}_l  \;,
%	(6)
\end{eqnarray}
or
\begin{eqnarray}
H^{}_\nu \left(\mu\right) = I^2_\nu  T^{}_l  H^{}_\nu \left(\Lambda\right) T^{}_l  \;,
%	(7)
\end{eqnarray}
corresponding to Majorana neutrinos or Dirac neutrinos, where  $T^{}_l = {\rm Diag} \{ I^{}_e, I^{}_\mu, I^{}_\tau \}$, and $I^{}_\beta$ (for $\beta = \kappa, \nu$) and $I^{}_\gamma$ (for $\gamma = e, \mu, \tau$)  are defined as
\begin{eqnarray}
I^{}_\beta \left( \mu \right) \hspace{-0.2cm}&=&\hspace{-0.2cm} \exp \left[ -\frac{1}{16\pi^2} \int^{\ln\left( \Lambda/\Lambda^{}_{\rm EW} \right)}_{\ln \left( \mu/\Lambda^{}_{\rm EW} \right)} \alpha^{}_\beta \left(t\right) {\rm d}t \right] \;,
\nonumber
\\
I^{}_\gamma \left( \mu \right) \hspace{-0.2cm}&=&\hspace{-0.2cm} \exp \left[ -\frac{C}{16\pi^2} \int^{\ln\left( \Lambda/\Lambda^{}_{\rm EW} \right)}_{\ln \left( \mu/\Lambda^{}_{\rm EW} \right)} y^2_\gamma \left(t\right) {\rm d}t \right] \;.
%	(8)
\end{eqnarray} 
Considering $y^2_e \ll y^2_\mu \ll y^2_\tau $ in the SM or MSSM together with our requirement of $\tan \beta \lesssim 30$ and the smallness of the loop factor $1/\left( 16\pi^2 \right)$, it is reasonable and safe to make the $\tau$-dominance approximations, i.e., $I^{}_e \simeq I^{}_\mu \simeq 1$ and $I^{}_\tau \simeq 1 + \Delta^{}_\tau $ with
%%%%%%%%%%%%%%%%%%%%%%%%%%%%%%%% figure 1 %%%%%%%%%%%%%%%%%%%%%%%%%%%%%%%%%%%%%%%
\begin{figure}[t!]
	\centering
	\includegraphics[width=\linewidth]{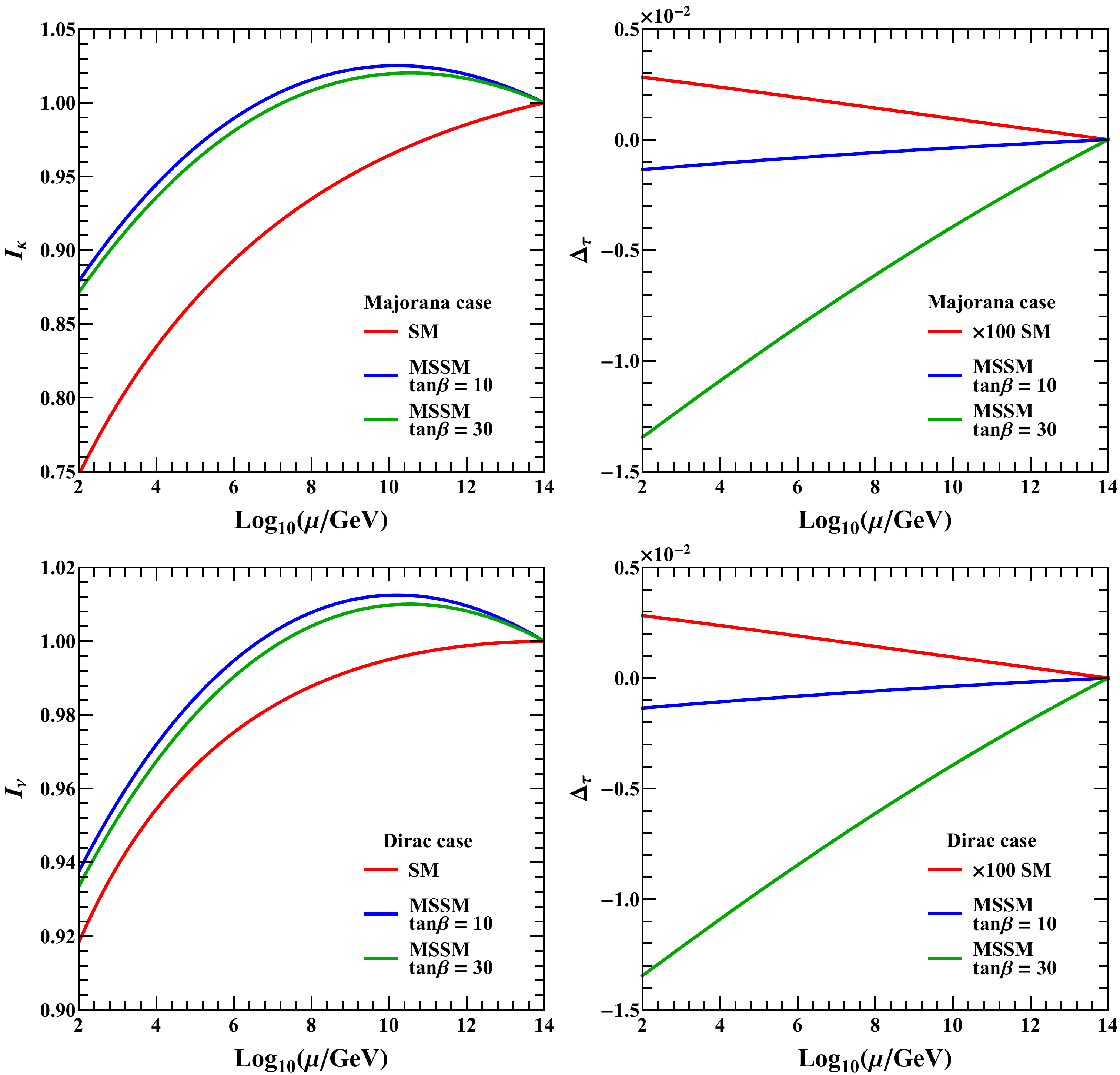}
	\caption{The values of $I^{}_\beta$ (for $\beta = \kappa$ or $\nu$) and $\Delta^{}_\tau$ against the energy scale $\mu$ in the Majorana case or the Dirac case with $\Lambda = 10^{14}$ GeV, where SM and MSSM with $\tan\beta = 10$ or $\tan\beta =30$ are considered. ``$\times 100$" in the right two panels means that the value of $\Delta^{}_\tau$ in the SM has been enhanced by a factor 100.}
	\label{loop-functions}
\end{figure}
%%%%%%%%%%%%%%%%%%%%%%%%%%%%%%%%%%%%%%%%%%%%%%%%%%%%%%%%%%%%%%%%%%%%%%%%%%%%%%%%%
\begin{eqnarray}
\Delta^{}_\tau \left( \mu \right) \simeq  -\frac{C}{16\pi^2} \int^{\ln\left( \Lambda/\Lambda^{}_{\rm EW} \right)}_{\ln \left( \mu/\Lambda^{}_{\rm EW} \right)} y^2_\tau \left(t\right) {\rm d}t
%	(9)
\end{eqnarray}
being a small quantity. Since the RGE evolution of gauge couplings, Yukawa coupling matrices of charged fermions and the Higgs self-coupling constant are essentially independent of $\kappa$ in the Majorana case and $Y^{}_\nu$ in the Dirac case, the values of $I^{}_\beta$ (for $\beta = \kappa, \nu$) and $\Delta^{}_\tau$ against the energy scale $\mu$ do not depend on initial inputs from the neutrino sector, as illustrated in Fig.~\ref{loop-functions}. As can be seen from Fig.~\ref{loop-functions}, the magnitude of $\Delta^{}_\tau$ is of $\mathcal{O} \left( 10^{-5} \right)$ at $\Lambda^{}_{\rm EW}$ in the SM, and it can be largely enhanced in the MSSM with a large value of $\tan\beta$. It can reach $\mathcal{O} \left( 10^{-3} \right)$ with $\tan\beta = 10$ or $\mathcal{O} \left( 10^{-2} \right)$ with $\tan\beta = 30$ at $\Lambda^{}_{\rm EW}$. Thus generally, $\Delta^{}_\tau$ can be regarded as a small quantity in the SM or MSSM with $\tan\beta \lesssim 30$. Note that the values of $\Delta^{}_\tau$ in the Majorana and Dirac cases are practically equal. The reason is that after neglecting $Y^{}_\nu Y^\dagger_\nu$ in the Dirac case, the one-loop RGEs for charged-fermion Yukawa coupling matrices and gauge couplings are the same in the Majorana and Dirac cases, and with the same initial inputs for these parameters, they evolve equally in these two cases. For the same reason, values of $I^{}_\kappa$ within the MSSM in the Majorana case are the square of the corresponding values of $I^{}_\nu$ in the Dirac case, as can be observed in Fig.~\ref{loop-functions}.

By the way, at the one-loop level, there are some interesting and exact relations which can be derived from Eqs.~(6)---(7). That is, 
\begin{eqnarray}
\left( m^{}_1   m^{}_2  m^{}_3 \right)^{}_\mu  \hspace{-0.2cm}&=&\hspace{-0.2cm}  I^3_\kappa I^2_e I^2_\mu I^2_\tau \left( m^{}_1   m^{}_2  m^{}_3 \right)^{}_\Lambda \;,
\nonumber
\\
\det \left[ U\left( \mu \right) \right]  \hspace{-0.2cm}&=&\hspace{-0.2cm} \pm \det \left[ U\left( \Lambda \right) \right]
%	(10)
\end{eqnarray}
in the Majorana case~\cite{Mei:2003gn,Zhu:2018dvj}, and
\begin{eqnarray}
\left( m^{}_1   m^{}_2  m^{}_3 \right)^{}_\mu \hspace{-0.2cm}&=&\hspace{-0.2cm}  I^3_\nu I^{}_e I^{}_\mu I^{}_\tau \left( m^{}_1   m^{}_2  m^{}_3 \right)^{}_\Lambda  \;,
\nonumber
\\
\left( \mathcal{J} \Delta m^2_{21} \Delta m^2_{31} \Delta m^2_{32} \right)^{}_{\mu} \hspace{-0.2cm}&=&\hspace{-0.2cm} I^6_\nu I^2_e I^2_\mu I^2_\tau \left( \mathcal{J} \Delta m^2_{21} \Delta m^2_{31} \Delta m^2_{32} \right)^{}_\Lambda
%	(11)
\end{eqnarray}
in the Dirac case~\cite{Xing:2017mkx}, where $\Delta m^2_{ij} \equiv m^2_i -m^2_j$ (for $i,j = 1, 2, 3$) are neutrino mass-squared differences and $\mathcal{J}$ is the Jarlskog invariant of CP violation~\cite{Jarlskog:1985ht}, defined as
\begin{eqnarray}
{\rm Im} \left( U^{}_{\alpha i} U^{}_{\beta j} U^\ast_{\alpha j} U^\ast_{\beta i} \right) = \mathcal{J} \sum^{}_\gamma \varepsilon^{}_{\alpha\beta\gamma} \sum^{}_k \varepsilon^{}_{ijk}
%	(12)
\end{eqnarray}
with the Greek and Latin subscripts running over $(e, \mu, \tau)$ and $(1, 2, 3)$, respectively. With the parametrization of the PMNS matrix $U$ given in Eq.~(5), the Jarlskog invariant is given by $\mathcal{J} = 1/8 \sin 2\theta^{}_{12} \sin 2\theta^{}_{13} \cos\theta^{}_{13} \sin 2\theta^{}_{23} \sin\delta $. Those relations in Eqs.~(10) and (11) are very interesting because they connect neutrino masses, phases in $U$ or the  Jarlskog invariant at $\mu$ and $\Lambda$ with each other via some simple ways without any approximation at the one-loop level. The relationships for neutrino masses in the Majorana and Dirac cases are slightly different, but both of them explicitly show that if one of the neutrinos is massless, it will remain massless at the one-loop level. Only after the two-loop effects are taken into account, can nonzero neutrino mass be generated for the initially massless neutrino~\cite{Davidson:2006tg,Xing:2020ezi}. The relationship for $\det \left[ U\right]$ in the Majorana case gives us a correlation between phases in $U$, but such a correlation is dependent on the parametrization of $U$. Within the parametrization given in Eq.~(5), $ \left( \phi^{}_e + \phi^{}_\mu + \phi^{}_\tau + \rho + \sigma \right)^{}_\mu - \left( \phi^{}_e + \phi^{}_\mu + \phi^{}_\tau + \rho + \sigma \right)^{}_\Lambda = 0$ or $\pi$ holds, and a similar relation $ \left( \phi^{}_e + \phi^{}_\mu + \phi^{}_\tau + \rho + \sigma - \phi \right)^{}_\mu - \left( \phi^{}_e + \phi^{}_\mu + \phi^{}_\tau + \rho + \sigma - \phi \right)^{}_\Lambda = 0$ or $\pi$ holds if the parametrization proposed in Ref.~\cite{Fritzsch:1997fw} and the phase matrices $P^{}_l$ and $P^{}_\nu$ are used. The differential forms of these relations for phases can be found in Refs.~\cite{Xing:2005fw,Ohlsson:2012pg}. The relationship for the Jarlskog invariant in the Dirac case is instructive. It transparently shows that if $\mathcal{J} = 0$ holds initially, it will keep vanishing during the RGE running, implying that in the Dirac case CP violation at a low energy scale can not be radiatively generated when there is no CP violation at the superhigh energy scale and vice versa, and if CP violation does exist, it will exist at any energy scale below the cutoff scale.

Now we are going to perturbatively solve Eqs.~(6) and (7) with the parametrization of $U$ given in Eq.~(5) 
%%%%%%%%%%%%%%%%%%%%%%%%%%%%%%%%%%%%% footnote %%%%%%%%%%%%%%%%%%%%%%%%%%%%%%%%%%
\footnote{It is worth pointing out that thanks to the $\tau$-dominance approximations, the running behaviors of lepton flavor mixing parameters are closely associated with the elements in the third row of $U$~\cite{Xing:2005fw}, so it is possible to make the results describing evolution behaviors more concise and simpler by choosing an appropriate parametrization of $U$ whose elements in the third row are as simple as possible, such as the parametrization put forward in Ref.~\cite{Fritzsch:1997fw}. But in this work, we just take the widely used parametrization shown in Eq.~(5).}
%%%%%%%%%%%%%%%%%%%%%%%%%%%%%%%%%%%%%%%%%%%%%%%%%%%%%%%%%%%%%%%%%%%%%%%%%%%%%%%%%
. Note that in the Majorana case, three unphysical phases $\phi^{}_\alpha$ (for $\alpha = e, \mu, \tau$) in $P^{}_l$ are all involved in the RGE running and have their own evolutions against the energy scale $\mu$, but in the Dirac case, two phases in $P^{}_\nu$ and one overall phase in $P^{}_l$ can be cancelled in $H^{}_\nu$ so that only two of the five unphysical phases take part in the RGE running. Thus in the Dirac case, we redefine $P^{}_l$ as $P^\prime \equiv {\rm Diag} \left\{ e^{{\rm i} \phi^{}_1}, e^{{\rm i} \phi^{}_2}, 1 \right\}$ and take $P^{}_\nu = 1$. It is convenient to define 
\begin{eqnarray}
&&\hspace{-1.5cm} \Delta \theta^{}_{ij} \equiv \theta^{}_{ij} \left( \mu \right) - \theta^{}_{ij} \left( \Lambda \right) \;, \quad \Delta \delta \equiv \delta \left( \mu \right) - \delta \left( \Lambda \right) \;, \quad \Delta \rho \equiv \rho \left( \mu \right) - \rho \left( \Lambda \right) \;,
\nonumber
\\
&&\hspace{-1.5cm} \Delta \sigma \equiv \sigma \left( \mu \right) - \sigma \left( \Lambda \right) \;, \quad \Delta \phi^{}_\alpha \equiv \phi^{}_\alpha \left( \mu \right) - \phi^{}_\alpha \left( \Lambda \right) \;,
%	(13)
\end{eqnarray}
with $ij = 12, 13, 23$ and $\alpha = e, \mu, \tau$ in the Majorana case, or 
\begin{eqnarray}
\Delta \theta^{}_{ij} \equiv \theta^{}_{ij} \left( \mu \right) - \theta^{}_{ij} \left( \Lambda \right) \;, \quad \Delta \delta \equiv \delta \left( \mu \right) - \delta \left( \Lambda \right) \;, \quad \Delta \phi^{}_k \equiv \phi^{}_k \left( \mu \right) - \phi^{}_k \left( \Lambda \right) \;,
%	(14)
\end{eqnarray}
with $ij = 12, 13, 23$ and $k = 1, 2$ in the Dirac case, to describe the evolution of the corresponding parameters. As long as $\Delta^{}_\tau$ is small, the corrections to angles and phases are generally also small and thus we treat $\Delta^{}_\tau$ and the quantities defined in Eq.~(13) or (14) as small perturbations in the leading order approximation. To make the relevant expressions concise, we can define $\Delta U \equiv U \left( \mu \right) - U \left( \Lambda \right)$ which is a function of the quantities defined in Eq.~(13) or (14) and satisfies 
\begin{eqnarray}
U\left( \mu \right) \Delta U^\dagger + \Delta U U^\dagger \left( \mu \right) = 0 \;, \quad U^\dagger \left( \mu \right) \Delta U  + \Delta U^\dagger U \left( \mu \right) = 0 \;,
%	(15)
\end{eqnarray}
at the leading order guaranteed by the unitarity of $U\left( \mu \right)$ and $U \left( \Lambda \right)$. After expanding $U \left( \Lambda \right)$ with respect to
quantities defined in Eq.~(13) or Eq.~(14), we can obtain the explicit expression of $\Delta U$ in terms of the low-energy parameters and quantities defined above, as shown in Appendix~A.

\subsection{The Majorana case}
Taking the $\tau$-dominance approximations and substituting $U \left( \Lambda \right) = U \left( \mu \right) - \Delta U$ into Eq.~(6), at the leading order of $\Delta^{}_\tau$ and $\Delta U$, we obtain
\begin{eqnarray}
D^{}_\nu \left( \mu \right) \simeq I^{}_\kappa \left[ D^{}_\nu \left( \Lambda \right) - \left( U^\dagger \Delta U \right) D^{}_\nu \left( \Lambda \right) - D^{}_\nu \left( \Lambda \right) \left( U^\dagger \Delta U \right)^T + \Delta^{}_\tau U^\dagger \left(\begin{matrix} 0 & 0 & M^{}_{e\tau} \\ 0 & 0 & M^{}_{\mu\tau} \\ M^{}_{e\tau} & M^{}_{\mu \tau} & 2 M^{}_{\tau\tau} \end{matrix}\right) U^\ast \right] \;,
%	(16)
\end{eqnarray}
with $M^{}_{\alpha\tau} \equiv \sum\limits^3_{i=1} m^{}_i \left( \Lambda \right) U^{}_{\alpha i} U^{}_{\tau i} $ (for $\alpha = e, \mu, \tau$) and $U$ is subject to the energy scale $\mu$. Here and hereafter, all the parameters without labelling an explicit scale are implied to be at the energy scale $\mu$. Making use of Eq.~(15), the diagonal and non-diagonal parts of Eq.~(16) respectively lead to
\begin{eqnarray}
m^{}_i \left( \mu \right) \simeq I^{}_\kappa \left[ 1 + 2\left( \Delta U^\dagger U \right)^{}_{ii} + 2 \Delta^{}_\tau \left| U^{}_{\tau i} \right|^2 \right] m^{}_i \left( \Lambda \right) \;,
%	(17)
\end{eqnarray}
with $i=1, 2 , 3$ and
\begin{eqnarray}
&&\hspace{-1.5cm} {\rm Re} \left[ \left( \Delta U^\dagger U \right)^{}_{ij} \right] - \Delta^{}_\tau \zeta^{-1}_{ij} {\rm Re} \left[ U^\ast_{\tau i} U^{}_{\tau j} \right] \simeq 0 \;,
\nonumber
\\
&&\hspace{-1.5cm} {\rm Im} \left[ \left( \Delta U^\dagger U \right)^{}_{ij} \right] - \Delta^{}_\tau \zeta^{}_{ij} {\rm Im} \left[ U^\ast_{\tau i} U^{}_{\tau j} \right] \simeq 0 \;,
%	(18)
\end{eqnarray}
where $\zeta^{}_{ij} \equiv \left( m^{}_i - m^{}_j \right) / \left( m^{}_i + m^{}_j \right)$ with $i,j = 1, 2, 3$ and $i>j$. Note that $m^{}_i \left( \mu \right) \simeq I^{}_\kappa m^{}_i \left( \Lambda \right)$ (for $i=1, 2, 3$) at the leading order revealed in Eq.~(17) have been used in Eq.~(18) to make $\zeta^{}_{ij}$ be expressed by $m^{}_i$ and $m^{}_j$ at $\mu$. Since Eq.~(15) infers that $\left( \Delta U^\dagger U \right)^{}_{ii}$ (for $i = 1, 2, 3$) are purely imaginary, with the help of the parametrization of $U$ given in Eq.~(5), we can easily obtain the relationships between $m^{}_i \left( \mu \right)$ and $m^{}_i \left( \Lambda \right)$ from Eq.~(17), namely 
\begin{eqnarray}
m^{}_1 \left( \mu \right) \simeq \hspace{-0.6cm}&&  I^{}_\kappa \left[ 1 + \Delta^{}_\tau \left( 2\sin^2\theta^{}_{12} \sin^2\theta^{}_{23} + 2\cos^2\theta^{}_{12} \sin^2\theta^{}_{13} \cos^2\theta^{}_{23} \right.\right.
\nonumber
\\
&& - \left.\left. \sin 2\theta^{}_{12} \sin\theta^{}_{13} \sin 2\theta^{}_{23} \cos\delta \right) \right] m^{}_1 \left( \Lambda \right) \;,
\nonumber
\\
m^{}_2 \left( \mu \right) \simeq \hspace{-0.6cm}&& I^{}_\kappa \left[ 1 + \Delta^{}_\tau \left( 2\cos^2\theta^{}_{12} \sin^2\theta^{}_{23} + 2\sin^2\theta^{}_{12} \sin^2\theta^{}_{13} \cos^2\theta^{}_{23} \right.\right.
\nonumber
\\
&& + \left.\left. \sin 2\theta^{}_{12} \sin\theta^{}_{13} \sin 2\theta^{}_{23} \cos\delta \right) \right] m^{}_2 \left( \Lambda \right) \;,
\nonumber
\\
m^{}_3 \left( \mu \right) \simeq \hspace{-0.6cm}&& I^{}_\kappa \left[ 1 + 2\Delta^{}_\tau \cos^2\theta^{}_{13}\cos^2\theta^{}_{23} \right] m^{}_3 \left( \Lambda \right) \;.
%	(19)
\end{eqnarray}
Hence $\left( m^{}_1 m^{}_2 m^{}_3 \right)^{}_\mu \simeq I^3_\kappa \left( 1 + 2\Delta\tau \right) \left( m^{}_1 m^{}_2 m^{}_3 \right)^{}_\Lambda $ can be easily achieved from Eq.~(17) or (19), which is just the relation given in Eq.~(10) with the $\tau$-dominance approximations. Considering the imaginary part of Eq.~(17) and those shown in Eq.~(18), there are totally nine independent linear equations which contain nine parameters $\Delta \theta^{}_{12}$, $\Delta \theta^{}_{13}$, $\Delta \theta^{}_{23}$, $\Delta \delta$, $\Delta \rho$, $\Delta \sigma$, $\Delta \phi^{}_e$, $\Delta \phi^{}_\mu$ and $\Delta \phi^{}_\tau$ describing the one-loop evolution of relevant flavor mixing angles and phase parameters. Therefore we can fully solve these nine linear independent equations to get the approximate analytical results of three flavor mixing angles, three CP-violating phases and three unphysical phases. Taking advantage of the explicit expression of $\Delta U$ in terms of the above nine parameters as given in Appendix~A, after some straightforward and lengthy calculations we can arrive at
%%%%%%%%%%%%%%%%%%%%%%%%%%%%%%%%%%% footnote 3 %%%%%%%%%%%%%%%%%%%%%%%%%%%%%%%%%%
\footnote{Here we only give analytical results for the physical parameters (i.e., three flavor mixing angles and three CP-violating phases) since we do not concern about the evolution of unphysical phases. But for completeness, analytical results for three unphysical phases are given in Appendix~B.}
%%%%%%%%%%%%%%%%%%%%%%%%%%%%%%%%%%%%%%%%%%%%%%%%%%%%%%%%%%%%%%%%%%%%%%%%%%%%%%%%%
:

\begin{eqnarray}
%%%%%%%%%%%%%%%%%%%%%%%%%%%%%% \Delta \theta^{}_{12} %%%%%%%%%%%%%%%%%%%%%%%%%
\Delta \theta^{}_{12} \simeq \hspace{-0.6cm}&& \frac{\Delta^{}_\tau}{2} \left\{ \vphantom{\frac{1}{1}} \sin{2\theta^{}_{12}} \sin^2{\theta^{}_{13}} \cos^2{\theta^{}_{23}} \left[ \zeta^{}_{32} \sin^2{ \left( \delta + \sigma \right) } - \zeta^{}_{31} \sin^2{ \left( \delta + \rho \right)}  + \zeta^{-1}_{32} \cos^2{\left( \delta + \sigma \right)} \right.\right.
\nonumber
\\
&& - \left. \zeta^{-1}_{31} \cos^2{\left( \delta + \rho \right)} \right] - \left[ \left( \sin^2{\theta^{}_{23}} - \sin^2{\theta^{}_{13}} \cos^2{\theta^{}_{23}} \right) \sin{2\theta^{}_{12}} - \sin{\theta^{}_{13}}\sin{2\theta^{}_{23}} \cos{2\theta^{}_{12}} \right.
\nonumber
\\
&& \times \left. \cos{\delta} \vphantom{\cos^2} \right] \left[ \zeta^{}_{21} \sin^2{\left( \rho - \sigma\right)} + \zeta^{-1}_{21}\cos^2{\left(\rho-\sigma\right)} \right]  + \sin{2\theta^{}_{23}} \sin{\theta^{}_{13}} \left[ \left( \zeta^{}_{31} -\zeta^{-1}_{31} \right)\sin^2{\theta^{}_{12}} \right.
\nonumber
\\
&& \times \left. \sin{(\delta+\rho)}\sin{\rho} + \left( \zeta^{}_{32} -\zeta^{-1}_{32} \right)\cos^2{\theta^{}_{12}}\sin{(\delta+\sigma)}\sin{\sigma} + \frac{1}{2} \left( \zeta^{}_{21} -\zeta^{-1}_{21} \right) \sin{2\left( \rho - \sigma\right)} \right.
\nonumber
\\
&& \times \left.\left. \sin{\delta} + \left( \zeta^{-1}_{31} \sin^2{\theta^{}_{12}} + \zeta^{-1}_{32} \cos^2{\theta^{}_{12}} \right)\cos{\delta}  \right] \vphantom{\frac{1}{1}} \right\} \;,
\nonumber
\\
%%%%%%%%%%%%%%%%%%%%%%%%%%%%%% \Delta \theta^{}_{13} %%%%%%%%%%%%%%%%%%%%%%%%%
\Delta \theta^{}_{13} \simeq \hspace{-0.6cm}&& -\frac{\Delta^{}_\tau}{2} \left\{ \frac{1}{2} \sin{2\theta^{}_{12}} \sin{2\theta^{}_{23}} \cos{\theta^{}_{13}} \left[ \left( \zeta^{-1}_{32} - \zeta^{-1}_{31} \right) \cos{\delta} + \left( \zeta^{}_{32} - \zeta^{-1}_{32} \right) \sin{\left( \delta + \sigma \right)} \sin{\sigma} \right. \right.
\nonumber
\\
&& - \left. \left( \zeta^{}_{31} - \zeta^{-1}_{31} \right) \sin{\left( \delta + \rho \right)} \sin{\rho}  \right]  + \sin{2\theta^{}_{13}} \cos^2{\theta^{}_{23}} \left[ \left( \zeta^{}_{32} \sin^2{\left( \delta + \sigma \right)} + \zeta^{-1}_{32} \cos^2{\left( \delta + \sigma \right)} \right) \right.
\nonumber
\\
&& \times \left.\left. \sin^2{\theta^{}_{12}} + \left( \zeta^{}_{31} \sin^2{\left( \delta + \rho \right)} + \zeta^{-1}_{31} \cos^2{\left( \delta + \rho \right)} \right) \cos^2{\theta^{}_{12}} \right] \vphantom{\frac{1}{1}} \right\} \;,
\nonumber
\\
%%%%%%%%%%%%%%%%%%%%%%%%%%%%%% \Delta \theta^{}_{23} %%%%%%%%%%%%%%%%%%%%%%%%%
\Delta \theta^{}_{23} \simeq \hspace{-0.6cm}&& -\frac{\Delta^{}_\tau}{2} \left\{ \frac{1}{2} \sin{2\theta^{}_{12}} \sin{\theta^{}_{13}} \cos^2{\theta^{}_{23}} \left[ \left( \zeta^{}_{31} - \zeta^{-1}_{31} \right) \cos{\left( \delta + 2\rho \right)} - \left( \zeta^{}_{32} - \zeta^{-1}_{32} \right) \cos{\left( \delta + 2\sigma \right)} \right. \right.
\nonumber
\\
&& + \left. \left( \zeta^{}_{32} - \zeta^{}_{31} + \zeta^{-1}_{32} - \zeta^{-1}_{31} \right) \cos{\delta} \right]  + \sin{2\theta^{}_{23}} \left[ \left( \zeta^{}_{32} \sin^2{\sigma} + \zeta^{-1}_{32} \cos^2{\sigma} \right) \cos^2{\theta^{}_{12}} \right.
\nonumber
\\
&& + \left.\left. \left( \zeta^{}_{31} \sin^2{\rho} + \zeta^{-1}_{31} \cos^2{\rho} \right) \sin^2{\theta^{}_{12}} \right] \vphantom{\frac{1}{1}} \right\} \;,
%	(20)
\end{eqnarray}
for three flavor mixing angles;
\begin{eqnarray}
%%%%%%%%%%%%%%%%%%%%%%%%%%%%%%% \Delta \delta %%%%%%%%%%%%%%%%%%%%%%%%%%%%%%%%%
\Delta \delta \simeq \hspace{-0.6cm}&& \frac{\Delta^{}_\tau}{2} \left\{ \vphantom{\frac{1}{1}} \sin{2 \theta^{}_{12}} \sin{\theta^{}_{13}} \cos{2\theta^{}_{23}} \cot{\theta^{}_{23}} \left[ \left( \zeta^{}_{32} - \zeta^{}_{31} \right)\sin{\delta} + \left( \zeta^{}_{32} - \zeta^{-1}_{32} \right) \cos{\left( \delta +\sigma \right)} \sin{\sigma} \right.\right.
\nonumber
\\
&& - \left.\left( \zeta^{}_{31} - \zeta^{-1}_{31} \right) \cos{\left( \delta +\rho \right)} \sin{\rho} \right] + \sin{2\theta^{}_{23}}\left( \frac{\sin{2\theta^{}_{12}}}{2\sin{\theta^{}_{13}}} - \frac{2\sin{\theta^{}_{13}}}{\sin{2\theta^{}_{12}}} \sin^4{\theta^{}_{12}} \right) \left[ \left( \zeta^{}_{31} - \zeta^{-1}_{31} \right) \right.
\nonumber
\\
&& \times \left. \cos{\left( \delta + \rho \right)} \sin{\rho} - \zeta^{-1}_{31} \sin{\delta} \right]  - \sin{2\theta^{}_{23}} \left( \frac{\sin{2\theta^{}_{12}}}{2\sin{\theta^{}_{13}}}  - \frac{2\sin{\theta^{}_{13}}}{\sin{2\theta^{}_{12}}} \cos^4{\theta^{}_{12}} \right) \left[ \left( \zeta^{}_{32} - \zeta^{-1}_{32} \right) \right.
\nonumber
\\
&& \times \left. \cos{\left( \delta + \sigma \right)} \sin{\sigma} - \zeta^{-1}_{32} \sin{\delta} \right] + \left( \zeta^{}_{21} - \zeta^{-1}_{21} \right) \sin{2\left( \rho - \sigma \right)} \left( \sin^2{\theta^{}_{23}} - \sin^2{\theta^{}_{13}} \cos^2{\theta^{}_{23}}
\right.
\nonumber
\\
&& - \left. \sin{\theta^{}_{13}} \sin{2\theta^{}_{23}} \cot{2\theta^{}_{12}} \cos{\delta} \right) + \left( \zeta^{}_{31} - \zeta^{-1}_{31} \right) \left[ \left( \sin^2{\theta^{}_{12}} \sin^2{\theta^{}_{13}} - \cos^2{\theta^{}_{12}} \right) \cos^2{\theta^{}_{23}} 
\right.
\nonumber
\\
&& \times \left. \sin{2\left( \delta + \rho \right)} + \sin^2{\theta^{}_{12}} \cos{2\theta^{}_{23}} \sin{2\rho}
\right]  + \left( \zeta^{}_{32} - \zeta^{-1}_{32} \right) \left[ \left( \cos^2{\theta^{}_{12}} \sin^2{\theta^{}_{13}} - \sin^2{\theta^{}_{12}} \right) 
\right.
\nonumber
\\
&& \times \left. \cos^2{\theta^{}_{23}} \sin{2\left( \delta + \sigma \right)} + \cos^2{\theta^{}_{12}} \cos{2\theta^{}_{23}} \sin{2\sigma}
\right] - \frac{2\sin{\theta^{}_{13}} \sin{2\theta^{}_{23}} \sin{\delta}}{\sin{2\theta^{}_{12}}} 
\nonumber
\\
&& \times \left. \left[ \zeta^{}_{21} \cos^2{\left( \rho - \sigma \right)} + \zeta^{-1}_{21} \sin^2{\left( \rho - \sigma \right)} \right] \vphantom{\frac{1}{1}} \right\} \;,
%	(21)
\end{eqnarray}
for the Dirac CP-violating phase; and
\begin{eqnarray}
%%%%%%%%%%%%%%%%%%%%%%%%%%%%%%%%% \Delta \rho %%%%%%%%%%%%%%%%%%%%%%%%%%%%%%%%%
\Delta \rho \simeq \hspace{-0.6cm}&& \frac{\Delta^{}_\tau}{2} \left\{ \vphantom{\frac{1}{1}} \sin{\theta^{}_{13}} \sin{\delta} \left[ \zeta^{}_{32} \sin{2\theta^{}_{23}} \cos{2\theta^{}_{12}} \cot{\theta^{}_{12}} - \zeta^{-1}_{32} \sin{2\theta^{}_{12}} \cos{2\theta^{}_{23}} \cot{\theta^{}_{23}} + \sin{2\theta^{}_{12}}  \right.\right.
\nonumber
\\
&& \times \left. \left( \zeta^{}_{31} \sin{2\theta^{}_{23}} + \zeta^{-1}_{31} \cos{2\theta^{}_{23}} \cot{\theta^{}_{23}} \right) + \left( \zeta^{}_{21} \cos^2{\left( \rho - \sigma \right)} + \zeta^{-1}_{21} \sin^2{\left( \rho - \sigma \right)} \right) \sin{2\theta^{}_{23}} \right.
\nonumber
\\
&& \times \left. \cot{\theta^{}_{12}} \right] - \left( \zeta^{}_{21} - \zeta^{-1}_{21} \right) \sin{2\left( \rho - \sigma \right)} \left[ \left( \sin^2{\theta^{}_{23}} - \sin^2{\theta^{}_{13}} \cos^2{\theta^{}_{23}}  \right) \cos^2{\theta^{}_{12}} - \frac{1}{2} \sin{\theta^{}_{13}} \right.
\nonumber
\\
&& \times \left. \sin{2\theta^{}_{23}} \cos{2\theta^{}_{12}} \cot{\theta^{}_{12}} \cos{\delta} \vphantom{\frac{1}{1}} \right] + \left( \zeta^{}_{31} - \zeta^{-1}_{31} \right) \left[ \left( \cot{\theta^{}_{23}}  - 2 \sin{2\theta^{}_{23}} \right) \sin{2\theta^{}_{12}} \sin{\theta^{}_{13}} \right.
\nonumber
\\
&& \times \left. \sin{\left( \delta + \rho \right)} \cos{\rho}  - \sin^2{\theta^{}_{12}} \cos{2\theta^{}_{23}} \sin{2\rho} + 2 \sin^2{\theta^{}_{13}} \cos^2{\theta^{}_{12}} \cos^2{\theta^{}_{23}} \sin{2\left( \delta + \rho \right)}   \right]
\nonumber
\\
&& - \left( \zeta^{}_{32} - \zeta^{-1}_{32} \right) \left[ \cos^2{\theta^{}_{12}} \cos{2\theta^{}_{23}} \sin{2\sigma} + \sin^2{\theta^{}_{13}} \cos^2{\theta^{}_{23}} \cos{2\theta^{}_{12}} \sin{2\left( \delta + \sigma \right)}  \right.
\nonumber
\\
&& + \left.\left. \left( \sin{2\theta^{}_{12}} \cos{2\theta^{}_{23}} \cot{\theta^{}_{23}} +  \sin{2\theta^{}_{23}} \cos{2\theta^{}_{12}} \cot{\theta^{}_{12}} \right) \sin{\theta^{}_{13}} \sin{\left( \delta + \sigma \right)} \cos{\sigma} \right] \vphantom{\frac{1}{1}} \right\}
\nonumber
\\
%%%%%%%%%%%%%%%%%%%%%%%%%%%%%%% \Delta \sigma %%%%%%%%%%%%%%%%%%%%%%%%%%%%%%%%%
\Delta \sigma  \simeq \hspace{-0.6cm}&& \frac{\Delta^{}_\tau}{2} \left\{ \left( \zeta^{}_{31} - \zeta^{-1}_{31} \right) \left[ \left( \sin{2\theta^{}_{12}} \cos{2\theta^{}_{23}} \cot{\theta^{}_{23}} - \sin{2\theta^{}_{23}} \cos{2\theta^{}_{12}} \tan{\theta^{}_{12}} \right) \sin{\theta^{}_{13}} \sin{\left( \delta + \rho \right)} \right.\right.
\nonumber
\\
&& \times \left. \cos{\rho} - \sin^2{\theta^{}_{12}} \cos{2\theta^{}_{23}} \sin{2\rho} + \sin^2{\theta^{}_{13}} \cos{2\theta^{}_{12}} \cos^2{\theta^{}_{23}} \sin{2\left( \delta + \rho \right)} \right] - \left( \zeta^{}_{32} - \zeta^{-1}_{32} \right) 
\nonumber
\\
&& \times \left[ \frac{1}{2} \left( \cot{\theta^{}_{23}} - 2\sin{2\theta^{}_{23}} \right) \sin{2\theta^{}_{12}} \sin{\theta^{}_{13}} \sin{\left( \delta + 2\sigma \right)} + \cos^2{\theta^{}_{12}} \cos{2\theta^{}_{23}} \sin{2\sigma} - 2\sin^2{\theta^{}_{12}} \right.
\nonumber
\\
&& \times \left. \sin^2{\theta^{}_{13}} \cos^2{\theta^{}_{23}} \sin{2\left( \delta + \sigma \right)} \vphantom{\frac{1}{1}} \right]  - \left( \zeta^{}_{21} - \zeta^{-1}_{21} \right) \sin{2\left( \rho - \sigma \right)} \left[ \vphantom{\frac{1}{1}} \left( \sin^2{\theta^{}_{23}} - \sin^2{\theta^{}_{13}} \cos^2{\theta^{}_{23}} \right) \right.
\nonumber
\\
&& \times \left. \sin^2{\theta^{}_{12}} - \frac{1}{2} \sin{\theta^{}_{13}} \sin{2\theta^{}_{23}} \cos{2\theta^{}_{12}} \tan{\theta^{}_{12}} \cos{\delta} \right] + \sin{\theta^{}_{13}} \sin{\delta} \left[ \vphantom{\frac{1}{1}} \left( \zeta^{}_{21} \cos^2{\left( \rho -\sigma \right)} \right.\right.
\nonumber
\\ 
&& + \left.\left. \zeta^{-1}_{21} \sin^2{\left( \rho - \sigma \right)} \right) \sin{2\theta^{}_{23}} \tan{\theta^{}_{12}} + \zeta^{}_{31} \sin{2\theta^{}_{23}} \cos{2\theta^{}_{12}} \tan{\theta^{}_{12}} + \zeta^{-1}_{31} \sin{2\theta^{}_{12}} \cos{2\theta^{}_{23}} \right.
\nonumber
\\
&& \times \left.\left. \cot{\theta^{}_{23}} - \frac{1}{2} \left( \zeta^{}_{32} + \zeta^{-1}_{32} \right) \sin{2\theta^{}_{12}} \cot{\theta^{}_{23}}
\right]  \right\}
%	(22)
\end{eqnarray}
for the two Majorana CP-violating phases. By means of the analytical results for flavor mixing angles and the Dirac CP-violating phase given in Eqs.~(20)---(21), we can gain the analytical result for the Jarlskog invariant $\mathcal{J}$ defined in Eq.~(12), that is 
\begin{eqnarray}
\mathcal{J} \left( \mu \right)  \simeq \hspace{-0.6cm}&& \left( 1 - \Delta^{}_\tau C^{\left(1\right)}_{\rm M} \right) \mathcal{J} \left( {\Lambda} \right) + \frac{1}{32} \Delta^{}_\tau \sin 2\theta^{}_{12} \sin 2\theta^{}_{13} \cos\theta^{}_{13} \sin 2\theta^{}_{23}  C^{\left(2\right)}_{\rm M}
%	(23)
\end{eqnarray}
with 
\begin{eqnarray}
C^{\left(1\right)}_{\rm M} = \hspace{-0.6cm}&& \cos^2 \theta^{}_{23} \left\{ \sin^2\theta^{}_{13} \left[ \left( \zeta^{}_{31} - \zeta^{-1}_{31} \right) \cos^2\theta^{}_{12} \cos 2\left( \delta + \rho \right) + \left( \zeta^{}_{32} - \zeta^{-1}_{32} \right) \sin^2\theta^{}_{12} \cos 2\left( \delta + \sigma \right) \right] \right.
\nonumber
\\
\hspace{-0.6cm}&& + \left( \cos^2\theta^{}_{12} - \sin^2\theta^{}_{13} \right) \left( \zeta^{}_{31} \cos^2\rho + \zeta^{-1}_{31} \sin^2\rho \right) + \left( \sin^2\theta^{}_{12} - \sin^2\theta^{}_{13} \right)   \left( \zeta^{}_{32} \cos^2\sigma \right.
\nonumber
\\
\hspace{-0.6cm}&& + \left.\left. \zeta^{-1}_{32} \sin^2\sigma \right) \right\} - \frac{1}{4} \sin 2\theta^{}_{12} \sin\theta^{}_{13} \left\{ \sin 2\theta^{}_{23} \cos\delta \left( \zeta^{}_{32} - \zeta^{}_{31} - 2\zeta^{}_{21} + \zeta^{-1}_{32} - \zeta^{-1}_{31} - 2\zeta^{-1}_{21} \right) \right.
\nonumber
\\
\hspace{-0.6cm}&& + \left. \left( 2\cot\theta^{}_{23} -3\sin 2\theta^{}_{23} \right) \left[ \left( \zeta^{}_{32} - \zeta^{-1}_{32} \right) \cos\left( \delta + 2\sigma \right) - \left( \zeta^{}_{31} - \zeta^{-1}_{31} \right) \cos\left( \delta + 2\rho \right) \right] \right\}   + \frac{1}{2} \left( \zeta^{}_{21}  \right.
\nonumber
\\
\hspace{-0.6cm}&& - \left. \zeta^{-1}_{21} \right) \sin\theta^{}_{13} \sin 2\theta^{}_{23} \left[ \left( \csc 2\theta^{}_{12} + \cos 2\theta^{}_{12} \cot 2\theta^{}_{12} \right) \cos\delta \cos 2\left( \rho - \sigma \right) - 2\cot 2\theta^{}_{12} \sin\delta \right.
\nonumber
\\
\hspace{-0.6cm}&& \times \left. \sin 2\left( \rho -\sigma \right) \right] + \cos 2\theta^{}_{12} \left( \sin^2\theta^{}_{23} - \sin^2\theta^{}_{13} \cos^2\theta^{}_{23} \right) \left[ \zeta^{}_{21} \sin^2\left( \rho - \sigma \right) + \zeta^{-1}_{21} \cos^2\left( \rho - \sigma \right) \right]
\nonumber
\\
\hspace{-0.6cm}&& + \left( \sin^2\theta^{}_{12} \cos 2\theta^{}_{23} - \cos^2\theta^{}_{23} \sin^2\theta^{}_{13} \cos^2\theta^{}_{12} \right) \left( \zeta^{}_{31} \sin^2\rho + \zeta^{-1}_{31} \cos^2\rho \right) + \left( \cos^2\theta^{}_{12} \cos 2\theta^{}_{23} \right. 
\nonumber
\\
\hspace{-0.6cm}&& - \left. \sin^2\theta^{}_{12} \sin^2\theta^{}_{13} \cos^2\theta^{}_{23} \right) \left( \zeta^{}_{32} \sin^2\sigma + \zeta^{-1}_{32} \cos^2\sigma \right) \;,
%	(24)
\end{eqnarray}
and
\begin{eqnarray}
C^{\left(2\right)}_{\rm M} = \hspace{-0.6cm}&& 2 \left( \zeta^{}_{21} - \zeta^{-1}_{21} \right) \sin 2\left( \rho - \sigma \right) \left[ \left( \sin^2\theta^{}_{23} - \sin^2\theta^{}_{13} \cos^2\theta^{}_{23} \right) \cos\delta - \sin\theta^{}_{13} \sin 2\theta^{}_{23} \cot 2\theta^{}_{12} \right]
\nonumber
\\
%\hspace{-0.6cm}&& + \left( \zeta^{}_{31} - \zeta^{-1}_{31} \right) \sin 2\rho \left\{ 2\left[ \sin^2\theta^{}_{12} \cos 2\theta^{}_{23} - \cos^2\theta^{}_{23} \left( \cos^2\theta^{}_{12} - \sin^2\theta^{}_{12} \sin^2\theta^{}_{13} \right) \right] \cos\delta \right.
%\nonumber
%\\
%\hspace{-0.6cm}&& - \left. \sin\theta^{}_{13} \left( \sin 2\theta^{}_{12} \cos 2\theta^{}_{23} \cot\theta^{}_{23} + \sin 2\theta^{}_{23} \sin^2\theta^{}_{12} \tan\theta^{}_{12} \right) + \frac{1}{2} \sin 2\theta^{}_{12} \sin 2\theta^{}_{23} \csc\theta^{}_{13} \right\}
%\nonumber
%\\
%\hspace{-0.6cm}&& + \left( \zeta^{}_{32} - \zeta^{-1}_{32} \right) \sin 2\sigma \left\{ 2\left[ \cos^2\theta^{}_{12} \cos 2\theta^{}_{23} - \cos^2\theta^{}_{23} \left( \sin^2\theta^{}_{12} - \cos^2\theta^{}_{12} \sin^2\theta^{}_{13} \right) \right] \cos\delta \right.
%\nonumber
%\\
%\hspace{-0.6cm}&& + \left. \sin\theta^{}_{13} \left( \sin 2\theta^{}_{12} \cos 2\theta^{}_{23} \cot\theta^{}_{23} + \sin 2\theta^{}_{23} \cos^2\theta^{}_{12} \cot\theta^{}_{12} \right) - \frac{1}{2} \sin 2\theta^{}_{12} \sin 2\theta^{}_{23} \csc\theta^{}_{13} \right\}
%\\
\hspace{-0.6cm}&& + \left( \zeta^{}_{31} - \zeta^{-1}_{31} \right) \sin 2\rho \left\{ 2\left[ \sin^2\theta^{}_{12} \cos 2\theta^{}_{23} - \cos^2\theta^{}_{23} \left( \cos^2\theta^{}_{12} - \sin^2\theta^{}_{12} \sin^2\theta^{}_{13} \right) \right] \cos\delta \right.
\nonumber
\\
\hspace{-0.6cm}&& - \left. \sin\theta^{}_{13}\sin 2\theta^{}_{23} \sin^2\theta^{}_{12} \tan\theta^{}_{12} \right\} + \left( \zeta^{}_{32} - \zeta^{-1}_{32} \right) \sin 2\sigma \left\{ 2\left[ \cos^2\theta^{}_{12} \cos 2\theta^{}_{23} - \cos^2\theta^{}_{23} \right.\right.
\nonumber
\\
\hspace{-0.6cm}&& \times \left.\left. \left( \sin^2\theta^{}_{12} - \cos^2\theta^{}_{12} \sin^2\theta^{}_{13} \right) \right] \cos\delta  + \sin\theta^{}_{13} \sin 2\theta^{}_{23} \cos^2\theta^{}_{12} \cot\theta^{}_{12}  \right\} +  \sin 2\theta^{}_{12} 
\nonumber
\\
\hspace{-0.6cm}&& \times \left( \sin\theta^{}_{13} \cos 2\theta^{}_{23} \cot\theta^{}_{23}  - \frac{\sin 2\theta^{}_{23}}{2\sin\theta^{}_{13}} \right) \left[ \left( \zeta^{}_{32} - \zeta^{-1}_{32} \right) \sin 2\sigma - \left( \zeta^{}_{31} - \zeta^{-1}_{31} \right) \sin 2\rho \right] \;.
%	(25)
\end{eqnarray}
Some immediate comments are in order.
\begin{itemize}
	\item All the results for $\Delta \theta^{}_{ij}$ (for $ij=12, 13, 23$), $\Delta \delta$, $\Delta \rho$ and $\Delta \sigma$ are proportional to $\Delta^{}_\tau$ so the relevant parameters at $\mu$ involved in these results can generally be replaced by their values at $\Lambda$ at the leading order level. The same observation is also true for terms proportional to $\Delta^{}_\tau$ in the results for $m^{}_i$ (for $i=1, 2, 3$) and $\mathcal{J}$.
	\item Since the signs of $\Delta^{}_\tau$ are opposite in the SM and MSSM which can be seen from Eq.~(9) and Fig.~\ref{loop-functions}, the running directions of flavor mixing angles and CP-violating phases, whose evolutions are determined by Eqs.~(20)---(22), are opposite in the SM and MSSM.
	\item With the help of Eq.~(22) and Eq.~(B.1) given in Appendix~B, it is easy to check that $\Delta \rho + \Delta \sigma + \Delta \phi^{}_e + \Delta \phi^{}_\mu + \Delta \phi^{}_\tau \simeq 0$ holds, a result consistent with the exact relation induced by Eq.~(10).
	\item In special cases where $\rho \left( \mu \right), \sigma \left( \mu \right) = 0, \pi/2$ or $\rho \left( \Lambda \right), \sigma \left( \Lambda \right) = 0, \pi/2$ are satisfied, $\Delta \delta$, $\Delta \rho$ and $\Delta \sigma$ are all proportional to $\sin \delta$, and $C^{(2)}_{\rm M}$ will vanish and thus lead to $\mathcal{J} \left( \mu \right) \propto \mathcal{J} \left( \Lambda \right)$. 
	%\textcolor{red}{Therefore if all the three CP-violating phases vanish initially, they can not be radiatively generated by the one-loop RGE effects, as well as the Jarlskog invariant.}
	\item As for the result of $\mathcal{J} \left( \mu \right)$, besides the term proportional to $\mathcal{J} \left( \Lambda \right)$, there exists an additional term which has no direct link to $\mathcal{J} \left( \Lambda \right)$. As expected, if $\delta \left( \Lambda^{}_{\rm EW} \right) = 0$ or $\delta \left( \Lambda \right) = 0$ is assumed, $\Delta \delta$ is generally nonvanishing and satisfies $\Delta \delta  \simeq \Delta^{}_\tau C^{(2)}_{\rm M}/4$, which infers $\mathcal{J} \left( \mu \right) \propto \Delta \delta$. Thus, even if $\mathcal{J} \left( \Lambda \right) = 0$ holds, $\mathcal{J} \left( \mu \right)$ may radiatively acquire a value via the one-loop RGE running.
	\item The results for flavor mixing angles, CP-violating phases and the Jarlskog invariant given in Eqs.~(20)---(25) are quite long, but they can be greatly simplified if one takes into account some special flavor symmetries (e.g., the $\mu$-$\tau$ reflection symmetry),  flavor mixing patterns (e.g., the tri-bimaximal mixing pattern) or specific neutrino mass spectrum. It is worth pointing out that these integral results with $\theta^{}_{13}$ taken to be zero can be directly achieved from the differential results for the mixing angles and CP-violating phases in Ref.~\cite{Antusch:2003kp} by integrating them with the assumption that all parameters are constant apart from the tau Yukawa coupling. But now it is well-known that $\theta^{}_{13}$ is not so small that $\theta^{}_{13} = 0$ is no longer a good approximation.
\end{itemize}

\subsection{The Dirac case}

Similarly, considering the $\tau$-dominance approximations and substituting $U \left( \Lambda \right) = U \left( \mu \right) - \Delta U$ into Eq.~(7), we arrived at
\begin{eqnarray}
D^2_\nu \left( \mu \right) \simeq I^2_\nu \left[ D^2_\nu \left( \Lambda \right) - \left( U^\dagger \Delta U \right) D^2_\nu \left( \Lambda \right) - D^2_\nu \left( \Lambda \right) \left( U^\dagger \Delta U \right)^\dagger + \Delta^{}_\tau U^\dagger \left(\begin{matrix} 0 & 0 & M^\prime_{e\tau} \\ 0 & 0 & M^\prime_{\mu \tau} \\
M^{\prime\ast}_{e \tau} & M^{\prime\ast}_{\mu \tau} & 2 M^{\prime}_{\tau \tau} \end{matrix}  \right) U \right] \;,
%	(26)
\end{eqnarray}
where $M^\prime_{\alpha \tau} \equiv \sum\limits^3_{i=1} m^2_i \left( \Lambda \right) U^{}_{\alpha i} U^\ast_{\tau i}$ (for $\alpha = e, \mu, \tau$) are defined. One can see that Eq.~(26) has a similar structure to Eq.~(16), and small differences between them are attributed to the different diagonalizations for a complex symmetric matrix (i.e., $M^{}_\nu$ in the Majorana case) and a Hermitian matrix (i.e., $H^{}_\nu = M^{}_\nu M^\dagger_\nu$ in the Dirac case). By means of the relations given in Eq.~(15), from the diagonal and non-diagonal parts of Eq.~(26), we obtain
\begin{eqnarray}
m^2_i \left( \mu \right) \simeq I^2_\nu \left( 1 + 2 \Delta^{}_\tau \left| U^{}_{\tau i} \right|^2 \right) m^2_i \left( \Lambda \right) \;,
%	(27)
\end{eqnarray}
with $i=1, 2, 3$ and 
\begin{eqnarray}
\left( \Delta U^\dagger U \right)^{}_{ij} - \Delta^{}_\tau \xi^{}_{ij} U^\ast_{\tau i} U^{}_{\tau j} \simeq 0 \;,
%	(28)
\end{eqnarray}
where $\xi^{}_{ij} \equiv \left( m^2_i + m^2_j \right)/\left( m^2_i - m^2_j \right)$ with $i,j=1,2,3$ and $i>j$, and $m^2_i \left( \mu \right) \simeq I^2_\nu  m^2_i \left( \Lambda \right)$ (for $i=1, 2, 3$) revealed in Eq.~(27) at the leading order have been used. Different from Eq.~(17), both the left-hand and the right-hand sides of Eq.~(27) are real, so there are only six independent linear equations given by the real and imaginary parts of Eq.~(28) for three flavor mixing angles and three phases. This is consistent with the fact that in the Dirac case, instead of five unphysical phases, only two unphysical phases participate in the RGE running. Therefore there are totally six unknown parameters, $\Delta \theta^{}_{12}$, $\Delta \theta^{}_{13}$, $\Delta \theta^{}_{23}$, $\Delta \delta$, $\Delta \phi^{}_1$ and $\Delta \phi^{}_2$ defined in Eq.~(14), which can be fully solved from the six independent linear equations extracted from Eq.~(28). Considering the parametrization of $U$ in Eq.~(5) with the redefined $P^\prime$ and $P^{}_\nu = 1$, Eq.~(27) leads us to
\begin{eqnarray}
m^{}_1 \left( \mu \right) \simeq \hspace{-0.6cm}&&  I^{}_\nu \left[ 1 + \frac{1}{2} \Delta^{}_\tau \left( 2\sin^2\theta^{}_{12} \sin^2\theta^{}_{23} + 2\cos^2\theta^{}_{12} \sin^2\theta^{}_{13} \cos^2\theta^{}_{23} \right.\right.
\nonumber
\\
&& - \left.\left. \sin 2\theta^{}_{12} \sin\theta^{}_{13} \sin 2\theta^{}_{23} \cos\delta \right) \vphantom{\frac{1}{1}} \right] m^{}_1 \left( \Lambda \right) \;,
\nonumber
\\
m^{}_2 \left( \mu \right) \simeq \hspace{-0.6cm}&& I^{}_\nu \left[ 1 + \frac{1}{2} \Delta^{}_\tau \left( 2\cos^2\theta^{}_{12} \sin^2\theta^{}_{23} + 2\sin^2\theta^{}_{12} \sin^2\theta^{}_{13} \cos^2\theta^{}_{23} \right.\right.
\nonumber
\\
&& + \left.\left. \sin 2\theta^{}_{12} \sin\theta^{}_{13} \sin 2\theta^{}_{23} \cos\delta \right)  \vphantom{\frac{1}{1}} \right] m^{}_2 \left( \Lambda \right) \;,
\nonumber
\\
m^{}_3 \left( \mu \right) \simeq \hspace{-0.6cm}&& I^{}_\nu \left[ 1 + \Delta^{}_\tau \cos^2\theta^{}_{13}\cos^2\theta^{}_{23} \right] m^{}_3 \left( \Lambda \right) \;.
%	(29)
\end{eqnarray}
As can be seen, $ \left( m^{}_1 m^{}_2 m^{}_3 \right)^{}_\mu \simeq I^3_\nu \left( 1 + \Delta^{}_\tau \right) \left( m^{}_1 m^{}_2 m^{}_3 \right)^{}_\Lambda  $ holds, and besides the overall factor (i.e., $I^{}_\kappa$ in the Majorana case and $I^{}_\nu$ in the Dirac case), the results given in Eq.~(29) are different from those given in Eq.~(19) just by a factor $1/2$ for the terms proportional to $\Delta^{}_\tau$. These relations and differences can be easily understood from the exact relations given in Eqs.~(10) and (11) with the $\tau$-dominance approximations. By the aid of the explicit expression of $\Delta U$ given in Appendix~A and the six linear independent equations extracted from Eq.~(28), the evolution behaviors of three flavor mixing angles and the Dirac CP-violating phase are given by
\begin{eqnarray}
\Delta \theta^{}_{12}  \simeq \hspace{-0.6cm}&& \frac{\Delta^{}_\tau}{2} \left[ \sin\theta^{}_{13} \sin{2\theta^{}_{23}} \cos\delta \left( \xi^{}_{21} \cos2\theta^{}_{12} +  \xi^{}_{31} \sin^2\theta^{}_{12} + \xi^{}_{32} \cos^2\theta^{}_{12} \right) \right.
\nonumber
\\
&& \left. + \sin2\theta^{}_{12} \sin^2\theta^{}_{13} \cos^2\theta^{}_{23} \left( \xi^{}_{21} + \xi^{}_{32} - \xi^{}_{31} \right) - \xi^{}_{21} \sin2\theta^{}_{12} \sin^2\theta^{}_{23} \right] \;,
\nonumber
\\
\Delta \theta^{}_{13} \simeq \hspace{-0.6cm}&& \frac{\Delta^{}_\tau}{4} \left[ \sin2\theta^{}_{12} \sin2\theta^{}_{23} \cos\theta^{}_{13} \cos{\delta} \left( \xi^{}_{31} - \xi^{}_{32} \right) - 2\sin2\theta^{}_{13} \cos^2\theta^{}_{23} \right.
\nonumber
\\
&& \times \left. \left( \xi^{}_{31} \cos^2\theta^{}_{12} + \xi^{}_{32} \sin^2\theta^{}_{12} \right) \right] \;,
\nonumber
\\
\Delta \theta^{}_{23} \simeq \hspace{-0.6cm}&& \frac{\Delta^{}_\tau}{2} \left[ \sin2\theta^{}_{12} \sin\theta^{}_{13} \cos^2\theta^{}_{23} \cos\delta \left( \xi^{}_{31} - \xi^{}_{32} \right) - \sin2\theta^{}_{23} \left( \xi^{}_{31} \sin^2\theta^{}_{12} \right.\right.
\nonumber
\\
&& + \left.\left. \xi^{}_{32} \cos^2\theta^{}_{12} \right) \right] \;,
%	(30)
\end{eqnarray}
and 
\begin{eqnarray}
\Delta \delta \simeq \hspace{-0.6cm}&& \Delta^{}_\tau \sin\delta \left\{ \sin2\theta^{}_{23} \left[ \frac{\sin2\theta^{}_{12}}{4\sin\theta^{}_{13}} \left( \xi^{}_{32} - \xi^{}_{31} \right) + \frac{\sin\theta^{}_{13}}{\sin2\theta^{}_{12}} \left( \xi^{}_{31} \sin^4\theta^{}_{12} - \xi^{}_{32} \cos^4\theta^{}_{12} \right)  \right] \right.
\nonumber
\\
&& \left. - \sin\theta^{}_{13} \left[ \xi^{}_{21} \frac{\sin2\theta^{}_{23}}{\sin2\theta^{}_{12}} - \frac{1}{2} \sin2\theta^{}_{12} \cos2\theta^{}_{23} \cot\theta^{}_{23} \left(\xi^{}_{32} -\xi^{}_{31} \right) \right] \right\} \;.
%	(31)
\end{eqnarray}
Again, the results for the two unphysical phases $\phi^{}_1$ and $\phi^{}_2$ are given in Appendix~B. With above results for $\Delta \theta^{}_{12}$, $\Delta \theta^{}_{13}$, $\Delta \theta^{}_{23}$ and $\Delta \delta$, we can establish a relation between the Jarlskog invariant at $\mu$ and $\Lambda$, viz.,
\begin{eqnarray}
\mathcal{J}\left( \mu \right)  \simeq \hspace{-0.6cm}&& \left( 1 - \frac{ \Delta^{}_\tau }{2} C^{}_{\rm D} \right) \mathcal{J} \left( \Lambda \right)
%	(32)
\end{eqnarray}
with
\begin{eqnarray}
C^{}_{\rm D} \simeq \hspace{-0.6cm}&& \cos 2\theta^{}_{13} \cos^2\theta^{}_{23} \left[ \xi^{}_{31} + \xi^{}_{32} + \left( \xi^{}_{31} + \xi^{}_{21} \right) \cos^2\theta^{}_{12} + \left( \xi^{}_{32} - \xi^{}_{21} \right) \sin^2\theta^{}_{12} \right]
\nonumber
\\
\hspace{-0.6cm}&& + \left( 1 - 3\sin^2\theta^{}_{23} \right) \left[ \left( \xi^{}_{31} + \xi^{}_{21} \right) \sin^2\theta^{}_{12} + \left( \xi^{}_{32} - \xi^{}_{21} \right) \cos^2\theta^{}_{12} \right] 
\nonumber
\\
\hspace{-0.6cm}&& + \left( 2\xi^{}_{21} + \xi^{}_{31} - \xi^{}_{32} \right) \sin 2\theta^{}_{12} \sin\theta^{}_{13} \sin 2\theta^{}_{23} \cos\delta  \;.
%	(33)
\end{eqnarray}
Though it is not obvious that Eq.~(32) coincides with the relation for the Jarlskog invariant given in Eq.~(11), one may use Eq.~(29) to check that the relation for $\mathcal{J}$ in Eq.~(11) actually leads to the results given by Eqs.~(32) and (33). All analytical results for flavor mixing angles, the Dirac CP-violating phase and the Jarlskog invariant in the Dirac case are much simpler than those in the Majorana case, owing to two less physical degrees of freedom in the former case. Some comments are as follows: 
\begin{itemize}
	\item Similar to that in the Majorana case, $\Delta \theta^{}_{ij}$ (for $ij=12, 13, 23$) and $\Delta \delta$ are all proportional to $\Delta^{}_\tau$, and thus the running directions of these parameters are opposite in the SM and MSSM. Generally, relevant flavor parameters at $\mu$ involved in these results can be replaced by their values at $\Lambda$, as well as those contained in the terms proportional to $\Delta^{}_\tau$ in the expressions of $m^{}_i$ (for $i = 1, 2, 3$) and $\mathcal{J}$.
	\item As shown in Eqs.~(31) and (32), $\Delta \delta \propto \sin\delta$ and $\mathcal{J} \left( \mu \right) \propto \mathcal{J} \left( \Lambda \right) $ hold. It means that $\delta$ and $\mathcal{J}$ can not be radiatively generated if they are initially vanishing. This observation differs from that in the Majorana case.
	\item Taking all the CP-violating phases to be vanishing and with the replacements $\zeta^{-1}_{ij} \to \xi^{}_{ij}$ (i.e., $m^{}_i \to m^2_i$ for $i,j = 1, 2, 3$) in the Majorana case, and meanwhile taking the Dirac CP-violating phase to be zero in the Dirac case, one may check that the corresponding results for flavor mixing angles in Eqs.~(20) and (30) are exactly the same. It is easy to understand these results, since after unphysical phases being rotated away from $M^{}_\nu$ and $H^{}_\nu$ in this case,  $M^{}_\nu$ in the Majorana case and $H^{}_\nu$ in the Dirac case are both real and can be diagonalized by a real orthogonal matrix, so all things are formally the same in the Majorana and Dirac cases except the eigenvalues of $M^{}_\nu$ and $H^{}_\nu$.
\end{itemize}

\section{Numerical analysis and discussion}
In the numerical analysis, we use the best-fit values of neutrino parameters obtained from the latest global analysis of currently available neutrino oscillation data~\cite{Capozzi:2020qhw,deSalas:2020pgw} including the T2K measurements of the Dirac CP-violating phase~\cite{Abe:2019vii}, namely,
\begin{eqnarray}
\sin^2 \theta^{}_{12} = \left\{ \begin{array}{l} 0.305 \\ 0.303 \end{array} \right.
\;, \quad \sin^2 \theta^{}_{13} = \left\{ \begin{array}{l} 0.0222 \\ 0.0223
\end{array} \right. \;, \quad \sin^2 \theta^{}_{23} = \left\{ \begin{array}{l} 0.545
\\ 0.551 \end{array} \right. \;,\quad \delta = \left\{ \begin{array}{l} 1.28\pi
\\ 1.52\pi \end{array} \right. \;,
%   (34)
\end{eqnarray}
and
\begin{eqnarray}
\delta m^2 = \left\{ \begin{array}{l} 7.34\times 10^{-5} ~{\rm eV^2} \\
7.34\times 10^{-5} ~{\rm eV^2} \end{array} \right. \;, \quad
\Delta m^2 = \left\{ \begin{array}{l} + 2.485 \times 10^{-3} ~{\rm eV^2} \\
-2.465 \times 10^{-3} ~{\rm eV^2} \end{array} \right. \;,
%   (35)
\end{eqnarray}
where both the normal neutrino mass ordering (upper values) and the
inverted one (lower values) have been taken into account, and the two
neutrino mass-squared differences are defined as $\delta m^2 \equiv
m^2_2 - m^2_1$ and $\Delta m^2 \equiv m^2_3 - \left(m^2_1 + m^2_2\right)/2$. These experimental data are all given at the eletroweak scale $\Lambda^{}_{\rm EW} \sim 100$ GeV.  We consider the following two neutrino mass spectra:
\begin{itemize}
	\item The normal mass ordering (NMO) with $m^{}_1 \left( \Lambda^{}_{\rm EW} \right) = 0.001$ eV. In the Dirac case, $\xi^{}_{ij} \left( \mu \right) \simeq \xi^{}_{ij} \left( \Lambda^{}_{\rm EW} \right) \simeq 1$ (for $ij = 21, 31, 32 $) hold pretty well, and thus the results for flavor mixing angles and the Dirac CP-violating phase can be largely simplified and their evolution behaviors are more transparent. But similar approximations for $\zeta^{}_{ij} \left( \mu \right)$ (for $ij = 21, 31, 32$) are not good in the Majorana case.
	\item The inverted mass ordering (IMO) with $m^{}_3 \left( \Lambda^{}_{\rm EW} \right) = 0.001$ eV. In the Majorana case, $\zeta^{-1}_{21} \left( \mu \right) \gg - \zeta^{-1}_{31} \left( \mu \right) \simeq -\zeta^{-1}_{32} \left( \mu \right) \simeq -\zeta^{}_{31} \left( \mu \right) \simeq -\zeta^{}_{32} \left( \mu \right) \simeq 1 \gg \zeta^{}_{21} \left( \mu \right) $ holds quite well. Similarly, the approximations $\xi^{}_{21} \left( \mu \right) \gg -\xi^{}_{31} \left( \mu \right) \simeq -\xi^{}_{32} \left( \mu \right) \simeq 1 $ are excellent in the Dirac case. These approximations can make results much simpler and clearer.
\end{itemize}
Here, we do not consider the case of nearly degenerate neutrino masses in which $\zeta^{-1}_{21} \left( \mu \right) $ or $\xi^{}_{21} \left( \mu \right)$ is so strongly enhanced that $\Delta \theta^{}_{12}$, $\Delta \delta$, $\Delta \rho$ and $\Delta \sigma$ containing $\zeta^{-1}_{21} \left( \mu \right) $ or $\xi^{}_{21} \left( \mu \right)$ are significantly enlarged and the approximations that they are small quantities become bad especially in the MSSM with a large $\tan\beta$. Hence the analytical results for $\Delta \theta^{}_{12}$, $\Delta \delta$, $\Delta \rho$ and $\Delta \sigma$ can remarkably deviate from the corresponding exact results in this case. Actually, we also confront this situation in the MSSM with a sizeable $\tan\beta$ for the inverted neutrino mass spectrum but it is not severe and thus acceptable when $\tan\beta \lesssim 30$. It is worth mentioning that the normal neutrino mass ordering is currently favored over the inverted one at the level of around $3\sigma$ indicated by a globle analysis of current neutrino oscillation data and the total neutrino mass is constrained to be $\sum m^{}_\nu < 0.12$ eV by some cosmology observations~\cite{Capozzi:2020qhw,deSalas:2020pgw,Vagnozzi:2017ovm,Aghanim:2018eyx}. The latter infers that nearly degenerate neutrino masses are disfavored by the CMB anisotropies at $2.4\sigma$ level or at $5.9\sigma$ level after the BAO data are added~\cite{Lattanzi:2020iik}. In the numerical analysis, we only exhibit the numerical results in the MSSM with $\tan\beta = 10$ or $30$ and neglect those in the SM because in the SM, the RGE effects are extremely small and hence all analytical results coincide with the exact ones very well even in the case where neutrino masses are nearly degenerate. But this does not mean that these small RGE-induced effects in the SM are inessential, on the contrary, sometimes they can play a greatly important role, such as establishing a direct connection between the CP-violating asymmetries at a superhigh energy scale and CP violation at the electroweak scale via the seesaw bridge~\cite{Xing:2020erm,Xing:2020ghj}.

To compute the evolution of flavor mixing parameters and the Jarlskog invariant from $\Lambda$ down to $\mu$, we choose the initial values of relevant parameters at $\Lambda$ in such a way that the best-fit values of $\theta^{}_{12}$, $\theta^{}_{13}$, $\theta^{}_{23}$, $\delta m^2$ and $\Delta m^2$ at $\Lambda^{}_{\rm EW}$ shown in Eqs.~(34) and (35) can be satisfied, and some given values of $\delta$ (together with $\rho$ and $\sigma$ in the Majorana case) and the lightest neutrino mass at $\Lambda$ or $\Lambda^{}_{\rm EW}$ are initially input or can be achieved. Therefore the initial inputs at $\Lambda$ may not be the same in different cases. With these initial inputs, we calculate both the exact results by numerically solving the RGEs and the approximate ones by means of the analytical results we have obtained above.

\subsection{Neutrino masses and flavor mixing angles}

%%%%%%%%%%%%%%%%%%%%%%%%%%%%%%%% figure 2 %%%%%%%%%%%%%%%%%%%%%%%%%%%%%%%%%%%%%%%
\begin{figure}[t]
	\centering
	\includegraphics[width=\linewidth]{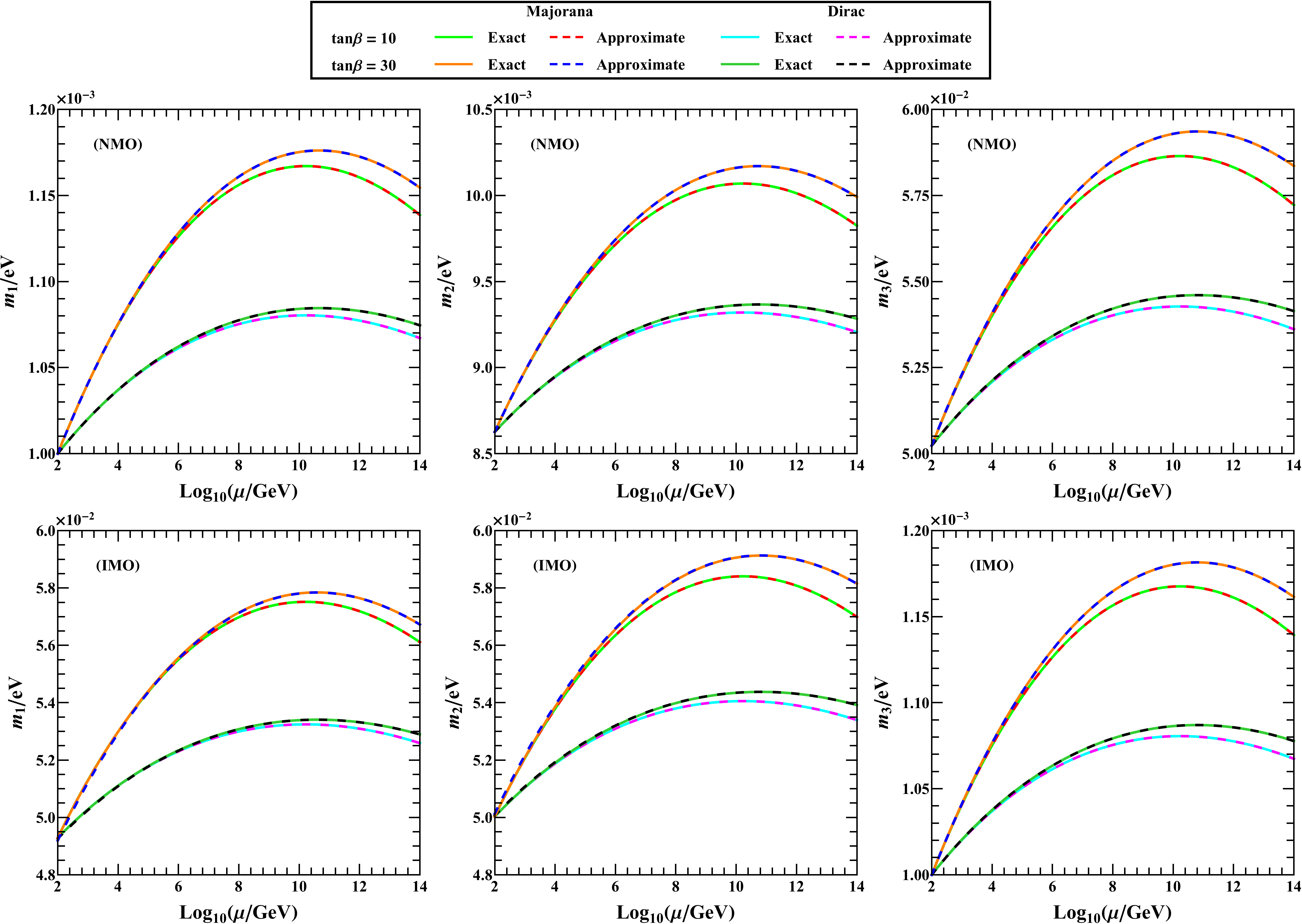}
	\caption{The evolution of $m^{}_i $ (for $i = 1, 2, 3$) against the energy scale $\mu$ in the Majorana case and the Dirac case with $\Lambda = 10^{14}$ GeV, where the MSSM with $\tan\beta = 10$ or $\tan\beta =30$ is considered and both the exact numerical results obtained by solving the complete RGEs and the approximate ones achieved from the analytical expressions given in Eq.~(19) or Eq.~(29) are illustrated.}
	\label{mass}
\end{figure}
%%%%%%%%%%%%%%%%%%%%%%%%%%%%%%%%%%%%%%%%%%%%%%%%%%%%%%%%%%%%%%%%%%%%%%%%%%%%%%%%%

\begin{figure}[h!]
	\centering
	\includegraphics[width=\linewidth]{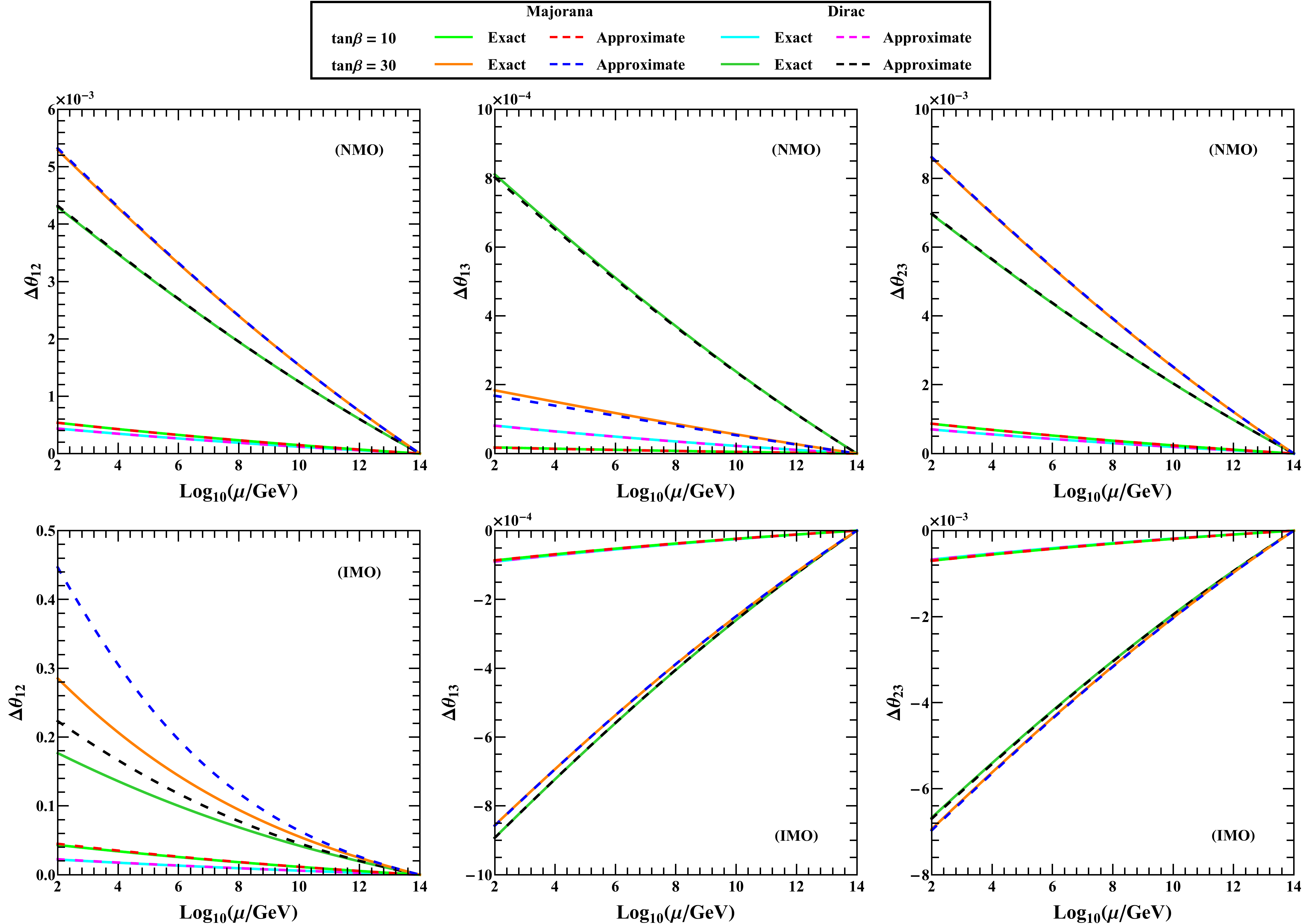}
	\caption{The evolution of $\Delta \theta^{}_{ij} $ (for $i = 12, 13, 23$) against the energy scale $\mu$ in the Majorana case and the Dirac case with $\Lambda = 10^{14}$ GeV, where the MSSM with $\tan\beta = 10$ or $\tan\beta =30$ is considered and both the exact numerical results obtained by solving the complete RGEs and the approximate ones achieved from the analytical expressions given in Eq.~(19) or Eq.~(29) are illustrated.}
	\label{angle}
\end{figure}

First, we compute the evolution of neutrino masses and flavor mixing angles. We require that $m^{}_1 \left( \Lambda^{}_{\rm EW} \right) = 0.001$ eV in the NMO case or $m^{}_3 \left( \Lambda^{}_{\rm EW} \right) = 0.001$ eV in the IMO case be satisfied, $\delta\left( \Lambda^{}_{\rm EW} \right)$ take its best-fit value shown in Eq.~(34), and $\rho \left( \Lambda \right) = \sigma \left( \Lambda \right) = 0$ be initially input in the Majorana case. The results for neutrino masses and flavor mixing angles are illustrated in Figs.~\ref{mass} and \ref{angle}, respectively. In particular, the values of neutrino masses and flavor mixing angles at $\Lambda^{}_{\rm EW}$ obtained with the help of the analytical expressions derived in section 2 are explicitly listed in Table~\ref{mass-angle}, where the corresponding numbers shown in the parentheses are the relative errors compared to the exact results acquired by numerically solving the one-loop RGEs.

%%%%%%%%%%%%%%%%%%%%%%%%%%%%%%%% figure 3 %%%%%%%%%%%%%%%%%%%%%%%%%%%%%%%%%%%%%%%
%%%%%%%%%%%%%%%%%%%%%%%%%%%%%%%%%%%%%%%%%%%%%%%%%%%%%%%%%%%%%%%%%%%%%%%%%%%%%%%%%

%%%%%%%%%%%%%%%%%%%%%%%%%%%%%%%%%%%%%% table 1 %%%%%%%%%%%%%%%%%%%%%%%%%%%%%%%%%%
\begin{table}[h!]
	\centering
	\caption{The values of $m^{}_i \left( \Lambda^{}_{\rm EW} \right)/ m^{}_i \left( \Lambda \right) $ (for $i=1, 2, 3$) and $\Delta \theta^{}_{ij} \left( \Lambda^{}_{\rm EW} \right)$ (for $ij= 12, 13, 23$) obtained by means of the analytical expressions derived in section 2 in the MSSM with $\tan\beta =10$ or $30$. The initial inputs at $\Lambda = 10^{14}$ GeV are chosen to guarantee that $m^{}_1 = 0.001$ eV or $m^{}_3 = 0.001$ eV and the best-fit values given in Eqs.~(34) and (35) can be achieved at $\Lambda^{}_{\rm EW} \simeq 100$ GeV, and in addition, $\rho = \sigma =0$ are initially input at $\Lambda$ in the Majorana case. The corresponding numbers given in the parentheses are the relative errors compared to the exact results obtained by numerically solving the one-loop RGEs.}
	\vspace{0.3cm}
	\resizebox{\textwidth}{!}{
		\begin{tabular}{c|p{3.4cm}<{\centering}p{3.4cm}<{\centering}|p{3.4cm}<{\centering}p{3.4cm}<{\centering}}
			\hline\hline
			& \multicolumn{2}{c|}{Normal neutrino mass ordering (NMO)} & \multicolumn{2}{c}{Inverted neutrino mass ordering (IMO)}
			\\
			\hline
			& $\tan\beta = 10$ & $\tan\beta = 30$ & $\tan\beta = 10$ & $\tan\beta = 30$
			\\
			\hline
			\multicolumn{5}{c}{Majorana neutrinos}
			\\
			\hline
			\multirow{2}{*}{$m^{}_1 \left( \Lambda^{}_{\rm EW} \right) / m^{}_1 \left( \Lambda \right)$} & $0.878$ & $0.866$ & $0.878$ & $0.867$
			\\
			& $\left( 7.4 \times 10^{-6} \right)$ & $\left( - 6.9 \times 10^{-5} \right)$ & $\left( -2.0 \times 10^{-5} \right)$ & $\left( -1.9 \times 10^{-3} \right)$ 
			\\
			\multirow{2}{*}{$m^{}_2 \left( \Lambda^{}_{\rm EW} \right) / m^{}_2 \left( \Lambda \right)$} & $0.878$ & $0.863$ & $0.878$ & $0.862$
			\\
			& $\left( 9.8 \times 10^{-6} \right)$ & $\left( -4.6 \times 10^{-5} \right)$ & $\left( 3.9 \times 10^{-5} \right)$ & $\left( 2.0 \times 10^{-3} \right)$ 
			\\
			\multirow{2}{*}{$m^{}_3 \left( \Lambda^{}_{\rm EW} \right) / m^{}_3 \left( \Lambda \right)$} & $0.878$ & $ 0.861$ & $0.878$ & $0.861$
			\\
			& $\left( 1.3 \times 10^{-5} \right)$ & $\left( 9.5 \times 10^{-5} \right)$ & $\left( 1.1 \times 10^{-5} \right)$ & $\left( -1.1 \times 10^{-4} \right)$ 
			\\
			\multirow{2}{*}{$\Delta \theta^{}_{12} \left( \Lambda^{}_{\rm EW} \right)$} & $ 5.36 \times 10^{-4}$ & $ 5.32 \times 10^{-3}$ & $ 4.49 \times 10^{-2}$ & $0.447$
			\\
			& $\left( -7.8 \times 10^{-4} \right)$ & $\left( 5.8 \times 10^{-3} \right)$ & $\left( 4.0 \times 10^{-2} \right)$ & $\left( 0.57 \right)$ 
			\\
			\multirow{2}{*}{$\Delta \theta^{}_{13} \left( \Lambda^{}_{\rm EW} \right)$} & $1.70 \times 10^{-5}$ & $ 1.68 \times 10^{-4}$ & $-8.63 \times 10^{-5}$ & $-8.58 \times 10^{-4}$
			\\
			& $\left( -4.1 \times 10^{-2} \right)$ & $\left( -8.6 \times 10^{-2} \right)$ & $\left( -3.7 \times 10^{-3} \right)$ & $\left( 1.4 \times 10^{-3} \right)$ 
			\\
			\multirow{2}{*}{$\Delta \theta^{}_{23} \left( \Lambda^{}_{\rm EW} \right)$} & $ 8.67 \times 10^{-4}$ & $8.61 \times 10^{-3}$ & $-7.01 \times 10^{-4}$ & $-6.97 \times 10^{-3}$
			\\
			& $\left( 3.5 \times 10^{-3} \right)$ & $\left(2.5 \times 10^{-3} \right)$ & $\left( 3.5 \times 10^{-3} \right)$ & $\left( 4.2 \times 10^{-3} \right)$
			\\
			\hline
			\multicolumn{5}{c}{Dirac neutrinos}
			\\
			\hline
			\multirow{2}{*}{$m^{}_1 \left( \Lambda^{}_{\rm EW} \right) / m^{}_1 \left( \Lambda \right)$} & $0.937$ & $0.931$ & $0.937$ & $0.931$
			\\
			& $\left(3.8 \times 10^{-6} \right)$ & $\left( -2.7 \times 10^{-5} \right)$ & $\left( -3.6 \times 10^{-6} \right)$ & $\left( -6.4 \times 10^{-4} \right)$ 
			\\
			\multirow{2}{*}{$m^{}_2 \left( \Lambda^{}_{\rm EW} \right) / m^{}_2 \left( \Lambda \right)$} & $0.937$ & $0.929$ & $0.937$ & $0.928$
			\\
			& $\left( 5.1 \times 10^{-6} \right)$ & $\left( -5.7 \times 10^{-6} \right)$ & $\left( 1.3 \times 10^{-5} \right)$ & $\left( 6.9 \times 10^{-4} \right)$ 
			\\
			\multirow{2}{*}{$m^{}_3 \left( \Lambda^{}_{\rm EW} \right) / m^{}_3 \left( \Lambda \right)$} & $0.937$ & $ 0.928$ & $0.937$ & $0.928$
			\\
			& $\left( 6.4 \times 10^{-6} \right)$ & $\left( 5.6 \times 10^{-5} \right)$ & $\left( 5.5 \times 10^{-6} \right)$ & $\left( -3.4 \times 10^{-5} \right)$ 
			\\
			\multirow{2}{*}{$\Delta \theta^{}_{12} \left( \Lambda^{}_{\rm EW} \right)$} & $ 4.35 \times 10^{-4}$ & $ 4.32 \times 10^{-3}$ & $ 2.25 \times 10^{-2}$ & $0.223$
			\\
			& $\left( -8.5 \times 10^{-4} \right)$ & $\left( 5.2 \times 10^{-3} \right)$ & $\left( 1.8 \times 10^{-2} \right)$ & $\left( 0.26 \right)$ 
			\\
			\multirow{2}{*}{$\Delta \theta^{}_{13} \left( \Lambda^{}_{\rm EW} \right)$} & $8.08 \times 10^{-5}$ & $ 8.03 \times 10^{-4}$ & $-8.99 \times 10^{-5}$ & $-8.93 \times 10^{-4}$
			\\
			& $\left( -5.8 \times 10^{-3} \right)$ & $\left( -1.1 \times 10^{-2} \right)$ & $\left( -3.8 \times 10^{-3} \right)$ & $\left( 4.0\times 10^{-4} \right)$ 
			\\
			\multirow{2}{*}{$\Delta \theta^{}_{23} \left( \Lambda^{}_{\rm EW} \right)$} & $ 7.01 \times 10^{-4}$ & $6.97 \times 10^{-3}$ & $-6.74 \times 10^{-4}$ & $-6.70 \times 10^{-3}$
			\\
			& $\left( 3.4 \times 10^{-3} \right)$ & $\left(2.7 \times 10^{-3} \right)$ & $\left( 3.5 \times 10^{-3} \right)$ & $\left( 4.1 \times 10^{-3} \right)$
			\\
			\hline\hline
	\end{tabular}}
	\label{mass-angle}
\end{table}
%%%%%%%%%%%%%%%%%%%%%%%%%%%%%%%%%%%%%%%%%%%%%%%%%%%%%%%%%%%%%%%%%%%%%%%%%%%%%%%%%

From Fig.~\ref{mass} and Table~\ref{mass-angle}, one can see that the approximate results for neutrino masses are consistent with the exact ones very well in all the cases we have considered. The evolution of neutrino masses is dominated by $m^{}_i \left( \mu \right) \simeq I^{}_{\beta} m^{}_i \left( \Lambda \right)$ (for $i=1, 2, 3$ and $\beta = \kappa$ or $\nu$). Considering the values of $I^{}_\kappa$ and $I^{}_\nu$ against $\mu$ shown in Fig.~\ref{loop-functions}, it is easy to understand that the evolution of neutrino masses in the Majorana case (or that with $\tan\beta = 30$) is slightly severer than that in the Dirac case (or that with $\tan\beta = 10$). The renormalization group corrections to the neutrino masses are larger in the SM than those in the MSSM, especially for Majorana neutrinos, which is indicated by $I^{}_\beta$ (for $\beta = \kappa$ or $\nu$) as it is shown in Fig.~\ref{loop-functions}.

As for the evolution of $\theta^{}_{ij}$ (for $i= 12, 13, 23$) shown in Fig.~\ref{angle} and Table~\ref{mass-angle}, the relative errors of approximate results for them are around $1\%$ or smaller at $\Lambda^{}_{\rm EW}$, except those for $\Delta \theta^{}_{12}$ with $\tan\beta = 30$ in the IMO case and $\Delta \theta^{}_{13}$ in the Majorana case with $\tan\beta =30$ and the normal neutrino mass ordering. Some discussions and comments on the evolution behaviors are as follows: 
\begin{itemize}
	\item Comparing the corresponding results with $\tan\beta = 10$ and $\tan\beta = 30$, such as those for Dirac neutrinos with the normal mass ordering in the MSSM where $\tan\beta = 10$ and $\tan\beta = 30$ are respectively considered, the results for $\Delta \theta^{}_{ij}$ (for $ij= 12, 13, 23$) with $\tan\beta = 30$ are about ten times larger than those with $\tan\beta = 10$ during the RGE running, mainly due to $\Delta \theta^{}_{ij} \propto \Delta^{}_\tau$ (for $ij = 12, 13, 23$) which are approximately proportional to $\tan^2\beta$.
	
	\item As can be seen from the left upper and lower panels of Fig.~\ref{angle}, the running of $\Delta \theta^{}_{12}$ can be largely enhanced in the IMO case, but the running direction keeps unchanged compared to that in the NMO case. The reason is that the expression for $\Delta \theta^{}_{12}$ given in Eq.~(20) or Eq.~(30) contains $\zeta^{-1}_{21}$ or $\xi^{}_{21}$ whose value can be strongly enhanced in the IMO case. To make it more distinct, we notice that $\Delta \theta^{}_{12}$ roughly approximates to $\Delta \theta^{}_{12} \simeq - 1/2 \Delta^{}_\tau \zeta^{-1}_{21} ({\rm or} ~\xi^{}_{21}) \sin 2\theta^{}_{12} \sin^2\theta^{}_{23}$ in the Majorana (or Dirac) case, where terms contain $\sin\theta^{}_{13}$ have been neglected owing to the smallness of $\theta^{}_{13}$. It is clear that $\Delta \theta^{}_{12}$ is always positive and enhanced by the largeness of $\zeta^{-1}_{21}$ or $\xi^{}_{21}$ in the IMO case. Especially, when $\tan\beta = 30$ is taken in the IMO case, $\Delta \theta^{}_{12}$ is extremely enlarged and no longer a small quantity, thus the approximate results for $\Delta \theta^{}_{12}$ in those cases severely deviate from the exact ones as shown in the left lower panel of Fig.~\ref{angle} and Table~\ref{mass-angle}.
	
	\item To understand the evolution of $\Delta \theta^{}_{13}$ shown in the middle panels of Fig.~\ref{angle}, one may consider the neutrino mass spectra and the smallness of $\theta^{}_{13}$ to simplify the analytical expression for $\theta^{}_{13}$ as $\Delta \theta^{}_{13} \simeq 1/2 \Delta^{}_\tau \sin2\theta^{}_{13} \cos^2\theta^{}_{23}$ in the IMO case for both Majorana and Dirac neutrinos, and $\Delta \theta^{}_{13} \simeq - 1/2 \Delta^{}_\tau \sin2\theta^{}_{13} \cos^2\theta^{}_{23}$ in the NMO case for Dirac neutrinos, where the sign difference is induced by different signs of $\zeta^{}_{3i}$ or $\xi^{}_{3i}$ (for $i = 1, 2$) in the normal and inverted neutrino mass ordering cases. It is apparent that the evolution of $\Delta \theta^{}_{13}$ is suppressed by the smallness of $\theta^{}_{13}$, and the running directions are opposite in the normal and inverted neutrino mass ordering cases but the absolute values of theirs are nearly equal for Dirac neutrinos, as shown in Fig.~\ref{angle} and Table~\ref{mass-angle}. Additionally, indicated by the approximate analytical results, the corresponding values for $\Delta \theta^{}_{13}$ in the IMO case for Majorana and Dirac neutrinos with the same $\tan\beta$ are negative and roughly equal, and this  can also be transparently seen in the middle-lower panel of Fig.~\ref{angle} and Table~\ref{mass-angle}. The evolution of $\Delta \theta^{}_{13}$ in the NMO case for Majorana neutrinos is relatively exotic since its value seems to be largely reduced compared to other cases, as can be seen in the middle-upper panel of Fig.~\ref{angle}. In fact, carefully checking the analytical expression of $\Delta \theta^{}_{13}$ given in Eq.~(20), one may discover that there is a large cancellation in this case, and this is also the reason why the relative error of the approximate result for $\Delta \theta^{}_{13}$ is large and reaches $-4\%$ with $\tan\beta = 10$ or $-8.6\%$ with $\tan\beta = 30$ at $\Lambda^{}_{\rm EW}$.
	
	\item The analytical results for $\Delta \theta^{}_{23}$ are consistent with the exact ones very well in all the cases under consideration and it is quite easy to understand its evolution behaviors in different cases, as shown in the right panels of Fig.~\ref{angle}. After the specific neutrino mass spectra are taken into account, the analytical expressions for $\Delta \theta^{}_{23}$ given in Eqs.~(20) and (30) can be further simplified to $\Delta \theta^{}_{23} \simeq -1/2 \Delta^{}_\tau \sin2\theta^{}_{23}$ (or $\Delta \theta^{}_{23} \simeq 1/2 \Delta^{}_\tau \sin2\theta^{}_{23}$) in the normal (or inverted) neutrino mass ordering case. Therefore, $\Delta \theta^{}_{23}$ is positive in the NMO case and negative in the IMO case, but its absolute values for these two neutrino mass spectra with the same $\tan\beta$ are roughly equal not only in the Majorana case but also in the Dirac case, as shown in the right upper and lower panels of Fig.~\ref{angle} and Table~\ref{mass-angle}. And with the same neutrino mass spectrum and $\tan\beta$, the values of $\Delta \theta^{}_{23}$ in the Majorana and Dirac cases are also nearly equal. As indicated by the right upper panel of Fig.~\ref{angle} and Table~\ref{mass-angle}, the results for $\Delta \theta^{}_{23}$ in the NMO case for Majorana and Dirac neutrinos slightly depart from each other to some extent. It is mainly because in the NMO case for Majorana neutrinos, $\zeta^{-1}_{32} \simeq 1 $ is not a good approximation.
\end{itemize}

From the above results and discussions, it is interesting to see that the evolution behaviors of neutrino masses and flavor mixing angles in the Majorana case even with initially vanishing Majorana CP-violating phases (i.e., $\rho \left( \Lambda \right) = \sigma \left(\Lambda \right) = 0$) can be distinguished from those in the Dirac case by their RGE running strengths~\cite{Xing:2006sp}. One may also consider the initially nonvanishing Majorana CP-violating phases in the Majorana case, and find that the cancellation in $\Delta \theta^{}_{13}$ is weakened or disappears in the NMO case.

\subsection{CP-violating phases and the Jarlskog invariant}

\begin{figure}[t!]
	\centering
	\includegraphics[width=\linewidth]{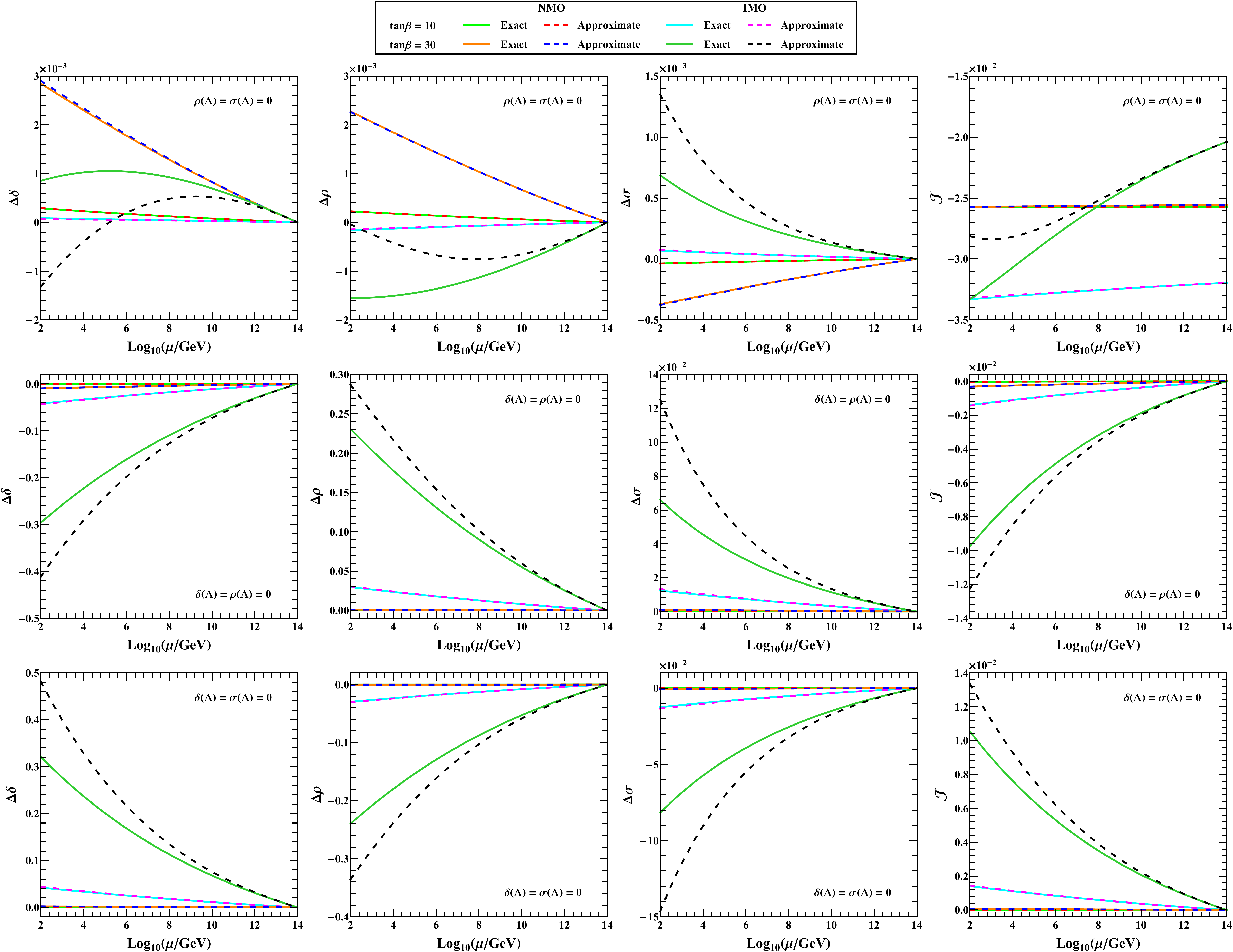}
	\caption{The evolution of three CP-violating phases (i.e., $\Delta \delta$, $\Delta \rho$ and $\Delta \sigma$) and the Jarlskog invariant $\mathcal{J}$ against the energy scale $\mu$ in the Majorana case with $\Lambda = 10^{14}$ GeV, where the MSSM with $\tan\beta = 10$ or $\tan\beta =30$ is considered and both the exact numerical results obtained by solving the complete RGEs and the approximate ones achieved from the analytical expressions given in Eqs.~(21)---(25) are illustrated. The results shown in each row correspond to one set of inputs for three CP-violating phases, as labelled in  each subfigure.}
	\label{phaseM}
\end{figure}

\begin{figure}[t!]
	\centering
	\includegraphics[width=\linewidth]{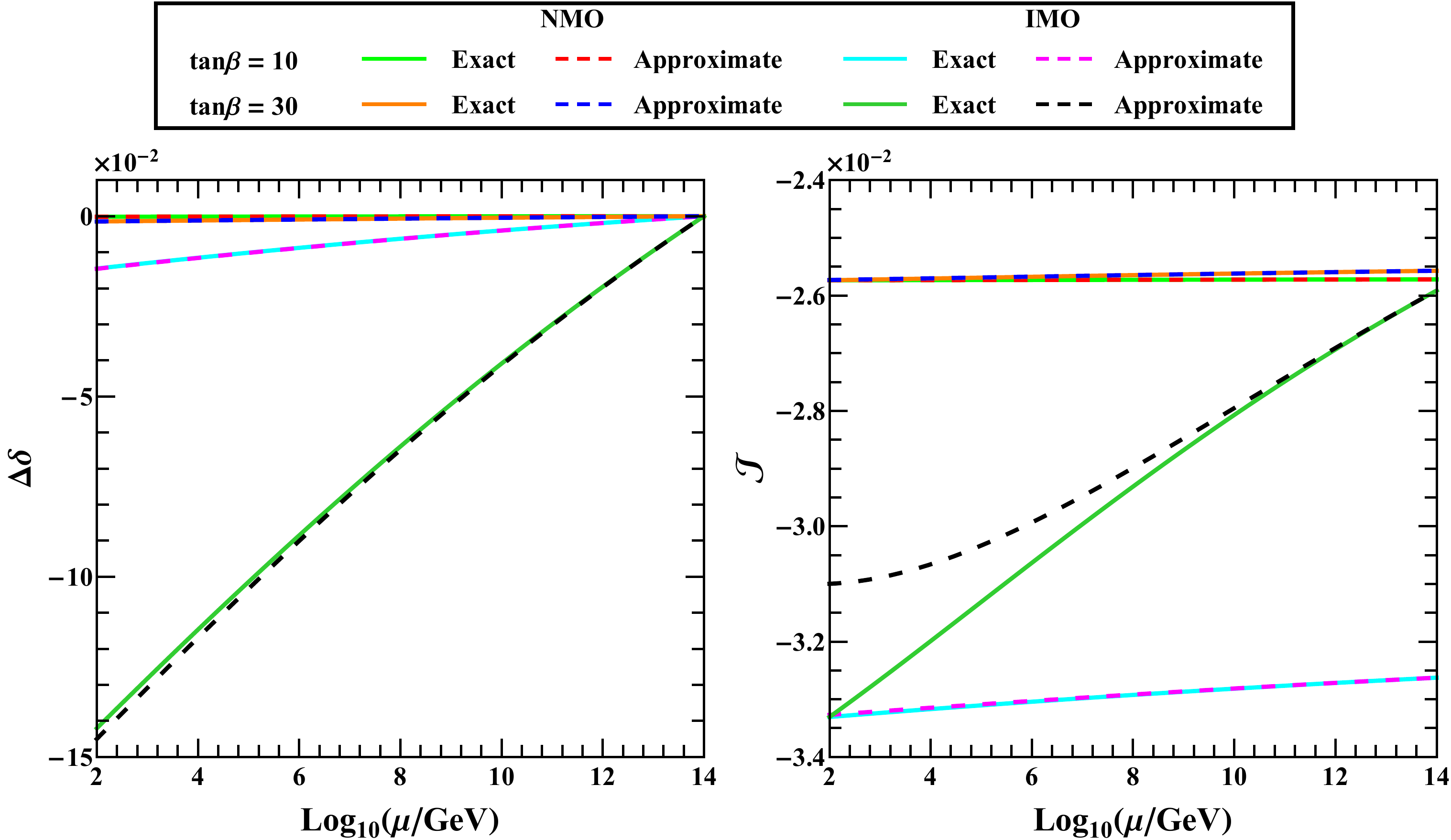}
	\caption{The evolution of the Dirac CP-violating phase $\Delta \delta$ and the Jarlskog invariant $\mathcal{J}$ against the energy scale $\mu$ in the Dirac case with $\Lambda = 10^{14}$ GeV, where the MSSM with $\tan\beta = 10$ or $\tan\beta =30$ is considered and both the exact numerical results obtained by solving the complete RGEs and the approximate ones achieved from the analytical expressions given in Eqs.~(31)---(33) are illustrated.}
	\label{phaseD}
\end{figure}

The evolution behaviors of CP-violating phases and the Jarlskog invariant in the Majorana case are much more complicated than those in the Dirac case as shown in Eqs.~(20) and (30), since in the Majorana case there are two additional Majorana CP-violating phases, whose evolution behaviors are entangled with that of the Dirac CP-violating phase. Thus as discussed in section 2, the Dirac CP-violating phase $\delta$ and the Jarlskog invariant $\mathcal{J}$  in the Majorana case can be radiatively generated via the one-loop RGE running unless both $\rho$ and $\sigma$ are initially equal to $0$ or $\pi/2$, while those in the Dirac case can not be radiatively generated by means of the one-loop RGE running. This is important and intuitive for us to distinguish between the evolution of relevant flavor parameters in the Majorana and Dirac cases, and understand CP violation at the eletroweak scale.  In this subsection, we are going to compare the evolution of $\delta$ and $\mathcal{J}$ in the Majorana and Dirac cases, and discuss the entanglements of three CP-violating phases in the Majorana case. For our purposes, we choose some sets of initial inputs in the Majorana case, to guarantee that at the electroweak scale $\Lambda^{}_{\rm EW}$, $m^{}_1 = 0.001$ eV or $m^{}_3 = 0.001$ eV, the best-fit values of two neutrino mass-squared differences and three flavor mixing angles can be achieved, and they satisfy one of the following requirements:
\begin{itemize}
	\item $\delta \left( \Lambda^{}_{\rm EW} \right)$ takes its best-fit value given in Eq.~(34), and $\rho \left( \Lambda \right) = \sigma \left( \Lambda \right) = 0$ are initially input;
	\item $\sigma \left( \Lambda^{}_{\rm EW} \right) = \pi/4$ can be achieved, and $\delta \left( \Lambda \right) = \rho \left( \Lambda \right) = 0$ are initially input;
	\item $\rho \left( \Lambda^{}_{\rm EW} \right) = \pi/4$ can be achieved, and $\delta \left( \Lambda \right) = \sigma \left( \Lambda \right) = 0$ are initially input.
\end{itemize}
In the Dirac case, we only consider the initial inputs, from which $m^{}_1 \left( \Lambda^{}_{\rm EW} \right) = 0.001$ eV or $m^{}_3 \left( \Lambda^{}_{\rm EW} \right) = 0.001$ eV can be satisfied, and two neutrino mass-squared differences, three flavor mixing angles and the Dirac CP-violating phase can take their best-fit values given in Eqs.~(34) and (35) at $\Lambda^{}_{\rm EW}$. With these initial inputs, the results for CP-violating phases and the Jarlskog invariant are plotted in Fig.~\ref{phaseM} for Majorana neutrinos and in Fig.~\ref{phaseD} for Dirac neutrinos. For illustration, the values of $\Delta \delta$, $\Delta \rho$, $\Delta \sigma$ and $\mathcal{J}$ at $\Lambda^{}_{\rm EW}$ in the Majorana case obtained with the help of Eqs.~(21)---(25) are listed in Table~\ref{phaseMt}, and those of $\Delta \delta$ and $\mathcal{J}$ at $\Lambda^{}_{\rm EW}$ in the Dirac case obtained by the aid of Eqs.~(31)---(33) are listed in Table~\ref{phaseDt}. Again, the corresponding numbers given in the parentheses are the relative errors compared to the exact results obtained by numerically solving the one-loop RGEs. At the first sight of Figs.~\ref{phaseM} and \ref{phaseD} together with Tables~\ref{phaseMt} and \ref{phaseDt}, one can find that results in the NMO case coincide with the exact results pretty well, no matter which type of neutrinos (Majorana or Dirac) is and how large the value of $\tan\beta$ (10 or 30) is, and the relative errors are of $\mathcal{O}\left(1\%\right)$ or smaller at $\Lambda^{}_{\rm EW}$. Since generally both the CP-violating phases and the Jarlskog invariant contain $\zeta^{-1}_{21}$ or $\xi^{}_{21}$ which can be largely enlarged in the IMO case, the results in the IMO case are worse, especially those with $\tan\beta =30$ in the Majorana case, but the relative errors of results with $\tan\beta = 10$ in the Majorana case or those in the Dirac case are mostly $\mathcal{O} \left(10\%\right)$ or smaller at $\Lambda^{}_{\rm EW}$, which are essentially acceptable. To understand the evolution behaviors of these CP-violating phases and the Jarlskog invariant, we first make some further approximations for those analytical formulas by taking into account the neutrino mass spectra and initial inputs under consideration, as well as the smallness of $\theta^{}_{13}$. Though these simplified versions may not be consistent with the original ones very well, they can illustrate the salient properties of the evolution. We begin with those in the Majorana case.
%%%%%%%%%%%%%%%%%%%%%%%%%%%%%%%% figure 4 %%%%%%%%%%%%%%%%%%%%%%%%%%%%%%%%%%%%%%%
%%%%%%%%%%%%%%%%%%%%%%%%%%%%%%%%%%%%%%%%%%%%%%%%%%%%%%%%%%%%%%%%%%%%%%%%%%%%%%%%%

%%%%%%%%%%%%%%%%%%%%%%%%%%%%%%%% figure 5 %%%%%%%%%%%%%%%%%%%%%%%%%%%%%%%%%%%%%%%
%%%%%%%%%%%%%%%%%%%%%%%%%%%%%%%%%%%%%%%%%%%%%%%%%%%%%%%%%%%%%%%%%%%%%%%%%%%%%%%%% 

%%%%%%%%%%%%%%%%%%%%%%%%%%%%%%%%%%%%%% table 2 %%%%%%%%%%%%%%%%%%%%%%%%%%%%%%%%%%
\begin{table}[h!]
	\centering
	\caption{The values of $\Delta \delta \left( \Lambda^{}_{\rm EW} \right)$, $\Delta \rho \left( \Lambda^{}_{\rm EW} \right) $, $\Delta \sigma \left( \Lambda^{}_{\rm EW} \right)$ and $\mathcal{J} \left( \Lambda^{}_{\rm EW} \right)$ obtained by means of the analytical expressions derived in section 2 for Majorana neutrinos in the MSSM with $\tan\beta =10$ or $30$. The initial inputs at $\Lambda = 10^{14}$ GeV are chosen to guarantee that $m^{}_1 = 0.001$ eV or $m^{}_3 = 0.001$ eV and the best-fit values of two neutrino mass-squared differences and three flavor mixing angles given in Eqs.~(34) and (35) can be achieved at $\Lambda^{}_{\rm EW} \simeq 100$ GeV, and in addition, three CP-violating phases are required to be (1) $\delta \left( \Lambda^{}_{\rm EW} \right) = 1.28\pi$ or $1.52\pi$ in the normal or inverted neutrino mass ordering case and $\rho \left( \Lambda \right) = \sigma \left( \Lambda \right) =0$; (2) $\sigma \left( \Lambda^{}_{\rm EW} \right) = \pi/4$ and $\delta \left( \Lambda \right) = \rho \left( \Lambda \right) =0$; (3) $\rho \left( \Lambda^{}_{\rm EW} \right) = \pi/4$ and $\delta \left( \Lambda \right) = \sigma \left( \Lambda \right) =0$. The corresponding numbers given in the parentheses are the relative errors compared to the exact results obtained by numerically solving the one-loop RGEs.}
	\vspace{0.3cm}
	\resizebox{\textwidth}{!}{
		\begin{tabular}{c|ccc|ccc}
			\hline\hline
			& \multicolumn{3}{c|}{$\tan \beta= 10$} &  \multicolumn{3}{c}{$\tan \beta = 30$}
			\\ \hline
			& $\rho \left( \Lambda \right) = \sigma \left( \Lambda \right) = 0$ & $ \delta \left( \Lambda \right) = \rho \left( \Lambda \right) = 0 $ &  $ \delta \left( \Lambda \right) = \sigma \left( \Lambda \right) = 0 $ & $\rho \left( \Lambda \right) = \sigma \left( \Lambda \right) = 0$ & $ \delta \left( \Lambda \right) = \rho \left( \Lambda \right) =0 $ &  $ \delta \left( \Lambda \right) = \sigma \left( \Lambda \right) =0 $
			\\ \hline
			\multicolumn{7}{c}{Normal neutrino mass ordering (NMO)}
			\\ \hline
			\multirow{2}{*}{$\Delta \delta \left( \Lambda^{}_{\rm EW} \right)$} & $2.92 \times 10^{-4}$ & $-9.47 \times 10^{-4}$ & $2.14 \times 10^{-4}$ & $2.90\times 10^{-3}$ & $-9.42 \times 10^{-3}$ & $2.12\times 10^{-3}$ 
			\\
			& $\left(6.9 \times 10^{-3}\right)$ & $\left(2.1 \times 10^{-3}\right)$ & $\left(-2.9 \times 10^{-4} \right)$ & $\left(2.0 \times 10^{-2}\right)$ & $\left(4.7 \times 10^{-3}\right)$ & $\left(5.8 \times 10^{-3}\right)$  
			\\
			\multirow{2}{*}{$\Delta \rho \left( \Lambda^{}_{\rm EW} \right)$} & $2.29 \times 10^{-4}$ & $1.15 \times 10^{-4}$ & $-1.06 \times 10^{-4}$ & $2.27 \times 10^{-3}$ & $1.16 \times 10^{-3}$ & $-1.06 \times 10^{-3}$
			\\
			& $\left( 2.8 \times 10^{-3} \right)$ & $\left(1.5 \times 10^{-3} \right)$ & $\left( -2.0 \times 10^{-3} \right)$ & $\left( -4.2 \times 10^{-3} \right)$ & $\left( 4.4 \times 10^{-2} \right)$ & $\left( 1.1 \times 10^{-2} \right)$
			\\ 
			\multirow{2}{*}{$\Delta \sigma \left( \Lambda^{}_{\rm EW} \right)$} & $-3.78 \times 10^{-5}$ & $1.14 \times 10^{-4}$ & $-4.55 \times 10^{-5}$ & $-3.78 \times 10^{-4}$ & $1.13 \times 10^{-3}$ & $-4.51 \times 10^{-4}$
			\\
			& $\left( 5.2 \times 10^{-3} \right)$ & $\left( 3.1 \times 10^{-3} \right)$ & $\left( -2.4 \times 10^{-3} \right)$ & $\left( 1.8 \times 10^{-2} \right)$ & $\left(3.0 \times 10^{-2} \right)$ & $\left( 9.8 \times 10^{-3} \right)$
			\\ 
			\multirow{2}{*}{$\mathcal{J} \left( \Lambda^{}_{\rm EW} \right)$} & $-2.57 \times 10^{-2}$ & $-3.16 \times 10^{-5}$ & $7.16 \times 10^{-6}$ & $-2.57 \times 10^{-2}$ & $-3.14 \times 10^{-4}$ & $7.14 \times 10^{-5}$
			\\
			& $\left( -6.3 \times 10^{-6} \right)$ & $\left( 1.9 \times 10^{-3} \right)$ & $\left( 3.5 \times 10^{-4} \right)$ & $\left( -2.7 \times 10^{-4} \right)$ & $\left( 2.4 \times 10^{-3} \right)$ & $\left( 1.2 \times 10^{-2} \right)$
			\\ \hline
			\multicolumn{7}{c}{Inverted neutrino mass ordering (IMO)}
			\\ \hline
			\multirow{2}{*}{$\Delta \delta \left( \Lambda^{}_{\rm EW} \right)$} & $6.42 \times 10^{-5}$ & $-4.39 \times 10^{-2}$ & $4.40 \times 10^{-2}$ & $-1.32 \times 10^{-3}$ & $-0.411$ & $0.482$
			\\
			& $\left( -0.25 \right)$ & $\left( 4.2 \times 10^{-2} \right)$ & $\left( 4.3 \times 10^{-2} \right)$ & $\left( -2.5 \right)$ & $\left( 0.39 \right)$ & $\left( 0.50 \right)$
			\\ 
			\multirow{2}{*}{$\Delta \rho \left( \Lambda^{}_{\rm EW} \right)$} & $-1.42 \times 10^{-4}$ & $3.06 \times 10^{-2}$ & $-3.07 \times 10^{-2}$ & $-4.47 \times 10^{-5}$ & $0.287$ & $-0.336$
			\\
			& $\left( -9.5 \times 10^{-2} \right)$ & $\left( 2.8 \times 10^{-2} \right)$ & $\left( 3.0 \times 10^{-2} \right)$ & $\left( -0.97 \right)$ & $\left( 0.24 \right)$ &  $\left( 0.40 \right)$
			\\ 
			\multirow{2}{*}{$\Delta \sigma \left( \Lambda^{}_{\rm EW} \right)$} & $7.61 \times 10^{-5}$ & $1.33 \times 10^{-2}$ & $-1.34 \times 10^{-2}$ & $1.35 \times 10^{-3}$ & $0.125$ & $-0.146$
			\\
			& $\left( 0.10 \right)$ & $\left( 7.5 \times 10^{-2} \right)$ & $\left( 7.3 \times 10^{-2} \right)$ & $\left( 0.97 \right)$ & $\left( 0.90 \right)$ & $\left( 0.79 \right)$
			\\ 
			\multirow{2}{*}{$\mathcal{J} \left( \Lambda^{}_{\rm EW} \right)$} & $-3.32 \times 10^{-2}$ & $-1.44 \times 10^{-3}$ & $1.45 \times 10^{-3}$ & $-2.81 \times 10^{-2}$ & $-1.22 \times 10^{-2}$ & $1.34 \times 10^{-2}$
			\\
			& $\left( -3.8 \times 10^{-3} \right)$ & $\left( 2.7 \times 10^{-2} \right)$ & $\left( 2.7 \times 10^{-2} \right)$ & $\left( -0.16 \right)$ & $\left( 0.26 \right)$ &  $\left( 0.27 \right)$
			\\
			\hline\hline
	\end{tabular}}
	\label{phaseMt}
\end{table}
%%%%%%%%%%%%%%%%%%%%%%%%%%%%%%%%%%%%%%%%%%%%%%%%%%%%%%%%%%%%%%%%%%%%%%%%%%%%%%%%%

\begin{center}
	{\bf (1) $\bm{\rho \left( \Lambda \right) = \sigma \left( \Lambda \right) = 0}$}
\end{center}

Eqs.~(21)---(25) can be simplified to 
\begin{eqnarray}
\Delta \delta \simeq \hspace{-0.6cm}&& \frac{\Delta^{}_\tau}{2} \sin\delta \left[ \left( \zeta^{-1}_{32} - \zeta^{-1}_{31} \right) \frac{\sin 2\theta^{}_{12} \sin 2\theta^{}_{23}}{2\sin\theta^{}_{13}} - 2\left( \zeta^{}_{32} - \zeta^{-1}_{32} \right) \sin^2\theta^{}_{12} \cos^2\theta^{}_{23} \cos\delta \right.
\nonumber
\\
&& - \left. \left( \zeta^{-1}_{32} \cos^4\theta^{}_{12} - \zeta^{-1}_{31} \sin^4\theta^{}_{12} + \zeta^{}_{21}  \right) \frac{2\sin\theta^{}_{13}\sin 2\theta^{}_{23}}{\sin 2\theta^{}_{12}} \right] \;,
\nonumber
\\
\Delta \rho \simeq \hspace{-0.6cm}&& \frac{\Delta^{}_\tau}{2} \sin\theta^{}_{13} \sin\delta \left[ \left( \zeta^{}_{31} \sin 2\theta^{}_{12} + \zeta^{}_{21} \cot\theta^{}_{12} + \zeta^{-1}_{32} \cos 2\theta^{}_{12} \cot\theta^{}_{12} \right) \sin 2\theta^{}_{23} \right.
\nonumber
\\
&& + \left. \left(  \zeta^{-1}_{31} - \zeta^{}_{32} \right) \sin 2\theta^{}_{12} \cos 2\theta^{}_{23} \cot\theta^{}_{23} \right] \;,
\nonumber
\\
\Delta \sigma \simeq \hspace{-0.6cm}&& \frac{\Delta^{}_\tau}{2} \sin\theta^{}_{13} \sin\delta \left[ \left( \zeta^{}_{32} - \zeta^{-1}_{32} \right) \sin 2\theta^{}_{23} \sin 2\theta^{}_{12} + \left( \zeta^{}_{21} + \zeta^{}_{31} \cos 2\theta^{}_{12} \right)  \sin 2\theta^{}_{23} \tan\theta^{}_{12} \right.
\nonumber
\\
&& - \left. \left( \zeta^{}_{32} - \zeta^{-1}_{31} \cos 2\theta^{}_{23} \right) \sin 2\theta^{}_{12} \cot\theta^{}_{23} \right] \;,
%	(36)
\end{eqnarray}
and $\mathcal{J} \left( \mu \right) \simeq  \left( 1 - \Delta^{}_\tau C^{(1)}_{\rm M}  \right) \mathcal{J} \left( \Lambda \right)$ with
\begin{eqnarray}
C^{(1)}_{\rm M} \simeq \hspace{-0.6cm}&& \left( \zeta^{}_{31} \cos^2\theta^{}_{12} + \zeta^{}_{32} \sin^2\theta^{}_{12} \right) \cos^2\theta^{}_{23} + \left( \zeta^{-1}_{31} \sin^2\theta^{}_{12} + \zeta^{-1}_{32} \cos^2\theta^{}_{12} \right) \cos 2\theta^{}_{23}
\nonumber
\\
&& + \zeta^{-1}_{21} \sin^2 \theta^{}_{23} \cos 2\theta^{}_{12} \;,
%	(37)
\end{eqnarray}
for the normal neutrino mass ordering; or 
\begin{eqnarray}
\Delta \delta \simeq \hspace{-0.6cm}&& \Delta^{}_\tau \sin\theta^{}_{13} \sin 2\theta^{}_{23} \cot 2\theta^{}_{12} \sin\delta \;,
\nonumber
\\
\Delta\rho \simeq \hspace{-0.6cm}&& -\frac{\Delta^{}_\tau}{2} \sin\theta^{}_{13} \sin 2\theta^{}_{23} \sin\delta \left( \cos 2\theta^{}_{12} \cot\theta^{}_{12} + \sin 2\theta^{}_{12} \right) \;,
\nonumber
\\
\Delta\sigma \simeq \hspace{-0.6cm}&& \frac{\Delta^{}_\tau}{2} \sin\theta^{}_{13} \sin 2\theta^{}_{23} \sin\delta \left( \sin 2\theta^{}_{12} - \cos 2\theta^{}_{12} \tan\theta^{}_{12} \right) \;,
%	(38)
\end{eqnarray}
and $\mathcal{J} \left( \mu \right) \simeq  \left( 1 - \Delta^{}_\tau C^{(1)}_{\rm M}  \right) \mathcal{J} \left( \Lambda \right)$ with
\begin{eqnarray}
\nonumber
\\
C^{(1)}_{\rm M} \simeq \hspace{-0.6cm}&& \zeta^{-1}_{21} \sin^2\theta^{}_{23} \cos 2\theta^{}_{12} \;,
%	(39)
\end{eqnarray}
for the inverted neutrino mass ordering. In both cases, $C^{(2)}_{\rm M} \simeq 0 $ holds.

\begin{center}
	{\bf (2) $\bm{ \delta \left( \Lambda \right) = \rho \left( \Lambda \right) = 0 }$}
\end{center}

Given $\sigma \left( \Lambda^{}_{\rm EW} \right) = \pi/4$, Eqs.~(21) and (22) can approximate to
\begin{eqnarray}
\Delta \delta \simeq \hspace{-0.6cm}&& \frac{\Delta^{}_\tau}{2} \left[ \left( \zeta^{}_{32} - \zeta^{-1}_{32} \right) \left( \cos^2\theta^{}_{12} \cos 2\theta^{}_{23} - \sin^2\theta^{}_{12} \cos^2\theta^{}_{23} - \frac{\sin 2\theta^{}_{12} \sin 2\theta^{}_{23}}{4\sin\theta^{}_{13}} \right) \right.
\nonumber
\\
&& - \left. \left( \zeta^{}_{21} - \zeta^{-1}_{21} \right) \sin^2\theta^{}_{23} \vphantom{\frac{1}{1}} \right] \;,
\nonumber
\\
\Delta\rho \simeq \hspace{-0.6cm}&& \frac{\Delta^{}_\tau}{2} \cos^2\theta^{}_{12} \left[ \left( \zeta^{}_{21} -\zeta^{-1}_{21} \right) \sin^2\theta^{}_{23} - \left( \zeta^{}_{32} - \zeta^{-1}_{32} \right) \cos 2\theta^{}_{23} \right] \;,
\nonumber
\\
\Delta\sigma \simeq \hspace{-0.6cm}&& \frac{\Delta^{}_\tau}{2} \sin^2\theta^{}_{12} \left[ \left( \zeta^{}_{21} - \zeta^{-1}_{21} \right) \sin^2\theta^{}_{23} - \left( \zeta^{}_{32} - \zeta^{-1}_{32} \right) \cos 2\theta^{}_{23} \cot^2\theta^{}_{12}  \right] \;,
%\nonumber
%\\
%C^{(2)}_{\rm M} \simeq \hspace{-0.6cm}&& 4\frac{\Delta \delta}{\Delta^{}_\tau} \;,
%	(40)
\end{eqnarray}
in the normal neutrino mass ordering case; or
\begin{eqnarray}
\Delta\delta \simeq \hspace{-0.6cm}&& \frac{\Delta^{}_\tau}{2} \zeta^{-1}_{21} \sin^2\theta^{}_{23} \;,
\nonumber
\\
\Delta\rho \simeq \hspace{-0.6cm}&& - \frac{\Delta^{}_\tau}{2} \zeta^{-1}_{21} \sin^2\theta^{}_{23} \cos^2\theta^{}_{12} \;,
\nonumber
\\
\Delta\sigma \simeq \hspace{-0.6cm}&& -\frac{\Delta^{}_\tau}{2} \zeta^{-1}_{21} \sin^2\theta^{}_{23} \sin^2\theta^{}_{12} \;,
%	(41)
\end{eqnarray}
in the inverted neutrino mass ordering case. In both cases, due to $\mathcal{J} \left( \Lambda \right) = 0$, the term proportional to $\mathcal{J} \left( \Lambda \right)$ in Eq.~(23) exactly vanishes, and thus the Jarlskog invariant is simplified to
\begin{eqnarray}
\mathcal{J} \left( \mu \right) \simeq \frac{1}{8} \sin 2\theta^{}_{12} \sin 2\theta^{}_{13} \cos\theta^{}_{13} \sin 2\theta^{}_{23} \Delta \delta \;,
%	(42)
\end{eqnarray} 
where $C^{(2)}_{\rm M} \simeq 4 \Delta\delta/\Delta^{}_\tau$, and $\Delta \delta$ is given by Eq.~(40) or Eq.~(41) in the NMO or IMO case. 
%It is unnecessary to consider $C^{(1)}_{\rm M}$ due to $\delta \left( \Lambda \right) =0$ (or $\mathcal{J} \left( \Lambda \right) = 0$), and $C^{(2)}_{\rm M} \simeq 4 \Delta\delta/\Delta^{}_\tau$ holds as discussed in section 2.

%%%%%%%%%%%%%%%%%%%%%%%%%%%%%%%%%%%%%% table 3 %%%%%%%%%%%%%%%%%%%%%%%%%%%%%%%%%%
\begin{table}[t!]
	\centering
	\caption{The values of $\Delta \delta \left( \Lambda^{}_{\rm EW} \right)$ and $\mathcal{J} \left( \Lambda^{}_{\rm EW} \right)$ obtained by means of the analytical expressions derived in section 2 for Dirac neutrinos in the MSSM with $\tan\beta =10$ or $30$. The initial inputs at $\Lambda = 10^{14}$ GeV are chosen to guarantee that $m^{}_1 = 0.001$ eV or $m^{}_3 = 0.001$ eV and the best-fit values given in Eqs.~(34) and (35) can be achieved at $\Lambda^{}_{\rm EW} \simeq 100$ GeV. The corresponding numbers given in the parentheses are the relative errors compared to the exact results obtained by numerically solving the one-loop RGEs.}
	\vspace{0.3cm}
	\resizebox{\textwidth}{!}{
		\begin{tabular}{c|p{3.4cm}<{\centering}p{3.4cm}<{\centering}|p{3.4cm}<{\centering}p{3.4cm}<{\centering}}
			\hline\hline
			& \multicolumn{2}{c|}{Normal neutrino mass ordering (NMO)} & \multicolumn{2}{c}{Inverted neutrino mass ordering (IMO)}
			\\
			\hline
			& $\tan\beta = 10$ & $\tan\beta = 30$ & $\tan\beta = 10$ & $\tan\beta = 30$
			\\
			\hline
			\multirow{2}{*}{$\Delta \delta \left( \Lambda^{}_{\rm EW} \right)$} & $-1.47 \times 10^{-4}$ & $ -1.46 \times 10^{-3}$ & $-1.46 \times 10^{-2}$ & $-0.145$
			\\
			& $\left( 3.0 \times 10^{-3} \right)$ & $\left( -1.3 \times 10^{-3} \right)$ & $\left( 4.6 \times 10^{-3} \right)$ & $\left( 2.1 \times 10^{-2} \right)$ 
			\\
			\multirow{2}{*}{$\mathcal{J} \left( \Lambda^{}_{\rm EW} \right)$} & $-2.57 \times 10^{-2}$ & $-2.57 \times 10^{-2}$ & $-3.33 \times 10^{-2}$ & $-3.10 \times 10^{-2}$
			\\
			& $\left( -5.9 \times 10^{-6} \right)$ & $\left( -2.2 \times 10^{-4} \right)$ & $\left( -1.1 \times 10^{-3} \right)$ & $\left( -6.9 \times 10^{-2} \right)$
			\\
			\hline\hline
	\end{tabular}}
	\label{phaseDt}
\end{table}
%%%%%%%%%%%%%%%%%%%%%%%%%%%%%%%%%%%%%%%%%%%%%%%%%%%%%%%%%%%%%%%%%%%%%%%%%%%%%%%%%

\begin{center}
	{\bf (3) $\bm{ \delta \left( \Lambda \right) = \sigma \left( \Lambda \right) = 0 }$}
\end{center}

With $\rho \left( \Lambda^{}_{\rm EW} \right) = \pi/4$, Eqs.~(21)---(22) can be simplified to
\begin{eqnarray}
\Delta\delta \simeq \hspace{-0.6cm}&& \frac{\Delta^{}_\tau}{2} \left( \zeta^{}_{21} - \zeta^{-1}_{21} \right) \sin^2\theta^{}_{23} \;,
\nonumber
\\
\Delta\rho \simeq \hspace{-0.6cm}&& -\frac{\Delta^{}_\tau}{2} \left( \zeta^{}_{21} - \zeta^{-1}_{21} \right) \sin^2\theta^{}_{23} \cos^2\theta^{}_{12} \;,
\nonumber
\\
\Delta\sigma \simeq \hspace{-0.6cm}&& - \frac{\Delta^{}_\tau}{2} \left( \zeta^{}_{21} - \zeta^{-1}_{21} \right) \sin^2\theta^{}_{23} \sin^2\theta^{}_{12} \;,
%	(43)
\end{eqnarray}
in the normal neutrino mass ordering case; and 
\begin{eqnarray}
\Delta\delta \simeq \hspace{-0.6cm}&& - \frac{\Delta^{}_\tau}{2} \zeta^{-1}_{21} \sin^2\theta^{}_{23} \;,
\nonumber
\\
\Delta\rho \simeq \hspace{-0.6cm}&& \frac{\Delta^{}_\tau}{2} \zeta^{-1}_{21} \sin^2\theta^{}_{23} \cos^2\theta^{}_{12} \;,
\nonumber
\\
\Delta\sigma \simeq \hspace{-0.6cm}&& \frac{\Delta^{}_\tau}{2} \zeta^{-1}_{21} \sin^2\theta^{}_{23} \sin^2\theta^{}_{12} \;,
%	(44)
\end{eqnarray}
in the inverted neutrino mass ordering case. In these two cases, the Jarlskog invariant is also determined by Eq.~(42) but $\Delta \delta$ involved in Eq.~(42) now is given by Eq.~(43) or Eq.~(44) in the NMO or IMO case.

For Dirac neutrinos, after some approximations are made, Eqs.~(31)---(33) can be simplified to 
\begin{eqnarray}
\Delta\delta \simeq \hspace{-0.6cm}&& \Delta^{}_\tau \sin\delta \left[ \left( \xi^{}_{32} - \xi^{}_{31} \right) \frac{\sin 2\theta^{}_{12} \sin 2\theta^{}_{23}}{4\sin\theta^{}_{13}}  - \left( \sin 2\theta^{}_{23} + \cos 2\theta^{}_{12} \right) \frac{\sin\theta^{}_{13}}{\sin 2\theta^{}_{12}} \right] \;,
\nonumber
\\
C^{}_{\rm D} \simeq \hspace{-0.6cm}&& 4\left( \cos^2\theta^{}_{23} - \sin^2\theta^{}_{23} \sin^2\theta^{}_{12} \right) \;,
%	(45)
\end{eqnarray}
in the normal neutrino mass ordering case; or
\begin{eqnarray}
\Delta\delta \simeq \hspace{-0.6cm}&& - \Delta^{}_\tau \xi^{}_{21} \frac{\sin\theta^{}_{13} \sin 2\theta^{}_{23}}{\sin 2\theta^{}_{12}} \sin\delta \;,
\nonumber
\\
C^{}_{\rm D} \simeq \hspace{-0.6cm}&& 2 \xi^{}_{21} \sin^2\theta^{}_{23} \cos 2\theta^{}_{12} \;,
%	(46)
\end{eqnarray}
in the inverted neutrino mass ordering case, where the Jarlskog invariant is governed by $\mathcal{J} \left( \mu \right) \simeq \left( 1 - \Delta^{}_\tau C^{}_{\rm D} /2 \right) \mathcal{J} \left( \Lambda \right)$.

Then based on Eqs.~(36)---(46), some discussions and comments on the evolution behaviors of these CP-violating phases and the Jarlskog invariant shown in Figs.~\ref{phaseM} and \ref{phaseD} or Tables~\ref{phaseMt} and \ref{phaseDt} are in order.

\begin{itemize}
	\item Since the RGE running effects are dominated by $\Delta^{}_\tau$, which is approximately proportional to the value of $\tan^2\beta$, in general the results with $\tan\beta = 30$ are nearly ten times larger than the corresponding results with $\tan\beta = 10$ as can be seen from Figs.~\ref{phaseM} and \ref{phaseD} or Tables~\ref{phaseMt} and \ref{phaseDt}. But this does not seem to be true for the results with $\rho \left( \Lambda \right) = \sigma \left( \Lambda \right) = 0$ in the IMO case. The formulas in Eq.~(38), suppressed by $\sin\theta^{}_{13}$, can well describe the evolution of $\Delta \delta$, $\Delta \rho $ and $\Delta \sigma$ with $\tan\beta =10$ in this case, but when $\tan\beta = 30$, an additional term proportional to $\Delta^{}_\tau \zeta^{-1}_{21} \sin 2\left( \rho -\sigma \right)$ gradually dominates the evolution of $\Delta \delta$, $\Delta \rho $ and $\Delta \sigma$ during the running. Since in the IMO case, $\zeta^{-1}_{21}$ is extremely large and $\sin 2\left( \rho -\sigma \right)$ at low energy scales with $\tan\beta = 30$ is not very small, this additional term becomes dominant over the evolution. In addition, it is the largeness of $\zeta^{-1}_{21}$ for Majorana neutrinos or $\xi^{}_{21}$ for Dirac neutrinos in the IMO case, together with the huge value of $\tan\beta$ dominating the strength of RGE running, that makes our approximations worse. Thus the results obtained from the analytical expressions with $\tan\beta = 30$ in the IMO case remarkably deviate from the corresponding exact results, as obviously shown in Figs.~\ref{phaseM} and \ref{phaseD} or Tables~\ref{phaseMt} and \ref{phaseDt}.
	
	\item Comparing the results with the inputs $\rho \left( \Lambda \right) = \sigma \left( \Lambda \right) = 0$ in the Majorana case shown in the first row of Fig.~\ref{phaseM} to those in the Dirac case illustrated in Fig.~\ref{phaseD}, one can find that both the running direction and strength of $\Delta \delta$ are different. These differences between Majorana and Dirac neutrinos in the IMO case are obviously shown by Eqs.~(38) and (46). However, the differences in the NMO case are not obviously indicated by Eqs.~(36) and (45), and each term in Eqs.~(36) and (45) needs to be carefully checked and compared. Nevertheless, one can conclude that the evolution behaviour of the Dirac CP-violating phase $\delta$ with the initially vanishing Majorana CP-violating phases in the Majorana case can be distinguished from that in the Dirac case not only by its RGE running strength but also by its running direction. Additionally, in the Majorana case, the Majorana CP-violating phases can be radiatively generated even they are initially vanishing, namely $\rho \left( \Lambda \right) = \sigma \left( \Lambda \right) = 0$. The evolution behaviors of the Jarlskog invariant $\mathcal{J}$ in the Majorana and Dirac cases are similar but the running effect on $\mathcal{J}$ with the inverted neutrino mass ordering in the Majorana case is stronger than that in the Dirac case due to $\zeta^{-1}_{21} > \xi^{}_{21}$, as indicated by Eqs.~(39) and (46) and clearly shown in Figs.~\ref{phaseM} and \ref{phaseD}.
	
	\item In Fig.~\ref{phaseM} and Table~\ref{phaseMt}, we have shown the results with different inputs for CP-violating phases in the Majorana case. It is evident that these three CP-violating phases are entangled with one another during the RGE running, implying that once there is a nonvanishing phase initially, the other two phases may be generated radiatively via the one-loop RGE running. If one wants to gain strong running effects to generate a sizeable phase, usually a large $\tan\beta$ and the nearly degenerate or inverted neutrino mass hierarchy should be taken into consideration~\cite{Luo:2005sq,Ohlsson:2012pg}, but under this circumstance, our analytical results are poor to describe the evolution of CP-violating phases. With the help of Eqs.~(40)---(44), it is easy to understand the evolution behaviors of three CP-violating phases and the Jarlskog invariant with the Dirac CP-violating phase vanishing initially (i.e., $\delta \left( \Lambda \right) = 0$), shown in the last two rows of Fig.~\ref{phaseM}. One may check that the running directions of three CP-violating phases in the case of $\delta \left( \Lambda \right) = \rho \left( \Lambda \right) = 0$ are all opposite to those in the case of $\delta \left( \Lambda \right) = \sigma \left( \Lambda \right) = 0$. And in the former case, $\Delta \delta < 0$, $\Delta \rho > 0$ and $\Delta \sigma > 0$ hold for both the normal and inverted neutrino mass hierarchies. As for the running strengths of these phases, Eqs.~(40)---(44) indicate that in the IMO case, the absolute values of $\Delta \delta$, $\Delta \rho$ and $\Delta \sigma$ with $\delta \left( \Lambda \right) = \rho \left( \Lambda \right) = 0$ are nearly equal to those with $\delta \left( \Lambda \right) = \sigma \left( \Lambda \right) = 0$; but in the NMO case, the former ones are slightly larger than the latter ones since there is an additional term proportional to $\zeta^{}_{32} - \zeta^{-1}_{32}$ in the formulas given by Eq.~(40) compared with those in Eq.~(43), which can enlarge the absolute values of $\Delta \delta$, $\Delta \rho$ and $\Delta \sigma$. These properties are explicitly illustrated by Fig.~\ref{phaseM} and Table~\ref{phaseMt}. Actually, the sizes of $\Delta \delta$, $\Delta \rho$ and $\Delta \sigma$ in the same case can also be understood by taking advantage of the analytical results given in Eqs.~(40)---(44). For example, in the NMO case with $\delta \left( \Lambda \right) = \sigma \left( \Lambda \right) = 0$, Eq.~(43) tells us that the absolute value of $\Delta \delta$ is the largest among those of $\Delta \delta$, $\Delta \rho$ and $\Delta \sigma$, which are different from one another only by some overall factors (namely $1$ for $\Delta \delta$, $\cos^2\theta^{}_{12}$ for $\Delta \rho$ and $\sin^2\theta^{}_{12}$ for $\Delta \sigma$). Finally, the evolution of $\mathcal{J}$ in the case of $\delta \left( \Lambda \right) = 0$ is controlled by that of $\Delta \delta$, namely $\mathcal{J} \left( \mu \right) \propto \Delta\delta$, as shown in Eq.~(42), indicating that the evolution of $\mathcal{J}$ is similar to $\Delta \delta$ in this case, and both $\delta$ and $\mathcal{J}$ can be radiatively generated by the one-loop RGEs even if they are initially vanishing.
\end{itemize}

\begin{figure}[t!]
	\centering
	\includegraphics[width=\linewidth]{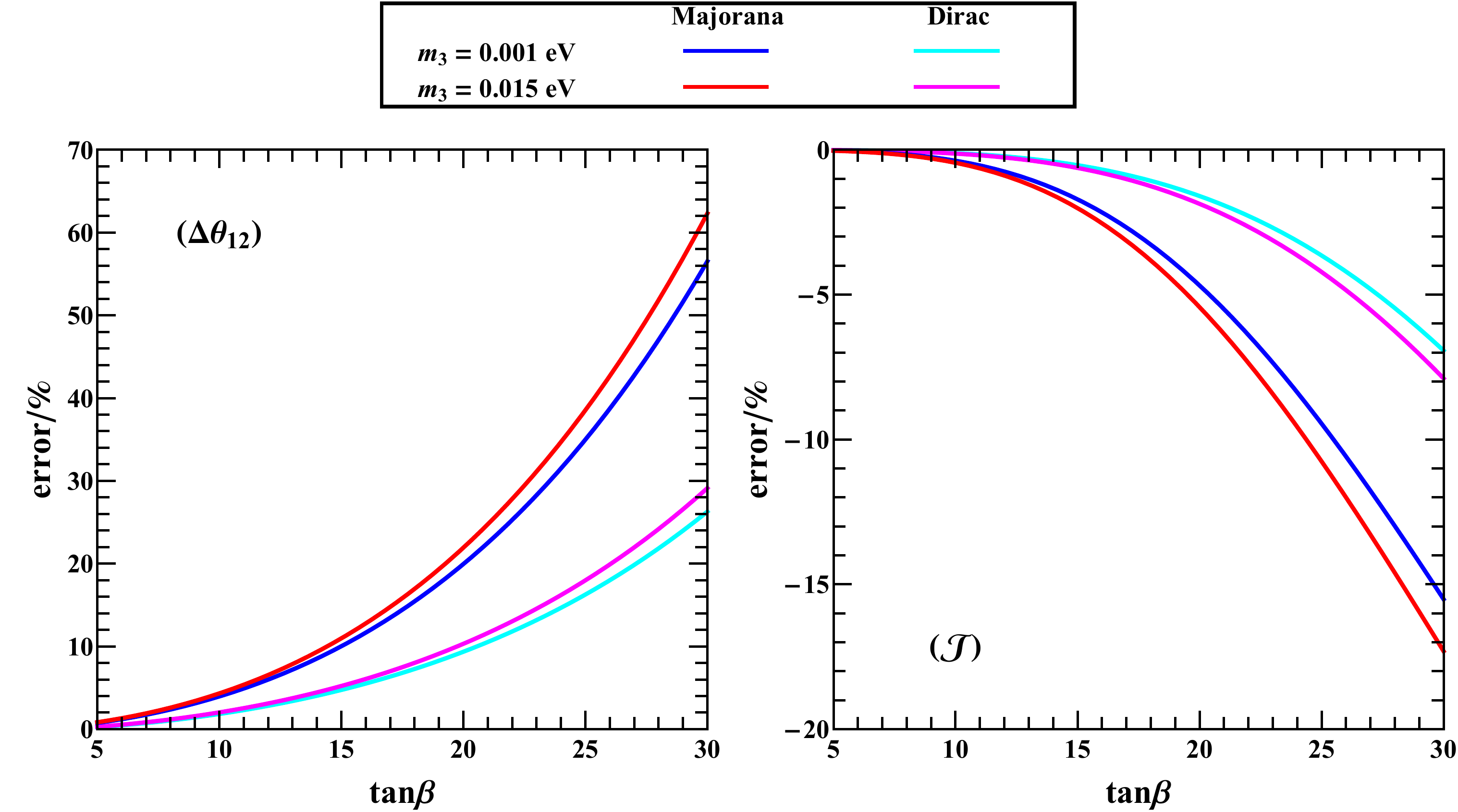}
	\caption{The relative errors for $\Delta \theta^{}_{12}$ and $\mathcal{J}$ at $\Lambda^{}_{\rm EW}$ against $\tan \beta$ for the inverted neutrino mass ordering in the MSSM with $\Lambda=10^{14}$ GeV, where both the Majorana and Dirac cases with $m^{}_3 = 0.001$ eV or $0.015$ eV are considered.}
	\label{etanb}
\end{figure}

For illustration, we repeat all the above numerical calculations with $m^{}_{1} \left( \Lambda^{}_{\rm EW} \right) = 0.03$ eV in the NMO case and $m^{}_{3} \left( \Lambda^{}_{\rm EW} \right) = 0.015$ eV in the IMO case, which are approximately the upper bounds constrained by the Planck data at $95\%$ CL~\cite{Aghanim:2018eyx}. The corresponding values of neutrino masses, flavor mixing angles, CP-violating phases and the Jarlskog invariant at $\Lambda^{}_{\rm EW}$ are listed in Tables~\ref{mass-angle-p}---\ref{phaseDt-p}. The values of $m^{}_i \left( \Lambda^{}_{\rm EW} \right) / m^{}_i \left( \Lambda \right)$ (for $i =1, 2, 3$) dominated by $I^{}_\beta$ ($\beta = \kappa$ or $\nu$) are basically unchanged. The results for flavor mixing angles and CP-violating phases in the NMO case are largely enhanced, since $\zeta^{-1}_{21}$ becomes much larger for $m^{}_{1} \left( \Lambda^{}_{\rm EW} \right) = 0.03$ eV. In comparison, those in the IMO case are only slightly enlarged. As discussed above and shown in Tables~\ref{mass-angle}---\ref{phaseDt-p} or Figs.~\ref{angle}---\ref{phaseD}, the approximate results for $\Delta \theta^{}_{12}$ and $\mathcal{J}$ remarkably deviate from the corresponding exact results in the IMO case especially with a sizeable $\tan \beta$. To make the size of these deviations against $\tan \beta$ more quantitative, we plot the relative errors for $\Delta \theta^{}_{12}$ and $\mathcal{J}$  at $\Lambda^{}_{\rm EW}$ against $\tan \beta$ for the inverted neutrino mass ordering within the MSSM in Fig.~\ref{etanb}, where both the Majorana and Dirac cases are considered and the initial values of relevant parameters at $\Lambda = 10^{14}$ GeV are chosen to achieve the best-fit values for the inverted neutrino mass ordering shown in Eqs.~(34) and (35), $m^{}_3\left(\Lambda^{}_{\rm EW} \right) = 0.001$ eV or $0.015$ eV and $\rho\left(\Lambda^{}_{\rm EW} \right) = \sigma \left(\Lambda^{}_{\rm EW} \right) = 0$. It is obvious that when $\tan \beta \lesssim 15$ in the Majorana case or $\tan \beta \lesssim 20$ in the Dirac case, the relative error for $\Delta \theta^{}_{12}$ is smaller than $10\%$, and similarly the relative error for $\mathcal{J}$ is smaller than $10\%$ when $\tan \beta \lesssim 25$ in the Majorana case or $\tan \beta \lesssim 30$ in the Dirac case.

%%%%%%%%%%%%%%%%%%%%%%%%%%%%%%%% figure 6 %%%%%%%%%%%%%%%%%%%%%%%%%%%%%%%%%%%%%%%
%%%%%%%%%%%%%%%%%%%%%%%%%%%%%%%%%%%%%%%%%%%%%%%%%%%%%%%%%%%%%%%%%%%%%%%%%%%%%%%%%

Before ending this section, it is worth remarking that in order to investigate general evolution behaviors of lepton flavor mixing parameters and the Jarlskog invariant by making full use of the analytical results obtained in section 2, we have only considered the initial inputs which are essentially independent of some specific textures of lepton Yukawa coupling matrices or flavor mixing patterns. These analytical results can certainly be applied to some fantastic models or flavor mixing patterns, such as the well-known tri-bimaximal mixing pattern~\cite{Harrison:2002er,Xing:2002sw,He:2003rm}, democratic mixing pattern~\cite{Fritzsch:1995dj,Fritzsch:1998xs,Fukugita:1998vn} and the $\mu$-$\tau$ reflection symmetry~\cite{Harrison:2002et,Xing:2015fdg}, which have been previously studied either with the differential RGEs~\cite{Mei:2005gp,Plentinger:2005kx,Luo:2005fc,Lin:2009sq} or with the special forms of their integral solutions~\cite{Xing:2015fdg,Xing:2017mkx,Huan:2018lzd,Huang:2020kgt}.

%%%%%%%%%%%%%%%%%%%%%%%%%%%%%%%%%%%%%% table 4 %%%%%%%%%%%%%%%%%%%%%%%%%%%%%%%%%%
\begin{table}[h!]
	\centering
	\caption{The values of $m^{}_i \left( \Lambda^{}_{\rm EW} \right)/ m^{}_i \left( \Lambda \right) $ (for $i=1, 2, 3$) and $\Delta \theta^{}_{ij} \left( \Lambda^{}_{\rm EW} \right)$ (for $ij= 12, 13, 23$) obtained by means of the analytical expressions derived in section 2 in the MSSM with $\tan\beta =10$ or $30$. The initial inputs at $\Lambda = 10^{14}$ GeV are chosen to guarantee that $m^{}_1 = 0.03$ eV or $m^{}_3 = 0.015$ eV and the best-fit values given in Eqs.~(34) and (35) can be achieved at $\Lambda^{}_{\rm EW} \simeq 100$ GeV, and in addition, $\rho = \sigma =0$ are initially input at $\Lambda$ in the Majorana case. The corresponding numbers given in the parentheses are the relative errors compared to the exact results obtained by numerically solving the one-loop RGEs.}
	\vspace{0.3cm}
	\resizebox{\textwidth}{!}{
		\begin{tabular}{c|p{3.4cm}<{\centering}p{3.4cm}<{\centering}|p{3.4cm}<{\centering}p{3.4cm}<{\centering}}
			\hline\hline
			& \multicolumn{2}{c|}{Normal neutrino mass ordering (NMO)} & \multicolumn{2}{c}{Inverted neutrino mass ordering (IMO)}
			\\
			\hline
			& $\tan\beta = 10$ & $\tan\beta = 30$ & $\tan\beta = 10$ & $\tan\beta = 30$
			\\
			\hline
			\multicolumn{5}{c}{Majorana neutrinos}
			\\
			\hline
			\multirow{2}{*}{$m^{}_1 \left( \Lambda^{}_{\rm EW} \right) / m^{}_1 \left( \Lambda \right)$} & $0.878$ & $0.866$ & $0.878$ & $0.867$
			\\
			& $\left( -6.0 \times 10^{-6} \right)$ & $\left( -1.2 \times 10^{-3} \right)$ & $\left( -2.2 \times 10^{-5} \right)$ & $\left( -2.0 \times 10^{-3} \right)$ 
			\\
			\multirow{2}{*}{$m^{}_2 \left( \Lambda^{}_{\rm EW} \right) / m^{}_2 \left( \Lambda \right)$} & $0.878$ & $0.863$ & $0.878$ & $0.862$
			\\
			& $\left( 2.1 \times 10^{-5} \right)$ & $\left( 9.4 \times 10^{-4} \right)$ & $\left( 4.2 \times 10^{-5} \right)$ & $\left( 2.2 \times 10^{-3} \right)$ 
			\\
			\multirow{2}{*}{$m^{}_3 \left( \Lambda^{}_{\rm EW} \right) / m^{}_3 \left( \Lambda \right)$} & $0.878$ & $ 0.861$ & $0.878$ & $0.861$
			\\
			& $\left( 1.4 \times 10^{-5} \right)$ & $\left( 2.7 \times 10^{-4} \right)$ & $\left( 1.0 \times 10^{-5} \right)$ & $\left( -1.7 \times 10^{-4} \right)$ 
			\\
			\multirow{2}{*}{$\Delta \theta^{}_{12} \left( \Lambda^{}_{\rm EW} \right)$} & $ 1.85 \times 10^{-2}$ & $ 0.184$ & $ 4.90 \times 10^{-2}$ & $0.487$
			\\
			& $\left( 8.3 \times 10^{-3} \right)$ & $\left( 0.14 \right)$ & $\left( 4.3 \times 10^{-2} \right)$ & $\left( 0.62 \right)$ 
			\\
			\multirow{2}{*}{$\Delta \theta^{}_{13} \left( \Lambda^{}_{\rm EW} \right)$} & $9.79 \times 10^{-5}$ & $ 9.36 \times 10^{-4}$ & $-4.97 \times 10^{-5}$ & $-5.07 \times 10^{-4}$
			\\
			& $\left( -1.1 \times 10^{-2} \right)$ & $\left( -5.2 \times 10^{-2} \right)$ & $\left( -1.2 \times 10^{-3} \right)$ & $\left( 2.6 \times 10^{-2} \right)$ 
			\\
			\multirow{2}{*}{$\Delta \theta^{}_{23} \left( \Lambda^{}_{\rm EW} \right)$} & $ -2.17 \times 10^{-3}$ & $2.16 \times 10^{-2}$ & $-1.22 \times 10^{-3}$ & $-1.21 \times 10^{-2}$
			\\
			& $\left( 3.1 \times 10^{-3} \right)$ & $\left(-1.5 \times 10^{-4} \right)$ & $\left( 3.7 \times 10^{-3} \right)$ & $\left( 5.8 \times 10^{-3} \right)$
			\\
			\hline
			\multicolumn{5}{c}{Dirac neutrinos}
			\\
			\hline
			\multirow{2}{*}{$m^{}_1 \left( \Lambda^{}_{\rm EW} \right) / m^{}_1 \left( \Lambda \right)$} & $0.937$ & $0.931$ & $0.937$ & $0.931$
			\\
			& $\left(3.6 \times 10^{-7} \right)$ & $\left( -3.4 \times 10^{-4} \right)$ & $\left( -4.3 \times 10^{-6} \right)$ & $\left( -6.8 \times 10^{-4} \right)$ 
			\\
			\multirow{2}{*}{$m^{}_2 \left( \Lambda^{}_{\rm EW} \right) / m^{}_2 \left( \Lambda \right)$} & $0.937$ & $0.929$ & $0.937$ & $0.928$
			\\
			& $\left( 8.2 \times 10^{-6} \right)$ & $\left( 2.8 \times 10^{-4} \right)$ & $\left( 1.4 \times 10^{-5} \right)$ & $\left( 7.5 \times 10^{-4} \right)$ 
			\\
			\multirow{2}{*}{$m^{}_3 \left( \Lambda^{}_{\rm EW} \right) / m^{}_3 \left( \Lambda \right)$} & $0.937$ & $ 0.928$ & $0.937$ & $0.928$
			\\
			& $\left( 6.8 \times 10^{-6} \right)$ & $\left( 8.8 \times 10^{-5} \right)$ & $\left( 5.5 \times 10^{-6} \right)$ & $\left( -4.2 \times 10^{-5} \right)$ 
			\\
			\multirow{2}{*}{$\Delta \theta^{}_{12} \left( \Lambda^{}_{\rm EW} \right)$} & $ 9.26 \times 10^{-3}$ & $ 9.20 \times 10^{-2}$ & $ 2.45 \times 10^{-2}$ & $0.243$
			\\
			& $\left( 3.2 \times 10^{-3} \right)$ & $\left( 6.3 \times 10^{-2} \right)$ & $\left( 2.0 \times 10^{-2} \right)$ & $\left( 0.29 \right)$ 
			\\
			\multirow{2}{*}{$\Delta \theta^{}_{13} \left( \Lambda^{}_{\rm EW} \right)$} & $1.42 \times 10^{-4}$ & $ 1.41 \times 10^{-3}$ & $-1.06 \times 10^{-4}$ & $-1.05 \times 10^{-3}$
			\\
			& $\left( -6.0 \times 10^{-3} \right)$ & $\left( 1.5 \times 10^{-2} \right)$ & $\left( -3.7 \times 10^{-3} \right)$ & $\left( 1.5\times 10^{-3} \right)$ 
			\\
			\multirow{2}{*}{$\Delta \theta^{}_{23} \left( \Lambda^{}_{\rm EW} \right)$} & $ 1.19 \times 10^{-3}$ & $1.18 \times 10^{-2}$ & $-7.96 \times 10^{-4}$ & $-7.91 \times 10^{-3}$
			\\
			& $\left( 3.3 \times 10^{-3} \right)$ & $\left(1.6 \times 10^{-3} \right)$ & $\left( 3.6 \times 10^{-3} \right)$ & $\left( 4.5 \times 10^{-3} \right)$
			\\
			\hline\hline
	\end{tabular}}
	\label{mass-angle-p}
\end{table}
%%%%%%%%%%%%%%%%%%%%%%%%%%%%%%%%%%%%%%%%%%%%%%%%%%%%%%%%%%%%%%%%%%%%%%%%%%%%%%%%%

%%%%%%%%%%%%%%%%%%%%%%%%%%%%%%%%%%%%%% table 5 %%%%%%%%%%%%%%%%%%%%%%%%%%%%%%%%%%
\begin{table}[h!]
	\centering
	\caption{The values of $\Delta \delta \left( \Lambda^{}_{\rm EW} \right)$, $\Delta \rho \left( \Lambda^{}_{\rm EW} \right) $, $\Delta \sigma \left( \Lambda^{}_{\rm EW} \right)$ and $\mathcal{J} \left( \Lambda^{}_{\rm EW} \right)$ obtained by means of the analytical expressions derived in section 2 for Majorana neutrinos in the MSSM with $\tan\beta =10$ or $30$. The initial inputs at $\Lambda = 10^{14}$ GeV are chosen to guarantee that $m^{}_1 = 0.03$ eV or $m^{}_3 = 0.015$ eV and the best-fit values of two neutrino mass-squared differences and three flavor mixing angles given in Eqs.~(34) and (35) can be achieved at $\Lambda^{}_{\rm EW} \simeq 100$ GeV, and in addition, three CP-violating phases are required to be (1) $\delta \left( \Lambda^{}_{\rm EW} \right) = 1.28\pi$ or $1.52\pi$ in the normal or inverted neutrino mass ordering case and $\rho \left( \Lambda \right) = \sigma \left( \Lambda \right) =0$; (2) $\sigma \left( \Lambda^{}_{\rm EW} \right) = \pi/4$ and $\delta \left( \Lambda \right) = \rho \left( \Lambda \right) =0$; (3) $\rho \left( \Lambda^{}_{\rm EW} \right) = \pi/4$ and $\delta \left( \Lambda \right) = \sigma \left( \Lambda \right) =0$. The corresponding numbers given in the parentheses are the relative errors compared to the exact results obtained by numerically solving the one-loop RGEs.}
	\vspace{0.3cm}
	\resizebox{\textwidth}{!}{
		\begin{tabular}{c|ccc|ccc}
			\hline\hline
			& \multicolumn{3}{c|}{$\tan \beta= 10$} &  \multicolumn{3}{c}{$\tan \beta = 30$}
			\\ \hline
			& $\rho \left( \Lambda \right) = \sigma \left( \Lambda \right) = 0$ & $ \delta \left( \Lambda \right) = \rho \left( \Lambda \right) = 0 $ &  $ \delta \left( \Lambda \right) = \sigma \left( \Lambda \right) = 0 $ & $\rho \left( \Lambda \right) = \sigma \left( \Lambda \right) = 0$ & $ \delta \left( \Lambda \right) = \rho \left( \Lambda \right) =0 $ &  $ \delta \left( \Lambda \right) = \sigma \left( \Lambda \right) =0 $
			\\ \hline
			\multicolumn{7}{c}{Normal neutrino mass ordering (NMO)}
			\\ \hline
			\multirow{2}{*}{$\Delta \delta \left( \Lambda^{}_{\rm EW} \right)$} & $-7.53 \times 10^{-4}$ & $-1.97 \times 10^{-2}$ & $1.86 \times 10^{-2}$ & $-5.74\times 10^{-3}$ & $-0.197$ & $0.191$ 
			\\
			& $\left(-3.4\times 10^{-2}\right)$ & $\left(1.4 \times 10^{-3}\right)$ & $\left(1.4 \times 10^{-2} \right)$ & $\left(-0.28\right)$ & $\left(0.15\right)$ & $\left(0.17\right)$  
			\\
			\multirow{2}{*}{$\Delta \rho \left( \Lambda^{}_{\rm EW} \right)$} & $3.77 \times 10^{-4}$ & $1.14 \times 10^{-2}$ & $-1.15 \times 10^{-2}$ & $2.41 \times 10^{-3}$ & $0.117$ & $-0.120$
			\\
			& $\left( -3.6 \times 10^{-2} \right)$ & $\left(1.1 \times 10^{-2} \right)$ & $\left( 1.1 \times 10^{-2} \right)$ & $\left( -0.38 \right)$ & $\left( 0.13 \right)$ & $\left( 0.16 \right)$
			\\ 
			\multirow{2}{*}{$\Delta \sigma \left( \Lambda^{}_{\rm EW} \right)$} & $-1.58 \times 10^{-4}$ & $5.29 \times 10^{-3}$ & $-4.99 \times 10^{-3}$ & $-2.15 \times 10^{-3}$ & $5.35 \times 10^{-2}$ & $-5.19 \times 10^{-2}$
			\\
			& $\left(4.8\times 10^{-2} \right)$ & $\left( 2.6 \times 10^{-2} \right)$ & $\left( 2.6 \times 10^{-2} \right)$ & $\left( 0.44 \right)$ & $\left(0.31 \right)$ & $\left( 0.30 \right)$
			\\ 
			\multirow{2}{*}{$\mathcal{J} \left( \Lambda^{}_{\rm EW} \right)$} & $-2.57 \times 10^{-2}$ & $-6.55\times 10^{-4}$ & $6.20 \times 10^{-4}$ & $-2.42 \times 10^{-2}$ & $-6.17 \times 10^{-3}$ & $6.15 \times 10^{-3}$
			\\
			& $\left( -7.8 \times 10^{-4} \right)$ & $\left( 6.8 \times 10^{-3} \right)$ & $\left( 1.2 \times 10^{-2} \right)$ & $\left( -5.8 \times 10^{-2} \right)$ & $\left( 8.5 \times 10^{-2} \right)$ & $\left( 0.13 \right)$
			\\ \hline
			\multicolumn{7}{c}{Inverted neutrino mass ordering (IMO)}
			\\ \hline
			\multirow{2}{*}{$\Delta \delta \left( \Lambda^{}_{\rm EW} \right)$} & $9.70 \times 10^{-5}$ & $-4.66 \times 10^{-2}$ & $4.71 \times 10^{-2}$ & $-2.74 \times 10^{-3}$ & $-0.432$ & $0.521$
			\\
			& $\left( -0.30 \right)$ & $\left( 4.6 \times 10^{-2} \right)$ & $\left( 4.8 \times 10^{-2} \right)$ & $\left( -3.0 \right)$ & $\left( 0.43 \right)$ & $\left( -0.92 \right)$
			\\ 
			\multirow{2}{*}{$\Delta \rho \left( \Lambda^{}_{\rm EW} \right)$} & $-2.45 \times 10^{-4}$ & $3.34 \times 10^{-2}$ & $-3.35 \times 10^{-2}$ & $1.57 \times 10^{-4}$ & $0.308$ & $-0.367$
			\\
			& $\left( -0.10 \right)$ & $\left( 3.0 \times 10^{-2} \right)$ & $\left( 3.3 \times 10^{-2} \right)$ & $\left( -1.1 \right)$ & $\left( 0.26 \right)$ &  $\left( 0.44 \right)$
			\\ 
			\multirow{2}{*}{$\Delta \sigma \left( \Lambda^{}_{\rm EW} \right)$} & $1.33 \times 10^{-4}$ & $1.44 \times 10^{-2}$ & $-1.46 \times 10^{-2}$ & $2.45 \times 10^{-3}$ & $0.133$ & $-0.161$
			\\
			& $\left( 0.11 \right)$ & $\left( 8.2 \times 10^{-2} \right)$ & $\left( 8.0 \times 10^{-2} \right)$ & $\left( 1.1 \right)$ & $\left(1.0 \right)$ & $\left( 0.86 \right)$
			\\ 
			\multirow{2}{*}{$\mathcal{J} \left( \Lambda^{}_{\rm EW} \right)$} & $-3.32 \times 10^{-2}$ & $-1.53 \times 10^{-3}$ & $1.54 \times 10^{-3}$ & $-2.75 \times 10^{-2}$ & $-1.28 \times 10^{-2}$ & $1.41 \times 10^{-2}$
			\\
			& $\left( -4.5 \times 10^{-3} \right)$ & $\left( 3.1 \times 10^{-2} \right)$ & $\left( 2.9 \times 10^{-2} \right)$ & $\left( -0.17 \right)$ & $\left( 0.29 \right)$ &  $\left( 0.28 \right)$
			\\
			\hline\hline
	\end{tabular}}
	\label{phaseMt-p}
\end{table}
%%%%%%%%%%%%%%%%%%%%%%%%%%%%%%%%%%%%%%%%%%%%%%%%%%%%%%%%%%%%%%%%%%%%%%%%%%%%%%%%%

%%%%%%%%%%%%%%%%%%%%%%%%%%%%%%%%%%%%%% table 6 %%%%%%%%%%%%%%%%%%%%%%%%%%%%%%%%%%
\begin{table}[t!]
	\centering
	\caption{The values of $\Delta \delta \left( \Lambda^{}_{\rm EW} \right)$ and $\mathcal{J} \left( \Lambda^{}_{\rm EW} \right)$ obtained by means of the analytical expressions derived in section 2 for Dirac neutrinos in the MSSM with $\tan\beta =10$ or $30$. The initial inputs at $\Lambda = 10^{14}$ GeV are chosen to guarantee that $m^{}_1 = 0.03$ eV or $m^{}_3 = 0.015$ eV and the best-fit values given in Eqs.~(34) and (35) can be achieved at $\Lambda^{}_{\rm EW} \simeq 100$ GeV. The corresponding numbers given in the parentheses are the relative errors compared to the exact results obtained by numerically solving the one-loop RGEs.}
	\vspace{0.3cm}
	\resizebox{\textwidth}{!}{
		\begin{tabular}{c|p{3.4cm}<{\centering}p{3.4cm}<{\centering}|p{3.4cm}<{\centering}p{3.4cm}<{\centering}}
			\hline\hline
			& \multicolumn{2}{c|}{Normal neutrino mass ordering (NMO)} & \multicolumn{2}{c}{Inverted neutrino mass ordering (IMO)}
			\\
			\hline
			& $\tan\beta = 10$ & $\tan\beta = 30$ & $\tan\beta = 10$ & $\tan\beta = 30$
			\\
			\hline
			\multirow{2}{*}{$\Delta \delta \left( \Lambda^{}_{\rm EW} \right)$} & $-4.28 \times 10^{-3}$ & $ -4.25 \times 10^{-2}$ & $-1.59 \times 10^{-2}$ & $-0.158$
			\\
			& $\left( -5.0 \times 10^{-4} \right)$ & $\left( -3.5 \times 10^{-2} \right)$ & $\left( 4.7 \times 10^{-3} \right)$ & $\left( 2.3 \times 10^{-2} \right)$ 
			\\
			\multirow{2}{*}{$\mathcal{J} \left( \Lambda^{}_{\rm EW} \right)$} & $-2.57 \times 10^{-2}$ & $-2.53 \times 10^{-2}$ & $-3.33 \times 10^{-2}$ & $-3.07 \times 10^{-2}$
			\\
			& $\left( -2.1 \times 10^{-4} \right)$ & $\left( -1.6 \times 10^{-2} \right)$ & $\left( -1.3 \times 10^{-3} \right)$ & $\left( -7.9 \times 10^{-2} \right)$
			\\
			\hline\hline
	\end{tabular}}
	\label{phaseDt-p}
	\vspace{0.5cm}
\end{table}
%%%%%%%%%%%%%%%%%%%%%%%%%%%%%%%%%%%%%%%%%%%%%%%%%%%%%%%%%%%%%%%%%%%%%%%%%%%%%%%%%

\section{Summary}

The RGEs as a powerful tool to establish a link between physical phenomena at the high and low energy scales have been extensively studied. In this work, working in the basis where the charged-lepton Yukawa matrix $Y^{}_l$ is diagonal and with the standard parametrization of the PMNS matrix $U$ given in Eq.~(5), we have analytically derived integral solutions to the one-loop RGEs for neutrino masses, flavor mixing angles, CP-violating phases and the Jarlskog invariant in the SM or MSSM for both Majorana and Dirac neutrinos. In addition, we have also gained some interesting and exact relations between neutrino masses or phases at the superhigh energy scale $\Lambda$ and those at a lower energy scale $\mu$ for Majorana neutrinos at the one-loop level, as well as the reltions between neutrino masses or the Jarlskog invariants at $\Lambda$ and $\mu$ for Dirac neutrinos.

Different from the differential form of RGEs for lepton flavor mixing parameters, our analytical results are of the integral form, which consist of two energy-dependent quantities, namely $I^{}_\beta$ (for $\beta = \kappa$ or $\nu$) and $\Delta^{}_\tau$, and the initial or final values of neutrino masses, flavor mixing angles and CP-violating phases at $\Lambda$ or $\mu$. Thus given the values of $I^{}_\beta$ (for $\beta = \kappa$ or $\nu$) and $\Delta^{}_\tau$ against the energy scale and the initial or final values of relevant flavor parameters, one can easily calculate the RGE effects on these parameters. Moreover, compared with some previous works~\cite{Bergstrom:2010id,Zhou:2014sya,Xing:2015fdg,Xing:2017mkx,Huan:2018lzd,Zhu:2018dvj,Huang:2020kgt,Xing:2020ezi}, we have acquired the most general and complete results for all the lepton flavor mixing parameters including the Jarlskog invariant in the standard parametrization of $U$ for both Majorana and Dirac neutrinos. Therefore, most results of the previous works can be easily reproduced by use of ours derived in this work with some specific assumptions.

We have also carried out the numerical analysis in the MSSM for Majorana and Dirac neutrinos. Both the approximate results calculated by using the analytical formulas and the exact results obtained by numerically solving the one-loop RGEs have been shown. One can see that our approximate results can well describe the evolution behaviors of neutrino masses, flavor mixing angles, CP-violating phases and the Jarlskog invariant in most cases, and thus make them easy to be analytically understood. But in the case of Majorana neutrinos with the inverted neutrino mass ordering and a sizeable $\tan\beta $, our approximate results may deviate from the exact ones to some extent. In particular, with our analytical expressions, the differences between evolution behaviors of lepton flavor mixing parameters for Dirac neutrinos and those for Majorana neutrinos with initially vanishing Majorana CP-violating phases are shown transparently, besides the entanglements among three CP-violating phases for Majorana neutrinos during the RGE running.

It is worth remarking that the integral solutions to the one-loop RGEs not only have great advantages of transparently describing the evolution behaviors of lepton flavor mixing parameters and the Jarlskog invariant, but also can be used to describe explicit radiative corrections to some fantastic flavor mixing patterns or mass textures with flavor symmetries. Of course, it may also be possible to establish an explicit connection between the phenomena of CP violation at low and high energy scales by means of such integral solutions. Thus our results are expected to be useful for model building at a superhigh energy scale so as to understand the true origin of neutrino masses and CP violation at the electroweak scale.

\section*{Acknowledgements}

I am greatly indebted to Prof. Zhi-zhong Xing for carefully reading this manuscript and giving many helpful suggestions. This research work is partly supported by the National Natural Science Foundation of China under grant No. 11775231 and grant No. 11835013.

\renewcommand{\theequation}{\thesection.\arabic{equation}}

\begin{appendices}
\setcounter{table}{0}
\setcounter{equation}{0}

\section{Expansion of the lepton flavor mixing matrix}
For simplicity, we define
\begin{eqnarray}
V \equiv \left(\begin{matrix} c^{}_{12}c^{}_{13} & s^{}_{12}c^{}_{13} & s^{}_{13}e^{-{\rm i}\delta} \cr -s^{}_{12}c^{}_{23} - c^{}_{12}s^{}_{13} s^{}_{23}e^{{\rm i}\delta} & c^{}_{12}c^{}_{23} - s^{}_{12}s^{}_{13}s^{}_{23} e^{{\rm i}\delta} & c^{}_{13}s^{}_{23} \cr s^{}_{12}s^{}_{23} - c^{}_{12} s^{}_{13}c^{}_{23}e^{{\rm i}\delta} & -c^{}_{12}s^{}_{23} - s^{}_{12}s^{}_{13} c^{}_{23}e^{{i}\delta} & c^{}_{13}c^{}_{23} \end{matrix}\right) \;.
%	(A1)
\end{eqnarray} 
Then the PMNS matrix $U$ can be written as $U=P^{}_l V P^{}_\nu$ in the Majorana case or $U = P^\prime V$ with $P^\prime \equiv {\rm Diag} \{ e^{{\rm i} \phi^{}_1}, e^{{\rm i} \phi^{}_2}, 1 \}$ in the Dirac case. We have defined the difference between $U \left( \Lambda \right)$ and $U \left( \mu \right)$ as $\Delta U$. Therefore, at the leading order, we have 
\begin{eqnarray}
\Delta U \simeq P^{}_l \Delta V P^{}_\nu + \Delta P^{}_l V P^{}_\nu + P^{}_l V \Delta P^{}_\nu \;,
%	(A2)
\end{eqnarray}
in the Majorana case; and 
\begin{eqnarray}
\Delta U \simeq P^\prime \Delta V + \Delta P^\prime V \;,
%	(A3)
\end{eqnarray}
in the Dirac case, where $V$, $P^{}_l$, $P^{}_\nu$ and $P^\prime$ are all at $\mu$, and $\Delta V \equiv V \left( \mu \right) - V \left( \Lambda \right)$, $\Delta P^{}_l \equiv P^{}_l \left( \mu \right) - P^{}_l \left( \Lambda \right)$, $\Delta P^{}_\nu \equiv P^{}_\nu \left( \mu \right) - P^{}_\nu \left( \Lambda \right)$ and $\Delta P^\prime \equiv P^\prime \left( \mu \right) - P^\prime \left( \Lambda \right)$ are defined. We can simultaneously expand $V\left( \Lambda \right)$, $P^{}_l \left( \Lambda \right)$ and $P^{}_\nu \left( \Lambda \right)$ or $V \left( \Lambda \right)$ and $P^\prime \left( \Lambda \right)$ with respect to the quantities defined in Eq.~(13) or Eq.~(14) and only keep the leading order terms. Then we obtain
\begin{eqnarray}
\Delta P^{}_l \simeq {\rm i}\left(\begin{matrix} \Delta \phi^{}_e e^{{\rm i} \phi^{}_e} & & \cr & \Delta \phi^{}_\mu e^{{\rm i} \phi^{}_\mu} & \cr & & \Delta \phi^{}_\tau e^{{\rm i} \phi^{}_\tau} \end{matrix} \right) \;, \quad \Delta P^{}_\nu \simeq {\rm i} \left(\begin{matrix} \Delta \rho e^{{\rm i} \rho} & & \cr & \Delta \sigma e^{{\rm i} \sigma} & \cr & & 0 \end{matrix}\right)\;,
%	(A4)
\end{eqnarray}
or 
\begin{eqnarray}
\Delta P^\prime \simeq {\rm i} \left(\begin{matrix} \Delta \phi^{}_1 e^{{\rm i} \phi^{}_1} & & \cr & \Delta \phi^{}_2 e^{{\rm i} \phi^{}_2} & \cr & & 0 \end{matrix} \right) \;,
%	(A5)
\end{eqnarray}
and the elements for $\Delta V$, that is
\begin{eqnarray}
\left( \Delta V \right)^{}_{11} = \hspace{-0.6cm}&& -\Delta \theta^{}_{12} s^{}_{12} c^{}_{13} - \Delta \theta^{}_{13} c^{}_{12} s^{}_{13} \;,
\nonumber
\\
\left( \Delta V \right)^{}_{12} = \hspace{-0.6cm}&& \Delta \theta^{}_{12} c^{}_{12} c^{}_{13} - \Delta \theta^{}_{13} s^{}_{12} s^{}_{13} \;,
\nonumber
\\
\left( \Delta V \right)^{}_{13} = \hspace{-0.6cm}&& \left( \Delta \theta^{}_{13} c^{}_{13} - {\rm i} \Delta \delta s^{}_{13} \right) e^{-{\rm i}\delta} \;,
\nonumber
\\
\left( \Delta V \right)^{}_{21} = \hspace{-0.6cm}&& \Delta \theta^{}_{12} \left( -c^{}_{12} c^{}_{23} + s^{}_{12} s^{}_{13} s^{}_{23} e^{{\rm i}\delta} \right) - \Delta \theta^{}_{13} c^{}_{12} c^{}_{13} s^{}_{23} e^{{\rm i} \delta} + \Delta \theta^{}_{23} \left( s^{}_{12} s^{}_{23} - c^{}_{12} s^{}_{13} c^{}_{23} e^{{\rm i} \delta} \right)
\nonumber
\\
\hspace{-0.6cm}&& - {\rm i} \Delta\delta c^{}_{12} s^{}_{13} s^{}_{23} e^{{\rm i}\delta} \;,
\nonumber
\\
\left( \Delta V \right)^{}_{22} = \hspace{-0.6cm}&& -\Delta \theta^{}_{12} \left( s^{}_{12} c^{}_{23} + c^{}_{12} s^{}_{13} s^{}_{23} e^{{\rm i}\delta} \right) - \Delta \theta^{}_{13} s^{}_{12} c^{}_{13} s^{}_{23} e^{{\rm i} \delta} - \Delta \theta^{}_{23} \left( c^{}_{12} s^{}_{23} + s^{}_{12} s^{}_{13} c^{}_{23} e^{{\rm i} \delta} \right)
\nonumber
\\
\hspace{-0.6cm}&& - {\rm i} \Delta\delta s^{}_{12} s^{}_{13} s^{}_{23} e^{{\rm i}\delta} \;,
\nonumber
\\
\left( \Delta V \right)^{}_{23} = \hspace{-0.6cm}&& - \Delta\theta^{}_{13} s^{}_{13} s^{}_{23} + \Delta\theta^{}_{23} c^{}_{13} c^{}_{23} \;,
\nonumber
\\
\left( \Delta V \right)^{}_{31} = \hspace{-0.6cm}&& \Delta \theta^{}_{12} \left( c^{}_{12} s^{}_{23} + s^{}_{12} s^{}_{13} c^{}_{23} e^{{\rm i}\delta} \right) - \Delta \theta^{}_{13} c^{}_{12} c^{}_{13} c^{}_{23} e^{{\rm i} \delta} + \Delta \theta^{}_{23} \left( s^{}_{12} c^{}_{23} + c^{}_{12} s^{}_{13} s^{}_{23} e^{{\rm i} \delta} \right)
\nonumber
\\
\hspace{-0.6cm}&& - {\rm i} \Delta\delta c^{}_{12} s^{}_{13} c^{}_{23} e^{{\rm i}\delta} \;,
\nonumber
\\
\left( \Delta V \right)^{}_{32} = \hspace{-0.6cm}&& \Delta \theta^{}_{12} \left( s^{}_{12} s^{}_{23} - c^{}_{12} s^{}_{13} c^{}_{23} e^{{\rm i}\delta} \right) - \Delta \theta^{}_{13} s^{}_{12} c^{}_{13} c^{}_{23} e^{{\rm i} \delta} - \Delta \theta^{}_{23} \left( c^{}_{12} c^{}_{23} - s^{}_{12} s^{}_{13} s^{}_{23} e^{{\rm i} \delta} \right)
\nonumber
\\
\hspace{-0.6cm}&& - {\rm i} \Delta\delta s^{}_{12} s^{}_{13} c^{}_{23} e^{{\rm i}\delta} \;,
\nonumber
\\
\left( \Delta V \right)^{}_{33} = \hspace{-0.6cm}&& - \Delta\theta^{}_{13} s^{}_{13} c^{}_{23} - \Delta\theta^{}_{23} c^{}_{13} s^{}_{23} \;,
%	(A6)
\end{eqnarray}
where $\theta^{}_{ij}$ (for $ij=12, 13, 23$), $\delta$, $\rho$, $\sigma$, $\phi^{}_\alpha$ (for $\alpha = e, \mu, \tau$) and $\phi^{}_i$ (for $i = 1, 2$) all take their values at the energy scale $\mu$.

\setcounter{table}{0}
\setcounter{equation}{0}
\section{Analytical results for unphysical phases}
In the Majorana case, the evolution of three unphysical phases is given by
\begin{eqnarray}
\Delta\phi^{}_e \simeq \hspace{-0.6cm}&& \frac{\Delta^{}_\tau}{2} \left\{ \left(\zeta^{}_{21} - \zeta^{-1}_{21} \right) \sin 2\left( \rho-\sigma \right) \left( \cos^2\theta^{}_{13} \cos^2\theta^{}_{23} - \cos 2\theta^{}_{23} - \sin\theta^{}_{13} \sin 2\theta^{}_{23} \cot 2\theta^{}_{12} \cos\delta \right)  \right.
\nonumber
\\
\hspace{-0.6cm}&& - \sin\theta^{}_{13} \sin 2\theta^{}_{23} \csc 2\theta^{}_{12} \sin\delta \left[ \zeta^{}_{21} \cos^2\left( \rho -\sigma \right) + \zeta^{-1}_{21} \sin^2\left( \rho -\sigma \right) \right] + \left( \zeta^{}_{31} - \zeta^{-1}_{31} \right)
\nonumber
\\
\hspace{-0.6cm}&& \times \left[ \sin^2\theta^{}_{12} \cos 2\theta^{}_{23} \sin 2\rho - \sin^2\theta^{}_{13} \cos 2\theta^{}_{12} \cos^2\theta^{}_{23} \sin 2\left( \delta + \rho \right) \right] + \left( \zeta^{}_{32} - \zeta^{-1}_{32} \right)  
\nonumber
\\
\hspace{-0.6cm}&& \times \left[ \cos^2\theta^{}_{12} \cos 2\theta^{}_{23} \sin 2\sigma + \sin^2\theta^{}_{13} \cos 2\theta^{}_{12} \cos^2\theta^{}_{23} \sin 2\left( \delta + \sigma \right) \right] + \sin 2\theta^{}_{12} \sin\theta^{}_{13}
\nonumber
\\
\hspace{-0.6cm}&& \times \cos 2\theta^{}_{23} \cot\theta^{}_{23}  \left[ \zeta^{}_{32} \cos\sigma \sin\left( \delta + \sigma \right) - \zeta^{}_{32} \sin\sigma \cos\left( \delta + \sigma \right) - \zeta^{}_{31} \cos\rho \sin\left( \delta + \rho \right) \right.
\nonumber
\\
\hspace{-0.6cm}&& + \left.   \zeta^{-1}_{31} \sin \rho \cos\left( \delta + \rho \right) \right] + \sin\theta^{}_{13} \sin 2\theta^{}_{23} \cos 2\theta^{}_{12}  \left[ \tan\theta^{}_{12} \left( \zeta^{}_{31}  \sin\rho \cos\left( \delta + \rho \right) \right.\right. 
\nonumber
\\
\hspace{-0.6cm}&& - \left.\left.\left. \hspace{-0.1cm} \zeta^{-1}_{31}  \cos\rho \sin\left( \delta + \rho \right) \right) + \cot\theta^{}_{12} \left( \zeta^{}_{32}  \sin\sigma \cos\left( \delta + \sigma \right) - \zeta^{-1}_{32} \cos\sigma \sin \left( \delta + \sigma \right) \right) \right] \right\} \;,
\nonumber
\\
\Delta\phi^{}_\mu \simeq \hspace{-0.6cm}&& \frac{\Delta^{}_\tau}{2} \left\{ \left( \zeta^{}_{31} - \zeta^{-1}_{31} \right) \cos^2\theta^{}_{23} \left[ \sin^2\theta^{}_{12} \sin 2\rho - \sin^2\theta^{}_{13} \cos^2\theta^{}_{12} \sin 2\left( \delta + \rho \right) \right] + \left( \zeta^{}_{32} - \zeta^{-1}_{32} \right) \right.
\nonumber
\\
\hspace{-0.6cm}&& \times \cos^2\theta^{}_{23} \left[ \cos^2\theta^{}_{12} \sin 2\sigma - \sin^2\theta^{}_{12} \sin^2\theta^{}_{13} \sin 2\left( \delta + \sigma \right) \right] + \frac{1}{2} \sin 2\theta^{}_{12} \sin\theta^{}_{13} \cot\theta^{}_{23}
\nonumber
\\
\hspace{-0.6cm}&& \times \left[ \left( \zeta^{}_{32} - \zeta^{-1}_{32} \right) \cos 2\theta^{}_{23} \sin\left( \delta + 2\sigma \right) - \left( \zeta^{}_{31} - \zeta^{-1}_{31} \right) \cos 2\theta^{}_{23} \sin\left( \delta + 2\rho \right) \right. 
\nonumber
\\
\hspace{-0.6cm}&& \left.\left. + \left( \zeta^{}_{32} + \zeta^{-1}_{32} -\zeta^{}_{31} - \zeta^{-1}_{31} \right) \sin\delta \right]\right\} \;,
\nonumber
\\
\Delta\phi^{}_\tau \simeq \hspace{-0.6cm}&& - \frac{\Delta^{}_\tau}{2} \left\{ \sin^2\theta^{}_{13} \cos^2\theta^{}_{23} \left[ \left( \zeta^{}_{31} - \zeta^{-1}_{31} \right) \cos^2\theta^{}_{12} \sin 2\left( \delta + \rho \right) + \left( \zeta^{}_{32} - \zeta^{-1}_{32} \right) \sin^2\theta^{}_{12} \sin 2\left( \delta + \sigma \right) \right] \right.
\nonumber
\\
\hspace{-0.6cm}&& + \sin^2\theta^{}_{23} \left[ \left( \zeta^{}_{31} - \zeta^{-1}_{31} \right) \sin^2\theta^{}_{12} \sin 2\rho + \left( \zeta^{}_{32} - \zeta^{-1}_{32} \right) \cos^2\theta^{}_{12} \sin 2\sigma \right] - \frac{1}{2} \sin 2\theta^{}_{12} \sin\theta^{}_{13}
\nonumber
\\
\hspace{-0.6cm}&& \times \left. \sin 2\theta^{}_{23} \left[ \left( \zeta^{}_{31} - \zeta^{-1}_{31} \right) \sin\left( \delta + 2\rho \right) - \left( \zeta^{}_{32} - \zeta^{-1}_{32} \right) \sin\left( \delta + 2\sigma \right) \right]  \right\} \;;
%	(B1)
\end{eqnarray}
and in the Dirac case, the results for two unphysical phases are
\begin{eqnarray}
\Delta\phi^{}_1 \simeq \hspace{-0.6cm}&& -\frac{\Delta^{}_\tau}{2} \sin\theta^{}_{13} \sin\delta \left[\xi^{}_{21} \frac{2\sin 2\theta^{}_{23}}{\sin 2\theta^{}_{12}} + \left( \xi^{}_{31} - \xi^{}_{32} \right) \sin 2\theta^{}_{12} \cot\theta^{}_{23} \right.
\nonumber
\\
\hspace{-0.6cm}&& - \left. \sin 2\theta^{}_{23} \left( \xi^{}_{31} \tan\theta^{}_{12} - \xi^{}_{32} \cot\theta^{}_{12} \right)  \vphantom{\frac{1}{1}} \right] \;,
\nonumber
\\
\Delta\phi^{}_2 \simeq \hspace{-0.6cm}&& -\frac{\Delta^{}_\tau}{2} \sin\delta \left[ 2\xi^{}_{21} \cos\theta^{}_{12} \tan\theta^{}_{13}  + \left( \xi^{}_{31} - \xi^{}_{32} \right) \sin 2\theta^{}_{12} \sin\theta^{}_{13} \cot\theta^{}_{23}  \right] \;,
%	(B2)
\end{eqnarray}
where $\theta^{}_{ij}$ (for $ij=12, 13, 23$), $\delta$, $\rho$, $\sigma$ and $\zeta^{}_{ij}$ or $\xi^{}_{ij}$ (for $ij = 21, 31, 32$) all take their values at $\mu$.

\end{appendices}

\end{document}